\documentclass[11pt,a4paper]{article}
\usepackage[T1]{fontenc}

\usepackage[dvipdfmx]{graphicx}
\usepackage{bmpsize}
\usepackage{jheppub}

\usepackage{comment}
\usepackage{here}

\newcommand{\al}[1]{\begin{align}#1\end{align}}

\newcommand{\bp}{\begin{pmatrix}}
\newcommand{\ep}{\end{pmatrix}}
\newcommand{\bb}{\begin{bmatrix}}
\newcommand{\eb}{\end{bmatrix}}

\newcommand{\fulltoday}{\number\day\space \ifcase\month\or
    January\or February\or March\or April\or May\or June\or
    July\or August\or September\or October\or November\or December\fi
    \space\number\year}

 \makeatletter
    
    \@addtoreset{equation}{section}
  \makeatother

\allowdisplaybreaks[3]

\usepackage{calc}
\newcounter{hours}\newcounter{minutes}
\makeatletter                   
\renewcommand*{\thehours}{\two@digits\c@hours}
\renewcommand*{\theminutes}{\two@digits\c@minutes}
\makeatother

\definecolor{mygrn}{rgb}{0,0.5,0}

\title{\boldmath Simultaneous interpretation of
$K$ and $B$ anomalies 
in terms of chiral-flavorful vectors}

\author[a,b]{Shinya Matsuzaki,}
\author[c,d]{Kenji Nishiwaki,}
\author[b,e,f,1]{Kei Yamamoto,\note{Corresponding author.}}

\affiliation[a]{Center for Theoretical Physics and College of Physics, Jilin University, Changchun, 130012, China.}
\affiliation[b]{Department of Physics, Nagoya University, Nagoya 464-8602, Japan.}
\affiliation[c]{School of Physics, Korea Institute for Advanced Study~(KIAS), Seoul 02455, Republic of Korea.}
\affiliation[d]{Ru{\dj}er Bo{\v{s}}kovi{\'c} Institute, Division of Theoretical Physics, Bijenicka cesta 54, 10000 Zagreb, Croatia.}
\affiliation[e]{Graduate School of Science, Hiroshima University, Higashi-Hiroshima 739-8526, Japan.}
\affiliation[f]{Physik-Institut, Universit\"at  Z\"urich, CH-8057 Z\"urich, Switzerland.}

\emailAdd{synya@hken.phys.nagoya-u.ac.jp}
\emailAdd{knishiw@irb.hr}
\emailAdd{keiy@hiroshima-u.ac.jp}

\abstract{
We address the presently reported significant flavor anomalies 
in the Kaon and $B$ meson systems such as the CP violating Kaon
decay ($\epsilon'/\epsilon$) and 
lepton-flavor universality violation in
$B$ meson decays ($R_{K^{(*)}},{R_{D^{(*)}}}$),  
by proposing flavorful and 
chiral vector bosons as the new physics constitution at around TeV scale.  
The chiral-flavorful vectors (CFVs) are introduced as a 63-plet of 
the global $SU(8)$ symmetry, identified as 
the one-family symmetry for left-handed quarks and leptons 
in the standard model (SM) forming the 8-dimensional vector. 
Thus the CFVs include massive gluons, vector leptoquarks, and 
$W',Z'$-type bosons, which are allowed to have 
flavorful couplings with left-handed quarks and leptons, and flavor-universal 
couplings to right-handed ones, where the latter arises from 
mixing with the SM gauge bosons. 
The flavor texture 
is assumed to possess  
a ``minimal'' structure to be consistent with the current flavor measurements 
on the $K$ and $B$ systems.
Among the presently reported significant flavor anomalies in the Kaon and $B$ meson systems 
($\epsilon'/\epsilon$, $R_{K^{(*)}}, {R_{D^{(*)}}}$), 
the first two $\epsilon'/\epsilon$ and $R_{K^{(*)}}$ anomalies can simultaneously be interpreted by the presence of CFVs;
the ${R_{D^{(*)}}}$ anomaly is predicted 
not to survive, due to the approximate $SU(8)$ flavor symmetry.
Remarkably, we find that 
as long as 
both of the $\epsilon'/\epsilon$ and $R_{K^{(*)}}$ anomalies 
persist beyond the SM,   
the CFVs 
predict the enhanced $K^+ \to \pi^+ \nu \bar{\nu}$ and $K_L \to \pi^0 \nu \bar{\nu}$ 
decay rates compared to the SM values, which will readily be explored  
by the NA62 and KOTO experiments, and they will also be explored 
in new resonance searches at the Large Hadron Collider.
}

\subheader{Report number: KIAS-P18045}

\begin{document}

\maketitle
\flushbottom


\section{Introduction}
Flavor physics is one of the most powerful probes for physics beyond the standard model (SM).
Recently, some discrepancies with the SM are reported in several observables in Kaon and $B$ meson decays.
Among them, 
one of the flavor changing neutral current 
observables,  the direct CP violation in $K \to \pi\pi$ decays, $\epsilon'/\epsilon$, 
has recently been paid a strong attention to 
because of the significant discrepancy between the SM prediction
\cite{Buras:2015yba, Kitahara:2016nld}
{~\footnote{
The results in the Dual QCD approach are supported by RBC-UKQCD lattice collaboration 
\cite{Buras:2015xba,Buras:2016fys}.
On the other hand, the study in the chiral perturbation theory predicts a consistent value 
with the experimental value \cite{Gisbert:2017vvj}.
}
using the first lattice calculation result reported by RBC-UKQCD collaboration \cite{Bai:2015nea} 
and the experimental data. 
The SM prediction can be estimated as \cite{Kitahara:2016nld}
$ 
\left(\epsilon'/\epsilon \right)_{\rm SM} = 
 (1.06 \pm 5.07) \times 10^{-4}
$,    
 which is around $2.8\sigma$ below 
the experimental data~\cite{Batley:2002gn,AlaviHarati:2002ye,Abouzaid:2010ny,Patrignani:2016xqp}, 
$ 
\left(
\epsilon'/\epsilon \right)_{\rm exp} = (16.6 \pm 2.3) \times 10^{-4}
$. 
In addition, 
the recent data on semi-leptonic $B$ decays also report the discrepancies.
Lepton-flavor universality (LFU) violating
observables 
$R_{K^{(*)}}={\text{Br}}[{\bar B}\to K^{(*)} \mu^+\mu^-]/{\text{Br}}[{\bar B}\to K^{(*)} e^+e^-]$ 
have been measured at around  $2.1$-$2.6 \sigma$ away from the SM prediction~\cite{Aaij:2014ora, Aaij:2017vbb}.
The discrepancies have been seen also in 
$R_{D^{(*)}}={\text{Br}}[B \to D^{(*)} \tau {\bar \nu}]
/{\text{Br}}[B \to D^{(*)} \ell {\bar \nu}]$ {(with 
$\ell$ standing for $e$ or $\mu$)}~{\cite{Lees:2012xj,Lees:2013uzd,Aaij:2015yra,Huschle:2015rga,Sato:2016svk,Hirose:2016wfn,Aaij:2017uff}}. 
Those flavor anomalies have nowadays made a fuss and urged theorists to make 
 viable conjectures and interpretations for the flavor structure hidden in 
a possible new physics (NP).

In this paper, we propose a new conjecture to simultaneously   
address these flavor anomalies in the Kaon and $B$ meson systems: 
it is a chiral-flavorful structure characterized by the presence of 
flavorful and chiral vector bosons (CFVs) as the new physics constitution at around TeV scale. 
The CFVs are introduced as adjoint representation of $SU(8)$ global symmetry, 
which is gauged by the left-handed gauges in the SM. 
Hence the CFVs can be classified into a generic set of vectors;  
massive gluon $G'$-like, vector leptoquark-like, $W'$ and $Z'$-like 
vectors~\footnote{
Such CFVs can be generated as composite particles 
arising due to an underlying strongly coupled (a hidden QCD) dynamics as proposed 
in~\cite{Matsuzaki:2017bpp}.
}.   

We find the characteristic feature 
in the present CFV scenario based on the one-family $SU(8)$ 
symmetry, by which the predictions in flavor physics 
are derived necessarily with a significant correlation 
between the $2 \leftrightarrow 3$ and $1 \leftrightarrow 2$ 
generation-transition processes.  
It is shown that there are allowed parameter regions 
which can address both of the discrepancies in $\epsilon'/\epsilon$ and in $R_{K^{(*)}}$, 
consistently with the several constraints on the flavor texture we adopt. 
New sizable enough contributions to $\epsilon'/\epsilon$ are 
produced from $G'$ and $Z'$-like CFVs 
via the $I=0$ amplitude through the QCD penguin {operators} and 
the $I=2$ one through the electroweak (EW) penguin {operators} {(with $I$ distinguishing isospin states)}, respectively, 
while those to the $R_{K^{(*)}}$ arise from the $Z'$-type and vector-leptoquark-type 
CFVs. 
Intriguingly enough as well, the CFVs do not give significant effect on 
the $R_{D^{(*)}}$ due to the one-family $SU(8)$ symmetry.

The CFVs turn out to also give the nontrivial predictions to 
the Kaon rare decays
${\text{Br}}[K_L \to \pi^{0} \nu \overline{\nu}]$ and ${\text{Br}}[K^+ \to \pi^{+} \nu \overline{\nu}]$, arising from the $Z'$-type and vector-leptoquark-type exchanges 
along with the strong correlation with the presence of the $R_{K^{(*)}}$ anomaly. 
These predictions will be explored by the NA62 and KOTO experiments, 
significantly in the correlation with the fate of $B$ anomalies as well as 
new-vector resonance searches at {the Large Hadron Collider (LHC)}, 
to be tested in the future high-luminosity phase.

This paper is structured as follows.  
In section~\ref{sec:CFV} we introduce the CFV model together with the proposed flavor texture 
and the couplings to 
the SM fermions based on the $SU(8)$ symmetry structure. 
Section~\ref{sec:flavor} provides the CFVs contributions to the flavor physics, 
including the $K$ and $B - \tau$ systems, and place the constraints 
on the model-parameter space, showing 
that the presently reported 
$\epsilon'/\epsilon$ and $R_{K^{(*)}}$
anomalies can 
be simultaneously accounted for 
by the CFVs, while the $R_{D^{(*)}}$ anomaly is predicted not to survive, due to the approximate $SU(8)$ flavor symmetry.
With the current phenomenological bounds at hand, 
in section~\ref{sec:future} we discuss the future prospect for the CFV scenario, 
aiming at the NA62 and KOTO experiments, in correlation 
with the fate of the $B$ anomalies in the future. 
Expected LHC signatures specific to the presence of the CFV resonances 
are also addressed. 
Finally, section~\ref{sec:summary} {is devoted} to summary and several discussions including 
the theoretical uncertainties in the present analysis. 
Explicit coupling formulae for the CFVs to the SM fermions 
are provided in appendix~\ref{rho-repr}, and effective-four fermion interaction forms 
relevant to the flavor physics study are given in appendix~\ref{4-fermi}.
In appendix~\ref{sec:global-fit}, we discuss a global significance of a bench mark point where
the best-fit value for $b \to s \mu^+ \mu^-$ is achieved and
the $\epsilon'/\epsilon$ anomaly can be addressed within $1\sigma$ confidence level (C.L.) consistently.

\section{Chiral-flavorful vectors}
\label{sec:CFV}

In this section we introduce the CFVs (hereafter symbolically denoted as $\rho$) 
and their generic interaction 
properties for the SM particles.

\subsection{The flavorful couplings}

The CFVs ($\rho$) couplings to the left-handed fermions in the SM are constructed in 
the one-family global-$SU(8)$ symmetric way as 
\begin{align} 
{\cal L}_{\rho f_Lf_L} 
&
{
= \sum_{i,j = 1}^{3} 
g_{\rho L}^{ij} \overline{f}_L^i \gamma^\mu 
\rho_\mu f_L^j 
}
\,, \label{rhoLff}
\end{align}
where $g_{\rho L}^{ij}$ denotes the {(hermitian)} couplings with the generation indices ($i,j$) {in the gauge eigenbases},  
and 
$f_L^i$ includes the {left-handed SM doublet quarks} ($q^{ic}_L=(u^{ic},d^{ic})_L^T$ 
with the QCD color index $c=r,g,b$) 
and {(left-handed)} lepton doublets ($l^i_L=(\nu^i, e^i)_L^T$) for the $i$th generation, which forms 
the 8-dimensional vector (the fundamental representation of the $SU(8)$) 
like
$f_L^i = (q^{i r}, q^{i g},q^{i b}, l^i)_L^T$.  
The CFV fields are embedded in the $8 \times 8$ matrix of the 
$SU(8)$ adjoint representation: $\rho_\mu = \sum_{A=1}^{63} \rho_\mu^A T^A$, 
where the $T^A$ stands for the $SU(8)$ generators, which are explicitly given 
in appendix~\ref{rho-repr} (Eqs.(\ref{A1})-(\ref{rho:assign})). 
To manifestly keep the SM gauge invariance in the coupling form of Eq.(\ref{rhoLff}), 
the CFVs are allowed to couple to the SM gauge fields, through the 
gauging of the global $SU(8)$ symmetry. 
It is reflected by the covariant 
derivative  
\begin{align} 
 D_\mu \rho_\nu = \partial_\mu \rho_\nu - i [{\cal V}_\mu, \rho_\nu] 
 \,, \label{D}
\end{align}
where the SM gauge fields $(G_\mu, W_\mu, B_\mu)$ for the 
$SU(3)_c \times SU(2)_W \times U(1)_Y$ symmetry 
are embedded in the $8 \times 8$ matrix form of 
${\cal V}_\mu$ as 
\begin{align} 
{\cal V}_\mu = 
\left( 
\begin{array}{c|c} 
{{\bf 1}_{2 \times 2} \otimes g_s G_\mu^a \frac{\lambda^a}{2}}
+ {\left( g_W  W_\mu \tau^{\alpha} 
+  \frac{1}{6} g_Y B_\mu \right) \otimes {\bf 1}_{3\times 3}}
& {{\bf 0}_{6 \times 2}} \\ 
\hline 
{{\bf 0}_{2 \times 6}} 
& 
g_W W_\mu^{{\alpha}} \tau^{{\alpha}} 
- \frac{1}{2} g_Y B_\mu \cdot {\bf 1}_{2 \times 2} 
\end{array}
\right) 
\,, \label{Vcal}
\end{align}
where $\lambda^a$ and $\tau^\alpha \equiv \sigma^\alpha/2$ $(\alpha=1,2,3)$ 
are Gell-Mann and (normalized) Pauli matrices, and 
$g_s, g_W$ and $g_Y$ the corresponding gauge couplings.  
One can easily check that the way of embedding the SM gauges in Eq.(\ref{Vcal}) 
manifestly ensures the SM gauge invariance when the CFVs couple to quarks 
and leptons as in Eq.(\ref{rhoLff}).  
It is then convenient to classify the CFVs ($\rho$) in the $SU(8)$ adjoint representation 
by the QCD charges as 
\begin{equation} 
{\rho = 
\left( 
\begin{array}{cc} 
(\rho_{QQ})_{6\times 6} & {(\rho_{QL})_{6 \times 2}} \\ 
{(\rho_{LQ})_{2 \times 6}} & (\rho_{LL})_{2 \times 2} 
\end{array}
\right)}
\,,    
\end{equation}      
where $\rho_{QQ}, \rho_{QL}(=\rho_{LQ}^\dag)$, and $\rho_{LL}$ include 
color-octet $\rho_{(8)}$ (of ``massive gluon $G'$ type''), 
-triplet $\rho_{(3)}$ (of ``vector-leptoquark type'')
\footnote{
Since the present CFV model includes the vector-leptoquarks on the 
TeV mass scale ($\rho^0_{(3)}$ and $\rho^\alpha_{(3)}$), which are flavor non-universally
coupled to quarks and leptons, one might worry about the rapid proton decay.  
{Here we therefore 
briefly make a comment on the proton decay via the vector leptoquarks in our scenario. 
The SM quantum numbers of $\rho^0_{(3)}$ and $\rho^\alpha_{(3)}$ are assigned as 
$(SU(3),SU(2),U(1))=(\bf{3},\bf{1},2/3),(\bf{3},\bf{3},2/3)$, which respectively correspond to $U_1$ and $U_3$ in the notation of the review~\cite{Dorsner:2016wpm}.
As mentioned in~\cite{Dorsner:2016wpm}, leptoquarks have a well-defined quantum number $F = 3B + L$ ($B/L$: the baryon/lepton number of the quark/lepton) and $\rho^0_{(3)}$ and $\rho^\alpha_{(3)}$ possess $F=0$, 
which means that they cannot form dimension-four diquark interactions.
Up to dimension-five operators, as discussed in~\cite{Assad:2017iib}, the two gauge invariant operators under the SM gauge interactions
with the Higgs doublet $H$,
$(\overline{q^c_L} H^\dagger) \gamma^\mu d_R \rho^0_{(3)\mu}/\Lambda$ and $(\overline{q^c_L} \tau^\alpha H^\dagger) \gamma^\mu d_R \rho^\alpha_{(3)\mu}/\Lambda$, seem to provide diquark interactions. 
However, in the hidden local symmetry approach~\cite{Bando:1984ej,Bando:1985rf,Bando:1987ym,Bando:1987br,Harada:2003jx} 
 on which the present CFV model is formulated based (as will be mentioned 
 in the next subsection), 
those operators and the resultant rapid proton decay 
can be prohibited,  
as in the case of the literature~\cite{Matsuzaki:2017bpp} (c.f., {Ref.}~\cite{Fukano:2015zua}).
Thereby, the leading effect to the proton decay would arise from  
dimension-six four-fermion operators, 
where we easily ensure a sufficiently long lifetime of the proton 
without fine tuning for the coefficients of four-fermion operators. 
Note also that the vector-leptoquarks could arise as composite particles, 
as in~\cite{Matsuzaki:2017bpp}, hence are embedded into an 
ultraviolet completed and asymptotic free gauge theory.
Thus, it would be possible to take $\sim 10^{16}\,\text{GeV}$ 
or higher as a cutoff scale (see {Ref.}~\cite{Matsuzaki:2017bpp}) for the present scenario.  
Thus, the CFV model can safely avoid the rapid proton decay.}
}, 
and -singlet $\rho_{(1)^{(\prime)}}$ (of ``$W'$ and/or $Z'$ type''), which can further be 
classified by the weak isospin charges ($\pm, 3$ for triplet and $0$ for singlet).  
Thus, decomposing the CFVs with respect to the SM charges, we find 
\begin{align}
 \rho_{QQ} = 
 &
 \left[ \sqrt 2 \rho_{(8)a}^{{\alpha}} \left( {\tau^\alpha} \otimes {\lambda^{{a}} \over 2} \right) 
 + {1 \over \sqrt 2} \rho_{(8)a}^0 \left( {\bf 1}_{2\times2} \otimes {\lambda^{{a}} \over 2} \right) \right] \notag \\[0.5em]
 & 
 + \left[ {1 \over 2} \rho_{(1)}^{{\alpha}} \left( {\tau^\alpha} \otimes {\bf 1}_{3\times3} \right) 
 + {1 \over 2\sqrt 3} \rho_{(1)'}^{{\alpha}} \left( {\tau^\alpha} \otimes {\bf 1}_{3\times3} \right) 
 + {1\over 4 \sqrt 3} \rho_{(1)'}^0 \Big( {\bf 1}_{2\times2} \otimes {\bf 1}_{3\times3} \Big) \right] \,, \notag \\[1em] 
 \rho_{LL} = 
 &
 {1 \over 2} \rho_{(1)}^{{\alpha}} \left( {\tau^\alpha} \right) - {\sqrt 3 \over 2} \rho_{(1)'}^{{\alpha}} \left( {\tau^\alpha} \right) - {\sqrt 3 \over 4} \rho_{(1)'}^0 \Big( {\bf 1}_{2\times2} \Big) \,, \notag \\[1em]
 \rho_{QL} = 
 &
 \rho_{(3){c}}^{{\alpha}} \left( {\tau^\alpha} \otimes {\bf e}_c \right) + {1 \over 2} \rho_{(3) {c}}^0 \Big( {\bf 1}_{2\times2} \otimes {\bf e}_c \Big) \,, \notag \\[1em]
 {\rho_{LQ}} = & \Big( {\rho_{QL}} \Big)^\dag \,, 
 \label{rho:assignment}  
\end{align} 
(for more details, see appendix~\ref{rho-repr} (Eqs.(\ref{A1})-(\ref{rho:assign}))),
 where ${\bf e}_c$ denotes the 3-dimensional eigenvector in the color space. 
The explicit expressions of the 
flavor-dependent CFV couplings to the SM fermions are listed in appendix~\ref{rho-repr} 
(Eqs.(\ref{eq:decrhouu1})-(\ref{eq:decrhouu8})).

\subsection{The flavor-universal couplings induced from vectorlike mixing with the
SM gauge bosons} 
As seen in the above, 
the one-family global $SU(8)$ symmetry is of course explicitly 
broken by the SM gauge interactions through the gauging in Eq.(\ref{D}), 
hence the CFV fields ($\rho$) generically 
mix with the SM-gauge boson fields (${\cal V}$). 
The interaction terms can 
arise as the form of the kinetic term mixing like   
\begin{align} 
{\cal L}_{\rho {\cal V}} 
= - \frac{1}{g_\rho} {\rm tr}[{\cal V}_{\mu\nu} \rho^{\mu\nu}] 
\,, \label{V-rho}
\end{align}
where ${\cal V}_{\mu\nu}= \partial_\mu {\cal V}_\nu - \partial_\nu {\cal V}_\mu 
- i [{\cal V}_\mu, {\cal V}_\nu]$ and 
$\rho_{\mu\nu}= D_\mu \rho_\nu - D_\nu \rho_\mu$.   
The $g_\rho$ has been introduced as 
the mixing strength common for all the SM gauges, as if 
a remnant of the $SU(8)$ symmetry {were} reflected, 
which can be justified when the CFVs are associated with 
gauge bosons of the spontaneously broken 
gauge symmetry, as in the case of the 
hidden-local symmetry approach~\cite{Bando:1984ej,Bando:1985rf,Bando:1987ym,Bando:1987br,Harada:2003jx}.   
The mixing in Eq.(\ref{V-rho}) induces the flavor-universal couplings 
for the CFVs to {both of the left-handed and right-handed quarks/leptons}, 
which scale as $\sim (g_{s,W,Y}^2/g_\rho)$.  
Since 
the electroweak precision tests  
have severely constrained the fermion couplings to new resonances, 
as a safety setup we may take the mixing strength $(1/g_\rho)$ to be 
much smaller than ${\cal O}(1)$, say 
\begin{align} 
g_\rho \sim {10} 
\,, \label{grho:size}
\end{align} 
which turns out to be 
a consistent size also for the flavor physics bound, as will be seen later. 
In that case, the mass shift among the CFVs arising from the mixing 
with the SM gauge bosons can be {safely neglected} (which is maximally about 5\% correction), so that the CFVs are almost degenerated to have the $SU(8)$ invariant mass $m_\rho$,  
which is set to be of order of TeV. 
Noting that at the on-shell of the CFVs, 
the mixing term in Eq.(\ref{V-rho}) gives rise to the mass mixing form, 
\begin{align} 
- \frac{2 m_\rho^2}{g_\rho} {\rm tr}[{\cal V}_\mu \rho^\mu],  
\label{mass-mixing} 
\end{align}
in which 
the ${\cal V}_\mu$ as in Eq.(\ref{Vcal}) couples with the SM fermion currents, 
one finds the perturbatively small and flavor-universal 
couplings, as listed in appendix~\ref{rho-repr} (Eq.(\ref{indirect:rho-coupling})).

\subsection{The flavor-texture {ansatz}} 
\label{subsection:texture}

Now we introduce the flavored texture for the $g_{\rho L}^{ij}$ in Eq.(\ref{rhoLff}) so that 
 the present flavor anomalies in $K$ and $B$ meson systems can be addressed. 
The proposed texture goes like~\footnote{
In the present study, we will not specify the origin of the 
flavor texture, though it might be 
derived by assuming some discrete symmetry among fermions, and so forth. 
} 
 \begin{equation} 
 g_{\rho L}^{ij} 
 = 
  {\begin{pmatrix} 
  0 & g_{ \rho L}^{12} & 0 \\ 
  (g_{\rho L}^{12})^* & 0 & 0 \\ 
  0 & 0 & g_{\rho L}^{33}  
  \end{pmatrix}^{ij} \,, }
\label{grhoL}
\end{equation} 
in which the hermiticity in the Lagrangian of Eq.(\ref{rhoLff})  
has been taken into account (i.e. $g_{\rho L}^{21} = (g_{\rho L}^{12})^*$ 
and $(g_{\rho L}^{33})^* = g_{\rho L}^{33}$). 
The size of the real part for $g_{\rho L}^{12}$ actually turns out to be constrained severely 
by the Kaon system measurements such as 
the indirect CP violation $\epsilon_K$, 
and $K_L \to \mu^+ \mu^-$, to be extremely tiny $(\lesssim {\cal O}(10^{-6}))$ 
(for instance, see Ref.~\cite{Buras:2015jaq}). 
In contrast, its imaginary part can be moderately larger, which 
will account for the presently reported $\epsilon'/\epsilon$ anomaly 
(deviated by about 3 sigma~\cite{Batley:2002gn,AlaviHarati:2002ye,Abouzaid:2010ny,Patrignani:2016xqp}).   
Hence we will take it to be pure imaginary: 
\begin{align} 
{\rm Re} [g_{\rho L}^{12}]=0\,, \qquad 
{\rm Im} [g_{\rho L}^{12}]  \equiv + g_{\rho L}^{12}
\qquad 
{\rm with} \quad  g_{\rho L}^{12} \in {\bf R}
\,
\,, 
\end{align} 
by which the new physics contributions will be vanishing for 
the $\epsilon_K$ and ${\rm Br}[K_L \to \mu^+ \mu^-]$ 
(for explicit formulae about those observables, e.g. 
see Refs.~\cite{Buras:2014sba,Buras:2015jaq}).

The base transformation among the {gauge- and flavor-eigenstates}
 can be made by rotating fields as 
(under {the} assumption that neutrinos are massless) 
 \al{
{(u_L)^i   = U^{iI} (u'_L)^I,\quad
(d_L)^i   = D^{iI} (d'_L)^I,\quad
(e_L)^i   = L^{iI} (e'_L)^I,\quad
(\nu_L)^i = L^{iI} (\nu'_L)^I,}
	\label{eq:definition_masseigenstates}
}
where $U$, $D$ and $L$ stand for $3 \times 3$ 
unitary matrices and the spinors with the prime symbol 
denote the fermions in the mass basis, which are specified by    
the capital Latin indices $I$ and $J$. 
The {Cabibbo--Kobayashi--Maskawa (CKM)} matrix is then given by $V_{\text{CKM}} \equiv U^\dag D$~\footnote{
The mixing between the CFVs and SM gauge bosons through 
Eq.(\ref{V-rho}) would actually generate corrections to the $V_{\rm CKM}$ 
by amount of ${\cal O}(m_W^2/m_\rho^2)$, which can be, however, neglected 
as long as the CFV mass is on the order of TeV. }. 
As in the literature~\cite{Bhattacharya:2016mcc}, 
to address several flavor anomalies recently reported 
in the measurements such as $R_{K^{(*)}}$ and $R_{D^{(*)}}$  
as well as to avoid severe constraints from flavor-changing neutral current processes 
among the first and second generations, 
we may take 
the mixing structures of $D$ and $L$ as 
\al{
D =
\begin{pmatrix}
{1} & 0 & 0 \\
0 &  \cos{\theta_D} & \sin{\theta_D} \\
0 & -\sin{\theta_D} & \cos{\theta_D} 
\end{pmatrix}, 
\quad\quad
L =
\begin{pmatrix}
{1} & 0 & 0 \\
0 &  \cos{\theta_L} & \sin{\theta_L} \\
0 & -\sin{\theta_L} & \cos{\theta_L}
\end{pmatrix},
\label{LDrotation}
} 
where we {recall} that 
the up-quark mixing matrix is automatically determined through $V_{\text{CKM}} = U^\dag D$.
In discussing the flavor phenomenologies, 
the following forms are useful to understand the flavor structure in $g_{\rho L}^{ij}$ (with $c \equiv g_{\rho L}^{12}/g_{\rho L}^{33} \in {\bf R}$): 
\al{
{X_{dd}} &\equiv
D^\dagger 
\begin{pmatrix} 0 & i \cdot c & 0 \\ 
- i\cdot c & 0  & 0 \\ 0 & 0 & 1 \end{pmatrix} D
= 
\begin{pmatrix}
0 & i \cdot c \cdot \cos\theta_D & i \cdot c \cdot \sin\theta_D \\
- i \cdot c \cdot \cos\theta_D &  \sin^2{\theta_D} & -\cos{\theta_D}\sin{\theta_D} \\
- i \cdot c \cdot \sin\theta_D & -\cos{\theta_D}\sin{\theta_D} & \cos^2{\theta_D}
\end{pmatrix}, \label{eq:M_transform_1} \\
{X_{ll}} &\equiv
L^\dagger \begin{pmatrix} 0 & i \cdot c & 0 \\ 
- i\cdot c & 0  & 0 \\ 0 & 0 & 1 \end{pmatrix} L
=
\begin{pmatrix}
0 & i \cdot c \cdot \cos\theta_L & i \cdot c \cdot \sin\theta_L \\
- i \cdot c \cdot \cos\theta_L &  \sin^2{\theta_L} & -\cos{\theta_L}\sin{\theta_L} \\
- i \cdot c \cdot \sin\theta_L & -\cos{\theta_L}\sin{\theta_L} & \cos^2{\theta_L}
\end{pmatrix}, \label{eq:M_transform_2} \\
{X_{uu}} &\equiv
U^\dagger \begin{pmatrix} 0 & i \cdot c & 0 \\ 
- i\cdot c & 0  & 0 \\ 0 & 0 & 1 \end{pmatrix} U
=
V_{\text{CKM}}
\begin{pmatrix}
0 & i \cdot c \cdot \cos\theta_D & i \cdot c \cdot \sin\theta_D \\
- i \cdot c \cdot \cos\theta_D &  \sin^2{\theta_D} & -\cos{\theta_D}\sin{\theta_D} \\
- i \cdot c \cdot \sin\theta_D & -\cos{\theta_D}\sin{\theta_D} & \cos^2{\theta_D}
\end{pmatrix}
V_{\text{CKM}}^\dagger, \label{eq:M_transform_3} \\
{X_{ld}} &\equiv
L^\dagger \begin{pmatrix} 0 & i \cdot c & 0 \\ 
- i\cdot c & 0  & 0 \\ 0 & 0 & 1 \end{pmatrix} D
=
\begin{pmatrix}
0 &  i \cdot c \cdot \cos\theta_D  & i \cdot c \cdot \sin_D \\
- i \cdot c \cdot \cos\theta_L &  \sin{\theta_L}\sin{\theta_D} & -\sin{\theta_L}\cos{\theta_D} \\
- i \cdot c \cdot \sin\theta_L & -\cos{\theta_L}\sin{\theta_D} &  \cos{\theta_L}\cos{\theta_D}
\end{pmatrix}, \label{eq:M_transform_4} \\
{X_{lu}} &\equiv
L^\dagger \begin{pmatrix} 0 & i \cdot c & 0 \\ 
- i\cdot c & 0  & 0 \\ 0 & 0 & 1 \end{pmatrix} U
=
\begin{pmatrix}
0 &  i \cdot c \cdot \cos\theta_D  & i \cdot c \cdot \sin_D \\
- i \cdot c \cdot \cos\theta_L &  \sin{\theta_L}\sin{\theta_D} & -\sin{\theta_L}\cos{\theta_D} \\
- i \cdot c \cdot \sin\theta_L & -\cos{\theta_L}\sin{\theta_D} &  \cos{\theta_L}\cos{\theta_D}
\end{pmatrix} V_{\text{CKM}}^\dagger,
\label{eq:M_transform_5} \\
{X_{ud}} &\equiv
U^\dagger \begin{pmatrix} 0 & i \cdot c & 0 \\ 
- i\cdot c & 0  & 0 \\ 0 & 0 & 1 \end{pmatrix}D
=
V_{\text{CKM}} D^\dag \begin{pmatrix} 0 & i \cdot c & 0 \\ 
- i\cdot c & 0  & 0 \\ 0 & 0 & 1 \end{pmatrix} D 
\notag \\ 
& = 
V_{\text{CKM}}
\begin{pmatrix}
0 & i \cdot c \cdot \cos\theta_D & i \cdot c \cdot \sin\theta_D \\
- i \cdot c \cdot \cos\theta_D &  \sin^2{\theta_D} & -\cos{\theta_D}\sin{\theta_D} \\
- i \cdot c \cdot \sin\theta_D & -\cos{\theta_D}\sin{\theta_D} & \cos^2{\theta_D}
\end{pmatrix}.
\label{eq:M_transform_6}
}
Note that in the limit where the mixing angles go to 0 or $\pi/2$,
the $2 \leftrightarrow 3$ transitions vanish, while
the $1 \leftrightarrow 2$ and $1 \leftrightarrow 3$ transitions remain accompanied by the factor $c$.\footnote{
We note that flavor-conserving interactions appear from the mixing between $V_\text{SM}$ and $\rho$,
which play important roles in $\epsilon'/\epsilon$ and the bound from the LHC (see appendix~\ref{rho-repr} for details).
}

\section{Contributions to 
flavor physics induced from CFV exchanges} 
\label{sec:flavor}

In this section we shall discuss the flavor physics constraints 
on the CFV-induced four-fermion contributions. 
The flavored-heavy CFVs exchanges generate effective four-fermion 
interactions at low-energy {${E_{\text{ref}}} \ll m_\rho={\cal O}({\rm TeV})$}. 
There the left-handed {current-current} interactions arise 
from both the flavorful coupling $g_{\rho L}^{ij}$ in Eq.(\ref{grhoL}) 
and flavor-universal couplings induced by mixing with the SM gauge bosons 
given in Eq.(\ref{indirect:rho-coupling}), 
while the right-handed {current-current} interactions only from 
the latter ones. 
Since the right-handed couplings are generated by mixing with 
the hypercharge gauge boson, having the form like $g_Y^2/g_\rho$ which 
is smaller than the left-handed coupling $g_W^2/g_\rho$, 
we may neglect the right-handed current interactions and keep only 
the leading order terms with respect to the gauge coupling expansion 
in evaluating the flavor physics contributions.

The effective four-fermion operators constructed from the left-handed {current-current} interactions   
are listed in appendix~\ref{4-fermi} (Eqs.(\ref{B5})-(\ref{eq:effective_operators})). 
Hereafter we shall consider a limit where 
all the CFVs are degenerated to have the $SU(8)$ invariant mass $m_\rho$, 
\begin{align} 
 M_{\rm CFVs} \simeq m_\rho 
 \,, 
 \label{degene}
\end{align}
namely, assuming that the possible split size by 
{$(g_{s,W,Y}/g_\rho) \, m_\rho$} is negligibly small due to the large $g_\rho$ 
as in Eq.(\ref{grho:size}).  
Nevertheless, the $1/g_\rho$ mixing term will significantly 
contribute to the flavor physics as long as the size 
of $g_\rho$ is less than a naive perturbative bound $\sim 4\pi$.~\footnote{
The present CFV model-setup is actually similar to 
the model proposed in Ref.~\cite{Matsuzaki:2017bpp}, 
in which the mixing strength with the SM gauge bosons, 
set by {$g_{s,W,Y}/g_\rho$}, was assumed to be ideally small 
(i.e. $g_\rho$ is {much} larger than the perturbative value $\sim 4 \pi$ 
because of the nonperturbative underlying dynamics). 
In the present model, the size of $g_\rho$ is taken to be 
$< 4 \pi$, so that 
the flavor-universal contribution will play somewhat a significant 
role in discussing the flavor limits, as will {be} seen in the later subsequent 
sections{.}}

\subsection{Flavor changing processes converting the third and second generations \label{sec:2and3-transition}}

The CFVs generate nonzero flavor-changing neutral-current 
contributions to the $b$ -- $s$ transition system and the lepton flavor violation
regarding the {third generation} charged leptons. 
Specifically, the relevant processes are:  
\begin{itemize}
 \item $\mathcal O_{2q2\ell (n)}$: $\bar B \to D^{(*)} \tau\bar\nu$ (only for $n=3$), $\bar B \to K^{(*)} \mu^+\mu^-$, $\bar B \to K^{(*)} \nu\bar\nu$, and $\tau\to\phi\mu$, 
 \item $\mathcal O_{4q(n)}$~~: $B_s^0$--$\bar B_s^0$ mixing, 
 \item $\mathcal O_{4\ell(n)}$~~:  $\tau\to 3\mu$,
\end{itemize}
{where $n$ stands for types of the contraction of {EW-$SU(2)$} indices shared by fermion fields in 
 dimension-six operators (see Eq.(\ref{B5})).}
These processes can be evaluated through the effective Hamiltonians for 
$b \to s \ell^{+} \ell^{-}$, $b \to s \nu \overline{\nu}$, $b \to c \tau^- \overline{\nu}$, $\tau \to \mu s \overline{s}$, $B^0_s \,(=s\overline{b}) \leftrightarrow \overline{B^0_s} \,(=b\overline{s})$, and $\tau^- \to \mu^- \mu^+ \mu^-$. 
\al{
{\cal H}_{\text{eff}}(b \to s e_I \overline{e}_J) &= - \frac{\alpha G_F}{\sqrt{2}\pi} V_{tb} V^\ast_{ts}
	\left[
	C_9^{IJ} (\overline{s'}_L \gamma_\mu b'_L) (\overline{e'}_{I} \gamma^\mu e'_{J}) +
	C_{10}^{IJ} (\overline{s'}_L \gamma_\mu b'_L) (\overline{e'}_{I} \gamma^\mu \gamma_5 e'_{J})
	\right], \\[0.5em]
{\cal H}_{\text{eff}}(b \to s \nu_I \overline{\nu}_J) &= - \frac{\alpha G_F}{\sqrt{2}\pi} V_{tb} V^\ast_{ts}
	C_L^{IJ} (\overline{s'}_L \gamma_\mu b'_L) (\overline{\nu'}_{I} \gamma^\mu (1-\gamma_5) \nu'_{J}), \\[0.5em]
{\cal H}_{\text{eff}}(b \to c \tau_I \overline{\nu}_J) &=  \frac{4 G_F}{\sqrt{2}} V_{cb}
	C_V^{IJ} (\overline{c'}_L \gamma_\mu b'_L) (\overline{e'}_{I} \gamma^\mu \nu'_{J}), \\ 
{\cal H}_{\text{eff}}(\tau \to \mu s \overline{s}) 
&= C_{\tau \mu}^{ss} 
	(\overline{\mu'}_L \gamma_\mu \tau'_L) (\overline{s'}_L \gamma^\mu s'_L), \\[5pt]
{\cal H}_{\text{eff}}(bs \leftrightarrow bs) 
&=	C_{bs}^{bs} 
	(\overline{s'}_L \gamma_\mu b'_L) (\overline{s'}_L \gamma^\mu b'_L), \\
\label{CFV-bsbs}
{\cal H}_{\text{eff}}(\tau^- \to \mu^- \mu^+ \mu^-) 
&=	C_{\tau  \mu}^{\mu\mu} 
	(\overline{\mu'}_L \gamma_\mu \tau'_L) (\overline{\mu'}_L \gamma^\mu \mu'_L), 
}   
where $\alpha$ and $G_F$ are the QED fine structure constant and 
the Fermi constant, respectively; 
the Wilson coefficients 
include both of the SM and the NP contributions like 
$C_X = C_X(\text{SM}) + C_X(\text{NP})$;   
the prime symbol attached on fermion fields stands for the mass eigenstates. 
As noted above, the CFVs dominantly couple to the left-handed currents, 
so that approximately enough, we have   
\al{
C_9^{IJ}(\text{NP}) = - C_{10}^{IJ}(\text{NP}).
}

Using Eq.(\ref{eq:definition_masseigenstates}) and Eq.(\ref{eq:effective_operators}) in appendix~\ref{4-fermi},  
we find the concrete expression for the NP contributions to the Wilson coefficients :  
\al{
C_9^{IJ}(\text{NP}) &= {-}
	\frac{\pi}{\sqrt{2} \alpha G_F V_{tb} V^\ast_{ts}}
	\left( C^{[1]}_{q_iq_j l_k l_l} + C^{[3]}_{q_i q_j l_k l_l} \right)\cdot 
	D^{\dag 2 i} D^{j3} L^{\dag I k} L^{l J}  
	\left( = - C_{10}^{IJ}(\text{NP}) \right), \\[0.5em]
C_L^{IJ}(\text{NP}) &= {-}
	\frac{\pi}{\sqrt{2} \alpha G_F V_{tb} V^\ast_{ts}}
	\left( C^{[1]}_{q_iq_j l_k l_l} - C^{[3]}_{q_i q_j l_k l_l} \right)\cdot 
	D^{\dag 2 i} D^{j3} L^{\dag I k} L^{l J},   \\[0.5em]
C_V^{IJ}(\text{NP}) &= 
	\frac{1}{2\sqrt{2} G_F V_{cb}}
	\left( 2 C^{[3]}_{q_i q_j l_k l_l} \right) 
	\cdot 
	U^{\dag 2 i} D^{j3} L^{\dag I k} L^{l J}  , \\  
C_{\tau \mu}^{ss} (\text{NP}) 
& = \left( C^{[1]}_{q_iq_j l_k l_l} + C^{[3]}_{q_i q_j l_k l_l} \right)\cdot 
	L^{\dag 2 k} L^{l 3} D^{\dag 2 i} D^{j 2}  , \\ 
C_{bs}^{bs}  (\text{NP}) 
&=	\left( C^{[1]}_{q_iq_j q_k q_l} + C^{[3]}_{q_i q_j q_k q_l} \right)\cdot 
	D^{\dag 2 i} D^{j 3} D^{\dag 2 k} D^{l 3},   \\ 
C_{\tau \mu}^{\mu\mu}  (\text{NP}) 
&=	2 \left( C^{[1]}_{l_i l_j l_k l_l} + C^{[3]}_{l_i l_j l_k l_l} \right)\cdot 
	L^{\dag 2 i} L^{j 3} L^{\dag 2 k} L^{l 2} .
}

Here, we comment on the renormalization group effects. The chirality-conserving dimension-six semi-leptonic operators (and also fully leptonic operators) take null effects from the QCD running of the Wilson coefficients due to the current conservation (see e.g., \cite{Gonzalez-Alonso:2017iyc}), while we need to 
take into account {of} nontrivial QCD-running effects on 
the fully hadronic ones in our estimation for the bound from the $B^0_s$-$\bar{B}^0_s$ mixing, as we will see 
in section~\ref{sec:B_s-mixing}. 
A short comment on the QED running effects 
will be provided in the summary section. 
In doing numerical analyses,  
for the SM gauge couplings  
we thus use the values evaluated at two-loop level, computed 
from the $Z$-boson mass scale values by running up to $m_\rho=1$ TeV,    
$g_Y^2(m_\rho) \simeq 0.129$, 
$g_W^2(m_\rho) \simeq 0.424$ 
and $g_s^2(m_\rho) \simeq 1.11$,  
with use of the electromagnetic couplings renormalized at the $Z$-boson mass scale  
($m_Z\simeq {91.2}$ GeV~\cite{Patrignani:2016xqp}), 
$\alpha(m_Z) = g_Y^2(m_Z) c_W^2/(4\pi) \simeq 1/128$~\cite{Patrignani:2016xqp} 
and the ($Z$-mass shell) Weinberg angle 
quantity $c_W^2 = m_W^2/m_Z^2 \simeq 0.778$, and 
the QCD coupling $\alpha_s(m_Z) = g_s^2/(4\pi) 
\simeq 0.118$~\cite{Patrignani:2016xqp}.
In addition, 
the pole mass of the top quark and the Higgs mass are selected as $173.15\,\text{GeV}$ and $125\,\text{GeV}$, 
respectively, and 
we refer to Refs.~\cite{Arason:1991ic,Ford:1992pn,Hamada:2012bp} for the form and formalism regarding the two-loop beta functions
(also to \cite{Degrassi:2012ry} for the boundary conditions).

\subsubsection{$\bar B \to K^{(*)} \ell^+\ell^-$}

Including the SM contributions, 
we write the effective Hamiltonian for $b \to s \ell^+\ell^-$, 
\begin{align}
{\cal H}_{\text{eff}}(b \to s \ell_{I}\bar\ell_{J}) 
 = 
 & - {\alpha G_F \over \sqrt 2 \pi} V_{tb} V_{ts}^* 
 \left( C_9^\text{SM} \delta^{IJ} + C_9^{IJ}(\text{NP}) \right) 
 \left(\bar s_L' \gamma^\mu b_L' \right) \left( \bar\ell_{I}' \gamma_\mu \ell_{J}' \right)  \notag \\
 & - {\alpha G_F \over \sqrt 2 \pi} V_{tb} V_{ts}^* 
 \left( C_{10}^\text{SM} \delta^{IJ} + C_{10}^{IJ}(\text{NP}) \right) 
 \left(\bar s_L' \gamma^\mu b_L' \right) 
 \left( \bar\ell_{I}' \gamma_\mu \gamma^5 \ell_{J}' \right)  \,,  
 \label{EQ:bsellellLag}
\end{align}
where 
$V_{tb} V_{ts}^* = -0.0405 \pm 0.0012$~\cite{Patrignani:2016xqp,Charles:2015gya}, 
and 
 $C_9^\text{SM} \simeq 0.95$ and $C_{10}^\text{SM} \simeq -1.00$, which are  
estimated 
at the $m_b$ scale, (see, {e.g.}, \cite{Bobeth:2013uxa}). 
The favored region for $C_9^{\mu\mu}(\text{NP}) 
= - C_{10}^{\mu\mu}(\text{NP})$ in the left-handed scenario 
is given at the $2\sigma$ level as 
\al{
-0.87 \leq C_9^{\mu\mu}(\text{NP}) = - C_{10}^{\mu\mu}(\text{NP}) \leq -0.36 \,, 
\label{EQlimit_bsmumu}
}
whereas the best fit point is $-0.61$.  
The quoted numbers have been taken from Ref.~\cite{Capdevila:2017bsm}
(see also~\cite{DAmico:2017mtc,Altmannshofer:2017yso,Geng:2017svp,Ciuchini:2017mik,Hiller:2017bzc,Celis:2017doq,Ghosh:2017ber,Alok:2017jaf,Alok:2017sui,Wang:2017mrd,Bardhan:2017xcc}), 
where all available {associated} data from LHCb, Belle, ATLAS and CMS were combined, 
which are also consistent with the current measurement for the 
(optimized) angular observable, so-called $P_5'$~\cite{DescotesGenon:2012zf,Aaij:2013qta,Aaij:2015oid,Abdesselam:2016llu,Wehle:2016yoi,ATLAS-CONF-2017-023,CMS-PAS-BPH-15-008,Aaij:2017vbb}.  

NP contributions to $b \to s \ell^+ \ell^-$ processes have been discussed in the context of new vector-boson interactions~\cite{
Altmannshofer:2013foa,Gauld:2013qba,Buras:2013qja,Altmannshofer:2014cfa,Biancofiore:2014wpa,Hiller:2014yaa,Crivellin:2015mga,Crivellin:2015lwa,Niehoff:2015bfa,Celis:2015ara,Greljo:2015mma,Niehoff:2015iaa,Altmannshofer:2015mqa,Falkowski:2015zwa,Carmona:2015ena,Chiang:2016qov,Altmannshofer:2016oaq,Boucenna:2016wpr,Boucenna:2016qad,Megias:2016bde,Celis:2016ayl,Altmannshofer:2016jzy,Crivellin:2016ejn,GarciaGarcia:2016nvr,Bhatia:2017tgo,Cline:2017lvv,Ko:2017lzd,Megias:2017ove,Kamenik:2017tnu,
Alonso:2017bff,Bonilla:2017lsq,Ellis:2017nrp,Alonso:2017uky,
Kawamura:2017ecz,King:2017anf,Chen:2017usq,Megias:2017vdg,Baek:2017sew,Bian:2017rpg,Lee:2017fin,Fuyuto:2017sys,DAmbrosio:2017wis,Dasgupta:2018nzt,Falkowski:2018dsl,Asadi:2018wea,Greljo:2018ogz}
and/or vector-leptoquark ones~\cite{
Calibbi:2015kma,Fajfer:2015ycq,Barbieri:2015yvd,Sahoo:2016pet,Hiller:2016kry,Barbieri:2016las,Buttazzo:2017ixm,Assad:2017iib,DiLuzio:2017vat,Calibbi:2017qbu,Cline:2017aed,Bordone:2017bld,Blanke:2018sro,
Azatov:2018knx,
Bordone:2018nbg,Sahoo:2018ffv}.
In our model, those NP contributions takes a hybrid form 
arising from the $Z'$-type and vector-leptoquark-type CFVs, as noted in Introduction.

\subsubsection{$\bar B \to K^{(*)} \nu\bar\nu$}

The effective Hamiltonian for $\bar B \to K^{(*)} \nu\bar\nu$ with 
the SM contribution included 
is given by 
\begin{align}
{\cal H}_{\text{eff}}(b \to s \nu_{{I}}\bar\nu_{{J}}) 
 = 
- {\alpha G_F \over \sqrt 2 \pi} V_{tb} V_{ts}^*  
\left( 
C_L^\text{SM} \delta^{IJ} +  C_L^{IJ}( \text{NP} )  
\right)\, 
 \left(\bar s_L' \gamma^\mu b_L' \right) \left( \bar\nu_{{I}}' \gamma_\mu (1-\gamma^5) \nu_{{J}}' \right)  \,, 
\end{align}
where the SM {contribution} is $C_L^\text{SM} \simeq -6.36$.  
The current upper bounds on the branching ratios of $\bar B \to K^{(*)} \nu\bar\nu$~\cite{Lees:2013kla,Lutz:2013ftz} place constraints on 
NP contributions~\cite{Buras:2014fpa}, so that at $90\%$ C.L. we find~\cite{Bhattacharya:2016mcc} 
\al{
-13 \sum_{{I}=1}^{3} \text{Re}[C_L^{{II}}(\text{NP})] 
+ \sum_{{I,J}=1}^{3} |C_L^{{IJ}} (\text{NP})|^2 \leq 473 \,.  
\label{EQlimit_bsnunu}
}

\subsubsection{$\tau\to\phi\mu$}

To this decay process, 
the SM (with three right-handed Dirac neutrinos introduced) 
 predicts so tiny lepton flavor violation highly suppressed 
by the neutrino mass scale. 
We may adopt the constraint obtained in~\cite{Bhattacharya:2016mcc} from the $90\%$ C.L. upper limit of $\mathcal{B}(\tau \to \mu \phi ) < 8.4 \times 10^{-8}$~\cite{Miyazaki:2011xe}, which reads 
\al{
\left|  C_{\tau \mu}^{ss} (\text{NP}) \right| < 0.019 \times \left( { m_{\rho} \over 1\,\text{TeV} } \right)^2~. 
\label{EQlimit_tauphimu}
}

\subsubsection{$\tau\to3\mu$}
\label{subsec:tauto3mu}

As in the case of the $\tau \to \mu \phi$ decay, 
the SM prediction is negligible in this process. 
The branching ratio for $\tau\to3\mu$ decay 
is then given by  
\begin{align}
 \text{Br}[\tau^- \to \mu^- \mu^+ \mu^-]
 = 
 {| C_{\tau \mu}^{\mu\mu}  (\text{NP})|^2} \times \frac{0.94}{4} \frac{m_\tau^5 \tau_\tau}{192\pi^3} \,,  
	\label{eq:BR_tauto3mu}
\end{align}
where {we checked the form of the branching ratio is consistent with those of 
Refs. \cite{Kuno:1999jp,Calcuttawala:2018wgo}.
$\tau_\tau$ represents the mean lifetime of the tau lepton ($\simeq 2.9 \times 10^{-13}\,\text{s}$~\cite{Patrignani:2016xqp}).} 
The factor $0.94$ came from the phase space suppression for the decay~\cite{Bhattacharya:2016mcc}.
The current upper bound at $90\%$ C.L. is placed to be~\cite{Hayasaka:2010np}:
\al{
\text{Br} [\tau^- \to \mu^- \mu^+ \mu^-] < 2.1 \times 10^{-8}. 
\label{EQlimit_tau3mu}
}

\subsubsection{$B^0_s$-$\bar{B}^0_s$ mixing: $\Delta M_{B_s}$ \label{sec:B_s-mixing}}

As noted above, 
the $B^0_s$-$\bar{B}^0_s$ mixing process 
would involve a bit more delicate deal than other semi-leptonic and fully leptonic 
decay processes,  
because nontrivial QCD corrections potentially come in.  
{To this process,
the effective Hamiltonian including the SM contribution is written like     
\begin{align}
{\cal H}_{\text{eff}}(bs \leftrightarrow bs) 
 =
 \left( \frac{G_F^2 m_W^2}{16\pi^2} (V_{tb} V_{ts}^*)^2 C_{VLL}^\text{SM} + 
 C_{bs}^{bs}(\text{NP}, \text{at } m_b^{\text{pole}})
 \right)
 ({\bar s}_L' \gamma^\mu b_L')\,({\bar s}_L' \gamma_\mu b_L')~, 
\label{effH-bsbs}
\end{align}
where $m_W = (80.379 \pm 0.012)\,\text{GeV}$~\cite{Patrignani:2016xqp},
{$m_b^{\text{pole}}$ means the pole mass of the bottom quark} and 
$C_{VLL}^\text{SM}$ is given as~{\cite{Bhattacharya:2016mcc,Alok:2017jgr}}
\begin{align}
C_{VLL}^\text{SM} = 4 \, {\eta_{2B}} \, S_0 (x_t), 
\label{CVLL-SM}
\end{align}
with $S_0(x_t)$ being  
the Inami--Lim function~\cite{Inami:1980fz}
\begin{align}
S_0 (x_t) = \frac{x_t}{4} \left[ 1 + \frac{9}{1 - x_t} - \frac{6}{(1 -  x_t)^2} - \frac{6 x_t^2 \log{x_t}}{( 1 -x_t )^3} \right].
\end{align}
In Eq.(\ref{CVLL-SM}) 
$x_t \equiv (\bar{m}_t(\bar{m}_t))^2/m_W^2$, 
in which 
$\bar{m}_t(\bar{m}_t)$ is 
the $\overline{\text{MS}}$ mass of the top quark, and  
${\eta_{2B}} \,(= 0.551)$ dictates the next-to-leading order (NLO) 
QCD correction~\cite{Buchalla:1995vs}. 
From the effective Hamiltonian in Eq.(\ref{effH-bsbs}), 
the mass difference is then evaluated as  
\al{
\Delta M_{B_s} = \frac{2}{3} M_{B_s} f_{B_s}^2 \hat{B}_{B_s} \left| \frac{G_F^2 m_W^2}{16\pi^2} (V_{tb} V_{ts}^*)^2 C_{VLL}^\text{SM} +  C_{bs}^{bs}{(\text{NP}, \text{at } m_b^{\text{pole}})} \right|.  
	\label{eq:DeltaMBs_Bsubs}
}

We shall first discuss the SM prediction (the first term in Eq.(\ref{eq:DeltaMBs_Bsubs})).  
We adopt the recently reported value for the $\overline{\text{MS}}$-top mass,
$\bar{m}_t(\bar{m}_t) = (162.1 \pm 1.0) \, \text{GeV}$, obtained 
in the NLO variant of the ABMP16 fit~\cite{Alekhin:2018pai}. 
For the $B_s$ mass we take  
$M_{B_s} = 5366.89(19)\,\text{MeV}$~\cite{Patrignani:2016xqp}. 
As to the $B_s$-decay constant and the bag parameter, 
we adopt the FLAG17 result, $f_{B_s}\sqrt{\hat B_{B_s}}=(274 \pm 8)$ 
MeV~\cite{Aoki:2013ldr,Aoki:2016frl}. 
The relevant CKM factor $(V_{tb} V_{ts}^*)^2 = \left|V_{tb} V_{ts}^*\right|^2 e^{-2 i \beta_s}$ 
can be simplified to 
$\left|V_{tb} V_{ts}^*\right|^2$, 
because of the small complex angle 
($\varphi^{c\bar{c}s}_s|_{\text{exp}} = -2 \beta_s = -0.030 \pm 0.033$~\cite{Amhis:2014hma}, $\varphi^{c\bar{c}s}_s|_{\text{SM}} = -0.03704 \pm 0.00064$~\cite{Hocker:2001xe,Charles:2004jd}). 
Since $C_{bs}^{bs}(\text{NP})$ turns out to be real in the CFV scenario, 
we can take the limit $\varphi^{c\bar{c}s}_s \to 0$ with good precision.
From the particle-data group-fit results 
for the magnitudes of all nine CKM elements 
(and the Jarlskog invariant) $\left|V_{tb} \right| = 0.999105 \pm 0.000032$ and $\left|V_{ts} \right| = 0.04133 \pm 0.00074$~\cite{Patrignani:2016xqp}, we find   
$\left|V_{tb} \right| \cdot \left| V_{ts}\right| \simeq 0.0413 \pm 0.000739$.
In evaluating the propagation of errors, we here simply
 ignored possible correlations between $\left|V_{tb} \right|$ and $\left| V_{ts}\right|$, which would be justified for estimating a conservative bound for new physics scenarios addressing the $R_{K^{\ast}}$ anomaly.

Combining all these values, we thus estimate the SM prediction~\footnote{ 
This estimated number is close to the result 
reported in~\cite{DiLuzio:2017fdq}, where 
$\Delta M_{B_s}^\text{SM} = (20.01 \pm 1.25) \, \text{ps}^{-1}$,   
which has been estimated by using 
the same FLAG17 variable for $f_{B_s} \sqrt{\hat{B}_{B_s}}$ 
as what we have used,  
while 
the $\overline{\text{MS}}$ mass of the top quark 
has been taken to be different from ours, 
$\bar{m}_t(\bar{m}_t) = 165.65(57) \, \text{GeV}$.  
(For details in other subtleties, see~\cite{DiLuzio:2017fdq}.)  
Use of their $\bar{m}_t(\bar{m}_t)$  
would yield $\Delta M_{B_s}^\text{SM} = (20.2 \pm 1.39) \, \text{ps}^{-1}$, 
in which 
the small enhancement in the error 
may originate from the increased value of $S_0(x_t)$.
}.  
\al{
\Delta M_{B_s}^\text{SM} = (19.7 \pm 1.35) \, \text{ps}^{-1}. 
\label{SMvalue_bsbs}
} 
On the other hand, 
the experimental value is~\cite{Amhis:2014hma}
\al{
\Delta M_{B_s}^{\text{exp}} = (17.757 \pm 0.021) \, \text{ps}^{-1}.
\label{EQlimit_bsbs}
} 

Looking at the SM prediction in Eq.(\ref{SMvalue_bsbs}), 
we should note that the theoretical uncertainty most dominantly comes 
from the input parameters  
$f_{B_s} \sqrt{\hat{B}_{B_s}}$, $V_{tb} V_{ts}^*$ and $\bar{m}_t(\bar{m}_t)$, 
where the errors from the first two quantities are much more dominant than the 
experimental uncertainty in Eq.(\ref{EQlimit_bsbs}), while  
the error for $\bar{m}_t(\bar{m}_t)$ is subdominant compared with the other two, 
though being still larger than the experimental uncertainty.

We next turn to estimation for the NLO-QCD correction to the NP contribution 
arising by the renormalization group evolution of the 
$C_{bs}^{bs}(\text{NP}, \text{at } m_\rho)$ with 
running down to the $m_b^{\rm pole}$ scale. 
The NLO-QCD running effect on the $C_{bs}^{bs}(\text{NP})$ 
can be evaluated 
by following the formalism given in Ref.~\cite{Becirevic:2001jj} (see also~\cite{Ciuchini:1997bw,Buras:2000if}) as
\al{
C_{bs}^{bs}(\text{NP}, \text{at } m_b^{\text{pole}}) \simeq \left( b_1^{(1,1)} + \eta \, (m_\rho) c_1^{(1,1)} \right) (\eta \, (m_\rho))^{a_1} \times
C_{bs}^{bs}(\text{NP}, \text{at } m_\rho),
}
where 
$m_b^{\rm pole} = 4.6\,\text{GeV}$, 
$\eta(m_\rho)  \equiv \alpha_s(m_\rho)/\alpha_s(m_t)$, 
$a_1 \,(= 0.286)$, $b_1^{(1,1)} \,(= 0.865)$ 
and $c_1^{(1,1)} \, (= -0.017)$~\cite{Becirevic:2001jj}. 
The exact form of the NLO-running QCD coupling~\cite{Gardi:1998qr,Deur:2016tte} 
together with {$m_t \simeq 173\,\text{GeV}$} (the pole mass of the top quark) and {$\Lambda_\text{QCD} \simeq 0.34\,\text{GeV}$} (the QCD confinement scale) \footnote{
{We used the approximated form for the Lambert $W$-function (in the `$-1$' branch)~\cite{RePEc:eee:matcom:v:53:y:2000:i:1:p:95-103} in this estimation.}
} yield  
$\eta \, (m_\rho = 1\,\text{TeV}) \simeq 0.771$. 
We thus have 
\al{
C_{bs}^{bs}(\text{NP}, \text{at } m_b^{\text{pole}}) \simeq 0.79 \times C_{bs}^{bs}(\text{NP}, \text{at } m_\rho = 1 \, \text{TeV}).
}
We will take into account of this net-NLO factor $\simeq 0.79$ 
in the later numerical calculations.
}

\subsubsection{$\bar B \to D^{(*)} \tau\bar\nu$}
\label{sec:RD-anomaly}

Remarkably, 
it turns out that the net effect of the charged CFVs on the $d \to u \ell\nu$ transition is 
almost vanishing, i.e., 
\begin{align} 
{
C_V^{IJ}({\rm NP})} \simeq 0 
\,. 
\end{align} 
This is because of 
the approximate degeneracy for CFVs 
(Eq.(\ref{degene})) as the consequence of 
the presence of the (approximate) one-family 
$SU(8)$ symmetry.  
Possible contributions arise from the mass difference in the charged CFVs and the $V_\text{SM}$-$\rho$ mixing effect, both of which are suppressed (controlled) by the factor $g_W^2/g_\rho^2$. 
In fact, for $g_\rho = {\cal O}(10)$ as in Eq.(\ref{grho:size})
 these contributions on $R_{D^{(*)}}$ do not exceed $5\%$ and thus it is not sufficient to account for ${\cal O}(10\%)$ deviations between the present data and the SM predictions in the ratios.

Generically, one could introduce extra interactions which 
would yield a sizable breaking effect for the $SU(8)$ degeneracy. 
Nevertheless, as a minimal setup we will not consider such 
extra terms in the present study, so as to keep the approximate 
{$SU(8)$} symmetry in Eq.(\ref{degene}) and hence the vanishing contribution 
to $R_{D^{(*)}}$.

At Belle~II, we may have potential to examine $R_{D^{(*)}}$ with a few $\%$ accuracy. 
Thus, if the discrepancy is reduced to a few $\%$ in future observation at Belle~II, it may point to the contributions from such small effects. 
See Eq.(\ref{eq:effective_operators}) in appendix~\ref{4-fermi} for explicit expression on $C_V^{IJ}(\text{NP})$.

\begin{figure}[ht]
\begin{center}
\includegraphics[width=0.45\columnwidth]{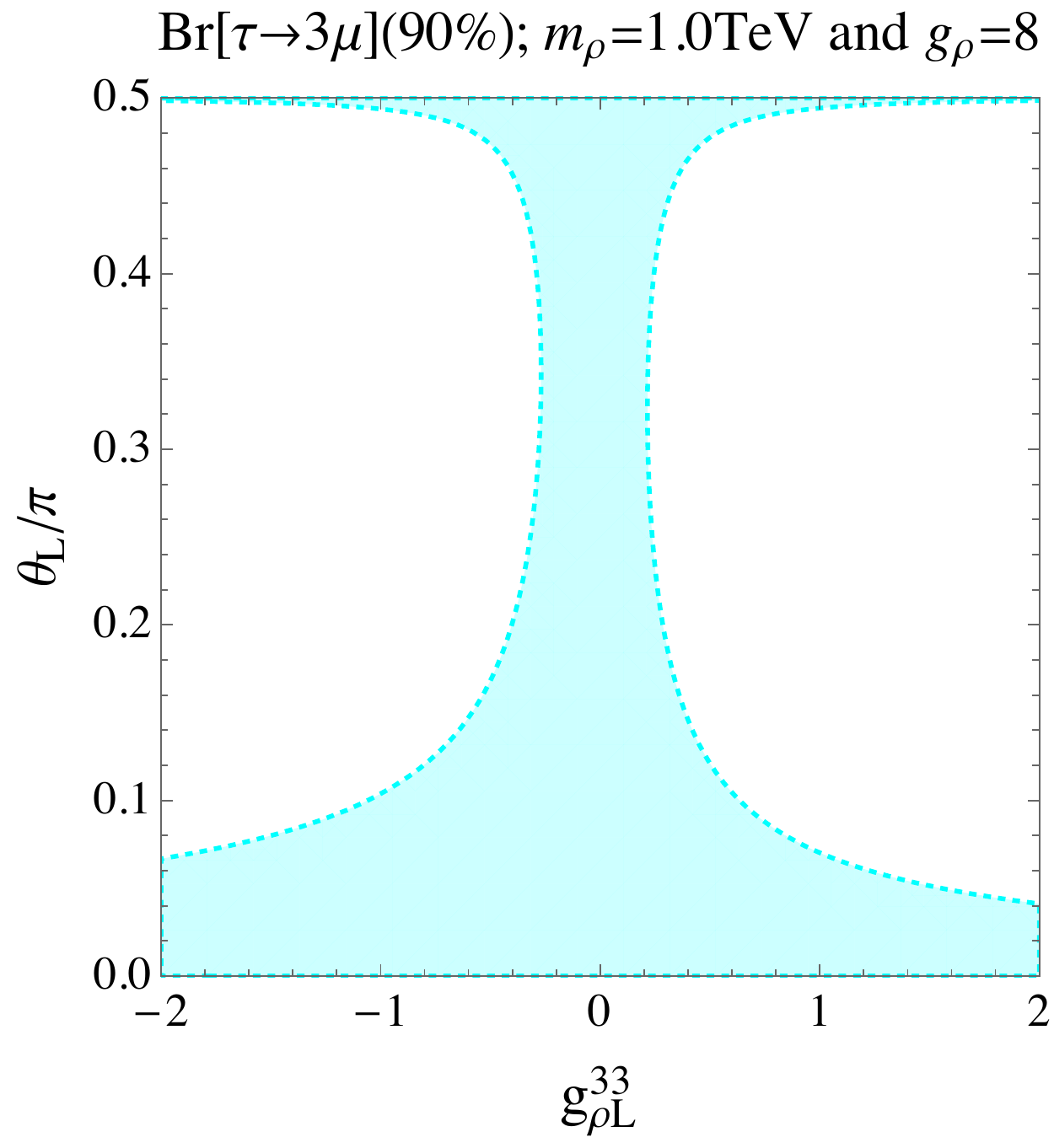}
\caption{
The $\tau \to 3 \mu$ decay constraint on the plane 
 {$(g_{\rho L}^{33}, \theta_L)$} for $m_\rho = 1$ TeV and $g_\rho = 8$. 
 The 90\% C.L. upper limit on the branching ratio has been 
 taken from Eq.(\ref{EQlimit_tau3mu}).  
 The shaded region is allowed at the 90\% C.L. read off from 
 Eq.(\ref{EQlimit_tau3mu}).  
}
\label{tau-3mu-bound}
\end{center}
\end{figure}

\begin{figure}[ht]
\begin{center}
\includegraphics[width=0.45\columnwidth]{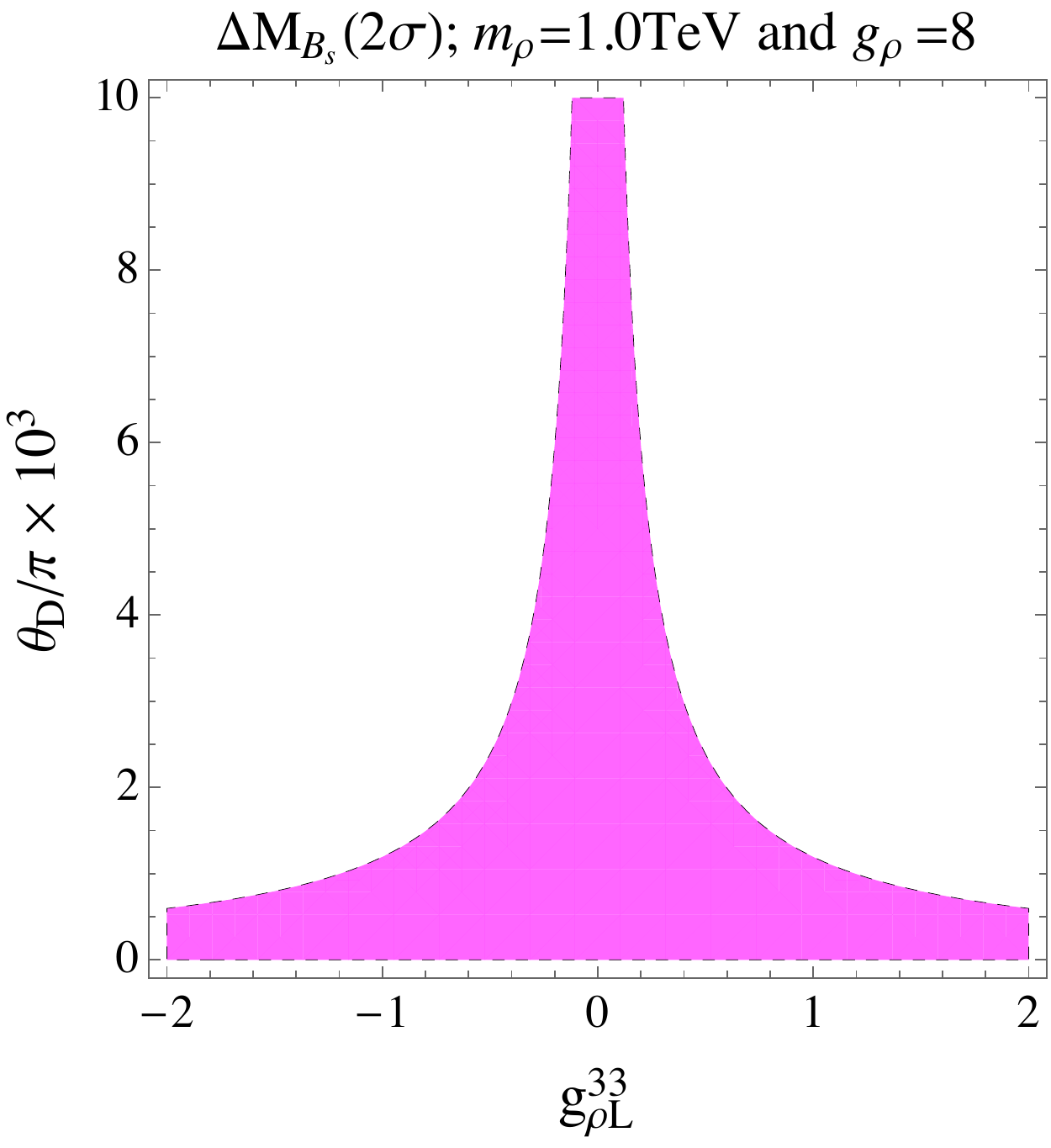}
\caption{
The $\Delta M_{B_s}$ constraint on the plane 
{$(g_{\rho L}^{33}, \theta_D)$} for $m_\rho = 1$ TeV and $g_\rho = 8$,  
where the shaded region is allowed at the {2$\sigma$} level read off from 
{Eqs.(\ref{EQlimit_bsbs}) and (\ref{SMvalue_bsbs}) 
with taking into account the
errors from the input variables $[ f_{B_s} \sqrt{\hat{B}_{B_s}}$, $V_{tb} V_{ts}^*$ and $\bar{m}_t(\bar{m}_t) ]$ as described in the text}.  
}
\label{deltaMs-bound}
\end{center}
\end{figure}

\begin{figure}[ht]
\begin{center}
\includegraphics[width=0.5\columnwidth]{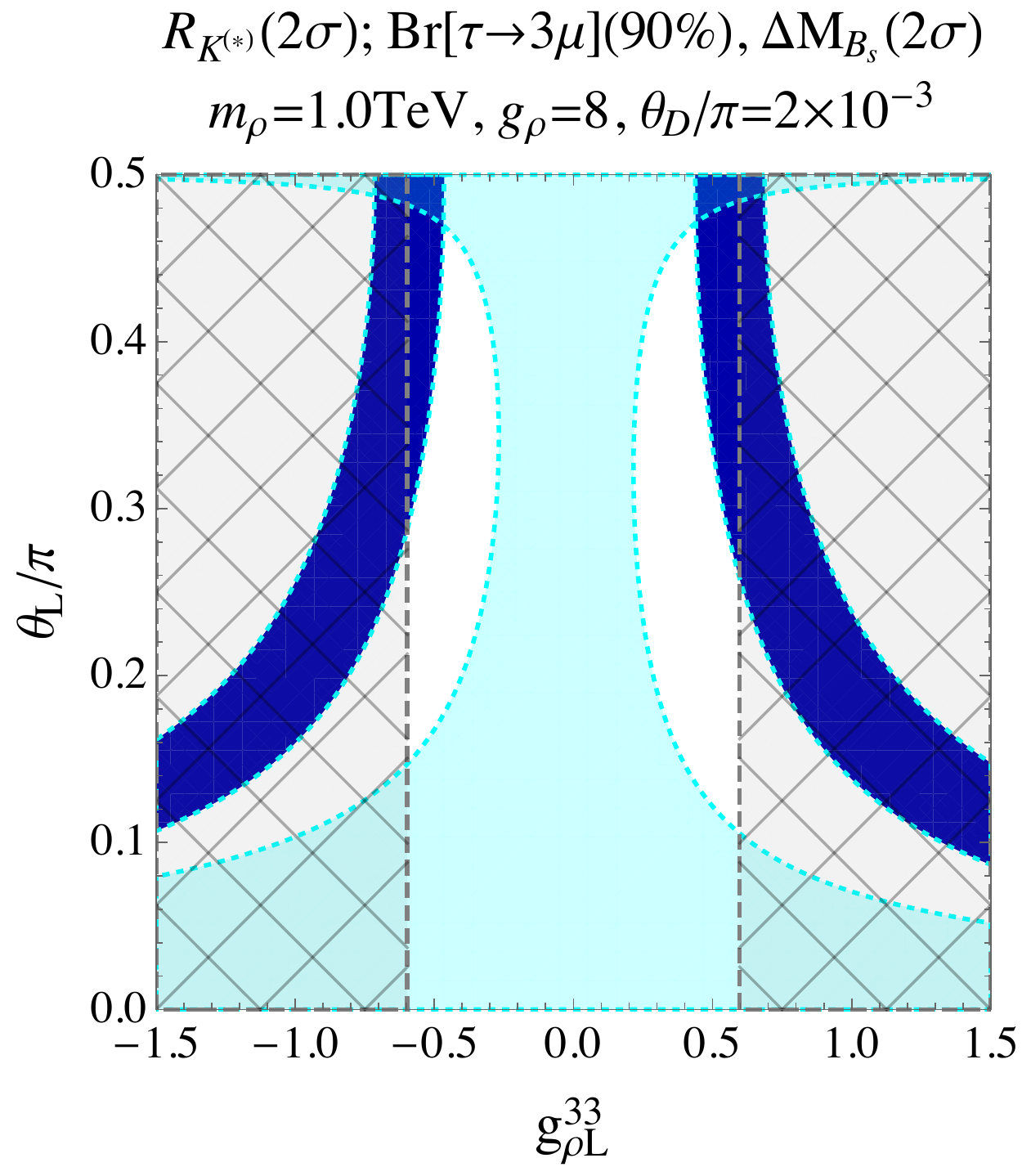}
\caption{
The region plot in the plane 
{$(g_{\rho L}^{33}, \theta_L)$} with {$\theta_D/\pi 
= 2 \times 10^{-3}$} fixed for $m_\rho = 1$ TeV and $g_\rho = 8$. 
The current $R_{K^{(*)}}$ anomaly can be explained in the thick-blue region
at the 2$\sigma$ level, while the cyan-shaded area represents the consistent region
with the current 90\% C.L. upper limit of $\text{Br} [\tau^- \to \mu^- \mu^+ \mu^-]$ {(see Fig.~\ref{tau-3mu-bound})}.
{The gray-hatched region is out of the {2$\sigma$-favored} area for $\Delta M_{B_s}$ (see Fig.~\ref{deltaMs-bound}).}
}
\label{B-tau-sys-cons-SU(8)inv-grhoL-rev}
\end{center}
\end{figure}

\subsubsection{$B -\tau$ system constraint}

The CFV contributions to the $2 \leftrightarrow 3$ transitions described as above 
are controlled by five parameters: $m_\rho$ in Eq.(\ref{degene}), 
$g_\rho$ in Eq.(\ref{grho:size}), $\theta_D, \theta_L$ in Eq.(\ref{LDrotation}) 
and $g_{\rho L}^{33}$ in Eq.(\ref{grhoL}).  
For a reference point, we take $m_\rho = 1$ TeV and $g_\rho = 8$, 
which will turn out to be consistent with the $K$ system bound later,   
and survey the allowed parameter region for the remaining couplings.

Figure~\ref{tau-3mu-bound} shows the region plot on the plane 
{$(g_{\rho L}^{33}, \theta_L)$} constrained by the 
 lepton flavor violating
 $\tau \to 3 \mu$ decay.   
The constraint from the  $\Delta M_{B_s}$ on  
the parameter space in the plane {$(g_{\rho L}^{33}, \theta_D)$
is depicted in Fig.~\ref{deltaMs-bound}.  
We have allowed the 2$\sigma$ deviation for the $\Delta M_{B_s}$ 
between the experimental and SM -predicted values, 
$\Delta M_{B_s}^{\rm exp} - \Delta M_{B_s}^{\rm SM}$}. 
This can be thought to be conservative 
because of the currently present 
 deviation $\gtrsim 1\sigma$, as captured from Eqs.(\ref{SMvalue_bsbs}) 
and (\ref{EQlimit_bsbs}).
In total, the favored parameter space in the plane 
{$(g_{\rho L}^{33}, \theta_L)$} with {$\theta_D/\pi 
= 2 \times 10^{-3}$} fixed~\footnote{
The larger $\theta_D$ case will be disfavored by the presence of 
the  $R_{K^{(*)}}$ anomaly.}  
is drawn in Fig.~\ref{B-tau-sys-cons-SU(8)inv-grhoL-rev}, 
where the overlapped domain (thick-blue {and cyan regions}) satisfies 
the 2$\sigma$ range for $C_9^{\mu\mu}$ {($= - C_{10}^{\mu\mu}$)} around the best fit point 
in Eq.(\ref{EQlimit_bsmumu}), hence explains the current $R_{K^{(*)}}$ anomaly 
at that level {consistently}. 
The range for the $g_{\rho L}^{33}$ has been restricted to {roughly $[-0.60, -0.47]$ or $[0.44, 0.60]$ (at $\theta_L \simeq \pi/2$)}, 
which is required by the $\Delta M_{B_s}$ bound for {$\theta_D/\pi =  2 \times 10^{-3}$}
read off from  Fig.~\ref{deltaMs-bound}.
The gray-hatched region in Fig.~\ref{B-tau-sys-cons-SU(8)inv-grhoL-rev} is out of the {2$\sigma$}-favored region for $\Delta M_{B_s}$,
which suggests that 
the favored region {is} limited to a muon-philic scenario with $\theta_L \sim \pi/2$ 
(in the mass-eigenstate basis).\footnote{ 
The favored flavor structure is 
thus similar to that proposed in Ref.~\cite{Bhattacharya:2016mcc,Cline:2017lvv}, 
where the NP is assumed to dominantly couple to 
the second-generation leptons and third-generation quarks 
in the mass-eigenstate basis.
Alternatively, we can consider that $(\nu_\mu, \mu)$ and $(t, b)$ are the ``third generation'' of the gauge eigenstates.
Whichever it is called, it does not make significant difference (occurring only in ways of parametrizing mixing angles)
in addressing the $R_{K^{(\ast)}}$ anomaly consistently.
It is also interesting to note the result obtained from the present CFV scenario 
looks quite different from 
the vector boson model in Ref.~\cite{Bhattacharya:2016mcc}, which is essentially because 
of the presence of mixing with the SM gauge bosons and the (approximate) $SU(8)$ 
symmetry constraining the coupling properties for 
$Z'$, $G'$ and vector leptoquarks.}

The constraints from $B \to K^{(*)} \nu \bar{\nu}$ and $\tau \to \mu \phi$ are irrelevant fully in the focused region 
in Fig.~\ref{B-tau-sys-cons-SU(8)inv-grhoL-rev}.
We provide a comment on the constraints from other three $2 \leftrightarrow 3$ observables, namely
$\bar{B} \to K^{(\ast)} \mu^\pm \tau^\mp$, $\Upsilon \to \mu^\pm \tau^\mp$ and $J/\psi \to \mu^\pm \tau^\mp$,
where they provide no additional bound in the focused region in Fig.~\ref{B-tau-sys-cons-SU(8)inv-grhoL-rev}.
Also, we have a discussion on how significant the best-fit point for $b \to s \mu^+ \mu^-$ in a global sense
by evaluating $\chi^2$ variable in a benchmark point in appendix~\ref{sec:global-fit}.

\subsection{Flavor changing processes converting the second and first generations} 

We next move on to the flavor constraints on the CFVs coming from 
the $K$ system. 
Looking at the flavor texture introduced in section~\ref{subsection:texture}, 
we find that 
the NP contributions to the $s$-$d$ transition observables, 
$\epsilon'/\epsilon$, $K \to \pi \nu {\bar \nu}$ 
and $K^0$-$\bar{K}^0$ mixing
($\Delta M_K$) are possibly generated.
Also, we focus on the $c$-$u$ transition, where observables in the $D^0$-$\bar{D}^0$ mixing
provide us with constraints on the scenario.
As discussed in the previous section, 
we first note that the down-sector rotation angle $\theta_D$ 
 is severely constrained by $B$ observables, 
most stringently by 
 $B^0_s$-$\overline{B^0_s}$ mixing, 
 to be almost vanishing, 
 \begin{align} 
 \theta_D \sim 0 
 \,, \label{thetaD} 
 \end{align}  
 (but should be nonzero to address $B$ anomalies like $b \to s \mu^+ \mu^-$), 
 while the lepton angle $\theta_L$ has to be 
\begin{align}  
 \theta_L 
 & \sim \frac{\pi}{2}  
 \,. \label{3-cons}
 \end{align}
In discussing the $K$ and $D$ systems, 
we shall take these conditions to survey the 
allowed parameter space for the CFVs.

\subsubsection{$K \to \pi \pi$ }

NP effects on the $K \to \pi \pi$ process have extensively been 
investigated in various context of scenarios beyond 
the SM~\cite{Blanke:2015wba,Buras:2015kwd, Tanimoto:2016yfy, Kitahara:2016otd, Endo:2016aws,Bobeth:2016llm,Endo:2016tnu,Crivellin:2017gks,Chobanova:2017rkj,Endo:2017ums,Bobeth:2017ecx, Haba:2018byj,Chen:2018ytc, Chen:2018vog,Buras:2018wmb}.   
To this process, in terms of effective operators,  
the contributions can be classified into  
i) $(V-A) \times (V-A)$, ii) $(V+A) \times (V+A)$, and 
iii) $(V-A) \times (V+A)$ current interaction types.  
Since in the present model, CFVs couple only to left-handed ($V-A$) current 
fermions with the generation conversion allowed, 
only the types of i) and iii) will be relevant. 
As to the type i) $(V-A) \times (V-A)$ interactions, characterized by 
called
$Q_2 = (\bar{s}' u')_{V-A} (\bar{u}' d')_{V-A}$ (charged current type),  
$Q_3 = (\bar{s}' d')_{V-A} \sum_{q'} (\bar{q}' q')_{V-A}$ (QCD penguin type) 
and 
$Q_9 = (\bar{s}' d')_{V-A} \sum_{q'} Q_{\rm em}^q (\bar{q}' q')_{V-A}$ (EW penguin type) 
operators, 
we see that the CFVs exchanges having only 
the flavored $g_{\rho L}^{ij}$ couplings do not generate any contributions,    
because of the third-generation-{philic} texture for the $g_{\rho L}^{ij}$ 
in Eq.(\ref{grhoL}) and the rotation matrix $D$ in Eq.(\ref{LDrotation}) 
 with the constraint on $\theta_D$ in Eq.(\ref{thetaD}) taken into account.  
Thus the nontrivial-leading terms to the type i) as well as the type iii) 
are generated necessarily  
along with the flavor-universal interactions suppressed by {$(g_{s,W,Y}/g_\rho)$},  
 where  only the neutral CFVs $\rho_{(1)}^3, \rho_{(1)'}^0$ and $\rho_{(8)}^0$  
 contribute to the effective four-fermion operators  
 (see Eq.(\ref{indirect:rho-coupling})). 
We thus find the relevant induced four-fermion operators like
\begin{align} 
{\cal H}_{\rm eff}
&= \sum_{j=1\textrm{-}10} C_j  \cdot Q_j 
\,, \notag \\ 
Q_1 &= (\bar{s}^{b \prime} u^{a \prime})_{V-A} (\bar{u}^{a \prime} d^{b \prime})_{V-A}
\,, &  
Q_2 &= (\bar{s}' u')_{V-A} (\bar{u}'d')_{V-A} 
\,, \notag \\ 
Q_3 &= (\bar{s}'d')_{V-A} \sum_{q' }(\bar{q}'q')_{V-A} 
\,, & 
Q_4 &= (\bar{s}^{b \prime} d^{a \prime})_{V-A} \sum_{q'} (\bar{q}^{a \prime} q^{b \prime})_{V-A} 
\,, \notag\\ 
Q_5 &= (\bar{s}' d')_{V-A} \sum_{q'} (\bar{q}'{q'})_{V+A} 
\,, & 
Q_6 &= (\bar{s}^{b \prime} d^{a \prime})_{V-A}
\sum_{q \prime} (\bar{q}^{a \prime} q^{b \prime})_{V+A} 
\,, \notag \\ 
Q_7 &= \frac{3}{2} (\bar{s}'d')_{V-A} \sum_{q'} Q_{em}^q (\bar{q}'q')_{V+A} 
\,,  &  
Q_8 &= \frac{3}{2} (\bar{s}^{b \prime} d^{a \prime})_{V-A} \sum_{q \prime} Q_{em}^q
 (\bar{q}^{a \prime} q^{b \prime})_{V+A} 
\,, \notag\\ 
Q_9
 &= \frac{3}{2}  (\bar{s}^{ \prime}d')_{V-A} \sum_{{q'}} Q_{em}^q (\bar{q}^{ \prime} q')_{V-A} 
\,,  &
Q_{10} 
&= \frac{3}{2}  (\bar{s}_L^{b  \prime} d_L^{a \prime})_{V-A}
\sum_{{q'}}  Q_{em}^q (\bar{q}^{a \prime} q^{b \prime})_{V-A} 
\,,
\end{align} 
 where the operator notation follows Refs.~\cite{Buras:2015yba,Buras:2015jaq}
 {[$(\bar{q}'q')_{V \pm A}$ being defined as ($\bar{q}' \gamma^\mu (1 \pm \gamma_5) q'$); $\gamma^\mu$ being suppressed in the above list]}, 
 {and $a,b$ stand for the color indices (with repeated ones being summed)}.  
The Wilson coefficients $C_{1 \textrm{-}10}$ are read off from Eq.(\ref{eq:effective_operators}) in appendix~\ref{4-fermi},  
which are interpreted as the ones evaluated at  
 the CFV mass scale {$m_\rho$}:  
\begin{align} 
C_1(m_\rho) &= 0
\,, &
C_2(m_\rho) &=  {-i \cdot \frac{1}{8} \frac{
g_W^2 g_{\rho L}^{12}}{m_\rho^2 g_\rho}}
\,, \notag \\ 
C_3(m_\rho) &=   i \cdot \frac{1}{24} \frac{
g_s^2 g_{\rho L}^{12}}{m_\rho^2 g_\rho} {+  i \cdot \frac{1}{8} \frac{
g_W^2 g_{\rho L}^{12} (-Y_q) }{m_\rho^2 g_\rho} -  i \cdot \frac{1}{144} \frac{
g_Y^2 g_{\rho L}^{12}}{m_\rho^2 g_\rho}}
\,, &
C_4(m_\rho) &=  - i \cdot \frac{1}{8} \frac{g_s^2 g_{\rho L}^{12}}{m_\rho^2 g_\rho}
\,, \notag \\ 
C_5(m_\rho) &=  i \cdot \frac{1}{24} \frac{g_s^2 g_{\rho L}^{12}}{m_\rho^2 g_\rho}
\,, &
C_6(m_\rho) &=  - i \cdot \frac{1}{8} \frac{g_s^2 g_{\rho L}^{12}}{m_\rho^2 g_\rho}
\,, \notag \\ 
C_7(m_\rho) &=  - i \cdot \frac{1}{36} \frac{g_Y^2 g_{\rho L}^{12}}{m_\rho^2 g_\rho}
\,, &
C_8(m_\rho) &= 0 
\,, \notag \\ 
C_9(m_\rho)
 &=  {i \cdot \frac{1}{12} \frac{
g_W^2 g_{\rho L}^{12}}{m_\rho^2 g_\rho}}
\,, &
C_{10}
(m_\rho) &= 0 
\,,
\label{wc}
\end{align}
{where $Y_q \, (=1/6)$ represents the weak hypercharge of the quark doublet.}

The {$K^0 \to \pi^0 \pi^0/\pi^+ \pi^-$} amplitudes, decomposed into the 
distinguished isospin states $(I=0,2)$ in the final state, 
are evaluated through the effective Hamiltonian as 
\begin{align} 
 A_I & \equiv \langle (\pi\pi)_I | {\cal H}_{\rm eff} { |K^0 \rangle }
 \notag\\ 
 & =  \sum_j  C_j({\mu}) \langle (\pi\pi)_I | Q_j({\mu}) { |K^0 \rangle }
 \equiv \sum_j C_j({\mu}) \langle Q_j({\mu})  \rangle_I 
\,, \label{AI}
\end{align}
{where $\mu$ represents a reference scale of the phenomenon.
In later numerical calculations, we put the values of the 
two-loop running couplings for ${\epsilon'}/{\epsilon}$.}

The CFV contributions to the direct CP violation in the $K \to \pi \pi$ processes
are evaluated {at the NLO perturbation in QCD and QED coupling expansions} as \cite{Kitahara:2016nld}
\begin{align} 
 \left(\frac{\epsilon'}{\epsilon}  \right)^{\rm CFVs} 
& =
\frac{\omega_{+}}{\sqrt{2} \left| \epsilon_{K}^{\textrm{exp}}\right|
  \textrm{Re} A_0^{\textrm{exp}}} 
  \langle \vec{Q}_{\epsilon'} ( \mu)^T\rangle 
  \hat{U} \left(\mu, m_{\rho} \right)
 \textrm{Im} \left[
  \vec{C}(m_{\rho}) \right],
\label{epp}
\end{align} 
where $\vec{C}(m_{\rho}) =
(C_1(m_{\rho}),C_2(m_{\rho}),C_3(m_{\rho}), \cdots  , C_{10}(m_{\rho}))^T
$,
${\textrm{Re}}A_0^{\textrm{exp}} = (3.3201 \pm 0.0018) \times 10^{-7} {\textrm{GeV}}$ \cite{Blum:2015ywa}, 
and
 $\omega_+|_{\rm SM} \equiv a \, {{\rm Re}A_2|_{\rm SM}}/{\rm Re}A_0|_{\rm SM} =
 4.53 
 \times 10^{-2}$ \cite{Cirigliano:2003gt,Buras:2015yba}.
The coefficients $\langle \vec{Q}_{\epsilon'} ( \mu)^T \rangle \hat{U} \left(\mu, m_{\rho} \right)$, which 
denote the evolution of the hadronic matrix elements from the scale $\mu$ to 
the NP scale $m_\rho$,  
are given in Ref. \cite{Kitahara:2016nld}, {where $\langle \vec{Q}_{\epsilon'} ( \mu)^T\rangle$ is defined as}
\begin{align}
{\langle \vec{Q}_{\epsilon'} ( \mu)^T\rangle \equiv
	\frac{1}{\omega_+} \langle \vec{Q}( \mu)^T\rangle_2 - \langle \vec{Q}( \mu)^T\rangle_0 (1 - \hat{\Omega}_{\text{eff}}).}
\end{align}
{The vector forms $\langle \vec{Q}( \mu)^T\rangle_{I}\,(I=0,2)$ are defined from $\langle Q_j({\mu})  \rangle_I$ like $\vec{C}(m_{\rho})$~\footnote{
{The values of $\langle \vec{Q}( \mu)^T\rangle_{I}$ and the form of $\hat{U} \left(\mu, m_{\rho}\right)$ are available in~\cite{Kitahara:2016nld}.}
}.
The factors for the isospin breaking correction are described in the matrix form,}
\begin{align}
{(1 - \hat{\Omega}_{\text{eff}})_{ij}
	=
\begin{cases}
0.852 & (i = j = 1-6), \\
0.983 & (i = j = 7-10), \\
0        & (i \not= j).
\end{cases}}
\end{align}
{Here the scale $\mu$ is set to be $1.3\,\text{GeV}$.}
In the LO analysis where $C_5(m_{\rho}), C_6(m_{\rho})$ and $C_7(m_{\rho})$
bring main effects on $C_6(m_c)$ and $C_8(m_c)$,  
we found that the contributions from QCD penguin $Q_6$ dominates in the $\epsilon'/\epsilon$, 
and the EW penguin $Q_8$ term yields about $60 \%$ contribution of them.

Actually, leptoquark-type CFVs ($\rho_{(3)}^{0,\alpha}$) would  
also contribute to the $\epsilon'/\epsilon$ at the one-loop level  
as discussed in Ref.~\cite{Bobeth:2017ecx}. 
However, in contrast to the literature, 
this kind of contributions are highly 
suppressed by a tiny $\theta_D$ in the present third-generation-philic scenario 
required by the constraint from the $B$ meson system, 
specifically from the $B^0_s - \bar{B}^0_s$ mixing (Eq.(\ref{thetaD})). 
This difference manifests the characteristic feature 
in the present CFV scenario based on the one-family $SU(8)$ 
symmetry, by which the predictions in flavor physics 
are derived necessarily with a significant correlation 
between the $2 \leftrightarrow 3$ and $1 \leftrightarrow 2$ 
transition processes, as will be more clearly seen later.

\subsubsection{$K^+ \to \pi^+ \nu \bar{\nu}$ and $K_L \to \pi^0 \nu \bar{\nu}$}

To these processes, the CFVs give contributions from  
the color-singlet $Z'$-like ($\rho_{(1)}^3$,
 $\rho_{(1)'}^{0}$) and the color-triplet vector leptoquark-like ($\rho_{(3)}^{3,0}$)  
 exchanges.  
Those CFVs exchange contributions are read off from Eq.(\ref{eq:effective_operators}) in 
appendix~\ref{4-fermi} as follows: 
\begin{align}
{\cal H}_{\text{eff}}(s \to d \nu \bar{\nu}) 
&\simeq   
	\left(- i \frac{7}{16} 
	\frac{ {g_{\rho L}^{12}} g_{\rho L}^{33} }{m_\rho^2} 
	\right)
	(\overline{s}'_L \gamma_\mu d'_L)
	(\overline{\nu}_{\mu L} \gamma^\mu \nu_{\mu L}) 
	\notag \\ 
	& 
	+ 
	\left( i \frac{1}{4} \frac{g_{\rho L}^{12}}{m_\rho^2} \frac{(g_W^2  +  g_Y^2/3)}{g_\rho} \right) 
	(\overline{s}'_L \gamma_\mu d'_L)
\sum_{l=e,\mu,\tau}
	(\overline{\nu}_{L}^l \gamma^\mu \nu_{L}^l) 
	\,,   \label{identity}
\end{align}
where we have taken into account $\theta_L \sim \pi/2$ (muon-philic condition in Eq.(\ref{3-cons})) in evaluating 
the contribution along with the flavorful coupling $g_{\rho L}^{33}$ (first line). 
The term in the first line comes from the $Z'$-type CFVs   
($\rho_{(1)}^3, \rho_{(1)^\prime}^0$) and the vector-leptoquark type ones $(\rho_{(3)}^{0,3})$   
-exchanges, while 
the one in the second line from the $Z'$-type ones. 
The dominant term actually comes from the vector-leptoquark type exchanges:  
the prefactor for the flavorful coupling term in the first line of Eq.(\ref{identity}) 
reads $7/16 = (-1/16)_{\rho_{(1)'}^0} + (1/2)_{\rho_{(3)}^{0,\alpha}}$   
(see Eq.(\ref{eq:effective_operators}))~\footnote{ 
A similar leptoquark scenario for addressing 
$K \to \pi \nu {\bar \nu}$ based on the third-generation-philic texture 
in light of the $R_{K^{(*)}}$ anomaly has been discussed 
in Ref.~\cite{Fajfer:2018bfj} where scalar leptoquarks at one-loop level 
play the game.}.

The branching ratios ${\rm Br}[K^+ \to \pi^+ \nu \bar{\nu}]$ 
 and ${\rm Br}[K_L \to \pi^0 \nu \bar{\nu}]$ 
 are computed as~\cite{Buras:2012dp,Buras:2015jaq}
\begin{align} 
 {\rm Br}[K^+ \to \pi^+ \nu \bar{\nu}]
 &= 
 \kappa_+ \left[ 
 \frac{1}{3}
 {\sum_{l=e, \mu, \tau}} 
 \left( \frac{{\rm Im} X^l_{\rm eff}}{\lambda^5} \right)^2 
 + \left( \frac{{\rm Re} X_{\rm eff}}{\lambda^5} {+ \frac{{\rm Re} \lambda_c}{\lambda}  P_c(X)} \right)^2  \right]
\,, \notag \\ 
 {\rm Br}[K_L \to \pi^0 \nu \bar{\nu}]
 &= 
\frac{1}{3} \kappa_L \sum_{l=e, \mu, \tau} 
\left( \frac{{\rm Im} X^l_{\rm eff}}{\lambda^5}  \right)^2
, 
\end{align} 
with $ \lambda = |V_{us}|=0.225$, {$P_c(X) = (9.39 \pm 0.31) \times 10^{-4}/\lambda^4 + (0.04 \pm 0.02)$, ${\rm Re} \lambda_c/\lambda \simeq -0.98$}
being the charm contribution,   
$\kappa_+=(5.157 \pm 0.025) \times 10^{-11} (\lambda/0.225)^8$ 
and $\kappa_L = (2.231 \pm 0.013) \times 10^{-10} 
(\lambda/0.225)^8$~\cite{Buras:2012dp,Buras:2015jaq}. 
Here the NP effects come in {$X^{(l)}_{\rm eff}$}, which are in the present 
CFV case 
numerically evaluated as  
\begin{align} 
{\rm Re} {X_{\rm eff}} 
&= - 4.83 \times 10^{-4} 
\,, \notag \\ 
{\rm Im} X_{\rm eff}^l 
& 
\simeq  
2.12 \times 10^{-4} - {2.46} \left( \frac{1\,{\rm TeV}}{m_\rho} \right)^2 
g_{\rho L}^{12}
\left[ 
 g^l  
- \frac{4}{7} \frac{(g_W^2+ g_Y^2/3)}{g_\rho}
\right]
\,,   \label{semi}
\end{align}
with $g^{l=e,\tau} =0$ and $g^{l=\mu} = g_{\rho L}^{33}$ {(see Eq.(\ref{identity}))}, 
in which we have quoted the values of SM predictions from Ref.~\cite{Endo:2016tnu} 
using the CKM{\footnotesize FITTER} result for the CKM elements.

\subsubsection{$K^0$-$\bar{K}^0$ mixing ($\Delta M_K$)}
 
The CFV contributions 
to the $\Delta M_K$, dominated by the flavored left-handed coupling $g_{\rho L}^{12}$,   
 are evaluated through 
the effective four-fermion interaction term 
(see Eq.(\ref{eq:effective_operators}) in appendix~\ref{4-fermi})
\begin{align} 
 {\cal H}_{\rm eff}(\Delta M_K) 
 &= C^{VLL}_{\rm NP}(m_\rho) \cdot (\bar{s}_L^\prime \gamma_\mu d_L^\prime)
 (\bar{s}_L^\prime \gamma^\mu d_L^\prime) 
 \,, 
 \notag\\ 
 C^{VLL}_{\rm NP}(m_\rho) &= 
- \frac{7 (g_{\rho L}^{12})^2}{32 m_\rho^2}
\,. 
\end{align}
This NP term  contributes to 
$\Delta M_K^{\rm NP}$ as 
(e.g. see Refs.\cite{Buras:2012fs,Buras:2015jaq})     
\begin{align} 
\Delta M_K^{\rm NP} 
&= 
2 {\rm Re}[M_{12}^K] 
\,, \notag\\ 
(M_{12}^K)^* 
&= 
\frac{1}{3} 
F_K^2 \hat{B}_K \, m_K \, \eta_2 \, \tilde{r} \cdot C^{VLL}_{\rm NP}({\mu}) 
\,,  
\label{MK}
\end{align} 
with the experimental values~\cite{Patrignani:2016xqp} 
$m_K = 0.497614$ GeV, $F_K=0.1561$ GeV, 
and 
$\hat{B}_K \simeq 0.764$, $\eta_2\simeq 0.5765$ 
and $\tilde{r}\simeq 1$.  
We may roughly neglect the 
small renormalization group evolution for the Wilson coefficient $C^{VLL}_{\rm NP}$ 
from $m_\rho$ scale down to ${\mu} \, (= 1.3\,\text{GeV})$ in Eq.(\ref{MK}), i.e., taking 
$C^{VLL}_{\rm NP}(m_\rho) \simeq C^{VLL}_{\rm NP}({\mu})$, {because 
the $\Delta M_K$ inevitably involves large theoretical uncertainties coming from
 long-distance contributions and 
the coefficient $C^{VLL}_{\rm NP}$   
cannot get drastic corrections such as a significant amplification for the isospin breaking effect during the running down, in contrast to the $\epsilon'/\epsilon$.
More on the uncertainties for the $\Delta M_K$ will 
be discussed in section~\ref{sec:summary}.

\subsubsection{$D^0$-$\bar{D}^0$ mixing \label{sec:D0D0bar}}

In this part, we discuss the CFV effect on the $D^0$-$\bar{D}^0$ mixing with taking account of the CP violation term.
We basically adopt the convention in~\cite{Bazavov:2017weg} for
the transition component of the $D^0$-$\bar{D}^0$ system, which is connected to the effective Hamiltonians $(\mathcal{H}^{\Delta C=2}, \mathcal{H}^{\Delta C=1})$ as
\al{
2 M_D \left( M_{12} - \frac{i}{2} \Gamma_{12} \right)
=
\langle D^0 | \mathcal{H}^{\Delta C=2} | \bar{D}^0 \rangle +
\sum_n \frac{ \langle D^0 | \mathcal{H}^{\Delta C=1} | n \rangle \langle n | \mathcal{H}^{\Delta C=1} | \bar{D}^0 \rangle}
{M_D-E_n+i\varepsilon},
}
where the mass eigenstate $| D^0_1 \rangle$ is almost identical with one of the CP eigenstates as 
$\langle D^0_1 | CP | D^0_1 \rangle \approx 1$.\footnote{
More concretely, $CP | D^0 \rangle = - | \bar{D}^0 \rangle$, $CP | \bar{D}^0 \rangle = - | D^0 \rangle$,
$ | D^0_1 \rangle = p | D^0 \rangle - q | \bar{D}^0 \rangle $,
$ | D^0_2 \rangle = p | D^0 \rangle + q | \bar{D}^0 \rangle $ with $|p|^2 + |q|^2 = 1$.
If there is no CP violation, $CP | D^0_{1 \atop 2} \rangle = \pm | D^0_{1 \atop 2} \rangle$.
}
Here, we skip to discuss the absorptive part $(\Gamma_{12})$ and
put $\Gamma_{12}^{\rm NP}=0$ as an assumption, since the mass scale of CFVs is of order of TeV, greater than 
the EW scale~\cite{Golowich:2009ii}.
$M_{12}$ is related to $\mathcal{H}^{\Delta C=2} = \sum_{i} C_i  \mathcal{O}^{\Delta C=2}_i$ as follows,
\al{
2 M_D M_{12} = \sum_{i} C_i (\mu) \langle D^0 | \mathcal{O}^{\Delta C=2}_i | \bar{D}^0 \rangle,
}
where the only $\mathcal{O}^{\Delta C=2}_1 = (\bar{c}'_L \gamma_\mu u'_L) (\bar{c}'_L \gamma^\mu u'_L)$ is relevant in the CFV scenario at tree level.
The latest allowed intervals at $95\%$ C.L. for three observables associated with the CP violation under the assumption of the existence of CP violation are read off from~\cite{HFLAV_charm2018} to be
\al{
y &\equiv \frac{\Delta \Gamma}{2 \Gamma_D}; \quad [0.46\%, 0.79\%],  \\
x &\equiv \frac{\Delta M}{\Gamma_D}; \quad [0.06\%, 0.70\%], \\
\phi_{12} &\equiv \text{arg}\left( \frac{M_{12}}{\Gamma_{12}} \right); \quad [-2.5^\circ, 1.7^\circ] \
(\leftrightarrow [-0.044, 0.030]),
}
where
$\Delta M \equiv M_1 - M_2$ and $\Delta \Gamma \equiv \Gamma_1 - \Gamma_2$.
Note that the absolute values of the variables,
\al{
y_{12} &\equiv \Gamma_{12} / \Gamma_D, &
x_{12} &\equiv 2 M_{12}/\Gamma_D,
	\label{eq:x12-and-y12}
}
are approximately the same
as the values of $y$ and $x$, respectively~\cite{Kagan:2009gb}.\footnote{
It is noted that $y_{12}$ and $x_{12}$ are defined as $y_{12} \equiv |\Gamma_{12}| / \Gamma_D$ and $x_{12} \equiv 2 |M_{12}|/\Gamma_D$
in the original paper~\cite{Bazavov:2017weg}.
}
We also note the relationship
\al{
x y = |x_{12}| |y_{12}| \cos{\phi_{12}}, 
	\label{eq:relation-xyphi12}
}
which tells us that $x$ and $y$ are the same sign since $\cos{\phi_{12}} > 0$
in the $95\%$ C.L. region~\cite{Kagan:2009gb}.
A constraint is put on the absolute value of the variable $x_{12}$ (which is complex in general) at $\mu_c \, (\simeq 1.3\,\text{GeV})$
through a NP contribution to the Wilson coefficient $C_i(\mu_c)$ as
\al{
x_{12}^\text{NP} = \frac{1}{M_D \Gamma_D} \sum_i C_i (\mu_c) \langle \mathcal{O}_i^{\Delta C=2} \rangle,
}
with
$M_D = 1.8648\,\text{GeV}$,
$(\Gamma_D)^{-1} = \tau_D = 0.4161\,\text{ps}$~\cite{Tanabashi:2018oca},
$\langle \mathcal{O}_1^{\Delta C=2} \rangle \, (\equiv \langle D^0 | {\cal O}_1^{\Delta C=2} | \bar{D}^0 \rangle) \sim 0.08\,\text{GeV}^4$~\cite{Bazavov:2017weg}.
Taking into account the assumption $\Gamma_{12}^{\rm NP}=0$,
we find that in Eq.(\ref{eq:relation-xyphi12}), 
the $y_{12} = y_{12}^{\rm SM}$ 
and the $\sin \phi_{12}$ can be expressed as a function of only the $x_{12}$ up to the SM factors:
\al{
\sin{\phi_{12}} = \sin\left[\text{arg}\left( \frac{M_{12}}{\Gamma_{12}} \right)\right]
= \frac{|\Gamma_{12}^\text{SM}|}{\Gamma_{12}^\text{SM}} \frac{\text{Im}[x_{12}]}{|x_{12}|}
= \frac{|\Gamma_{12}^\text{SM}|}{\Gamma_{12}^\text{SM}} \sin\left[\text{arg}\left( x_{12} \right)\right],
}
where we used the relation in Eq.(\ref{eq:x12-and-y12}).

Here, we shall make comments on the current status on estimate for the SM prediction.
\begin{itemize}
\item
The short-distance effects of $x$ and $y$ in the SM are estimated in~\cite{Golowich:2005pt}, which are negligible compared with the
following long-distance effects.
\item
The ratio $x^\text{SM}/y^\text{SM}$ and the magnitude of $y^\text{SM}$ have been evaluated as
$-1 \lesssim x^\text{SM}/y^\text{SM} \lesssim -0.1$ and $|y^\text{SM}| \sim 1\%$, respectively~\cite{Falk:2004wg,Falk:2001hx}.
Thereby, $|x^\text{SM}|$ currently lies in a range $0.1\% \lesssim |x^\text{SM}| \lesssim 1\%$.
The relative sign between $x$ and $y$ has been reported to be minus, 
which seems to be a tension to the observed result (both of $x$ and $y$ being positive).
\item
In our analysis, we choose $x^\text{SM} \sim + 1\% \text{ or } +0.1\%$ and $y^\text{SM} \sim +0.8\%$,
which may be reasonable when we take account of possible uncertainties.
Now, $y \simeq y^\text{SM} \sim 0.8\%$ (since we assume $y^\text{NP} = 0$), which can be within the $95\%$ C.L. region.
Thus, the sign of $x$ is determined to be positive from the relation in Eq.(\ref{eq:relation-xyphi12}).
\end{itemize}

\begin{figure}[t]
\centering
\includegraphics[width=0.45\columnwidth]{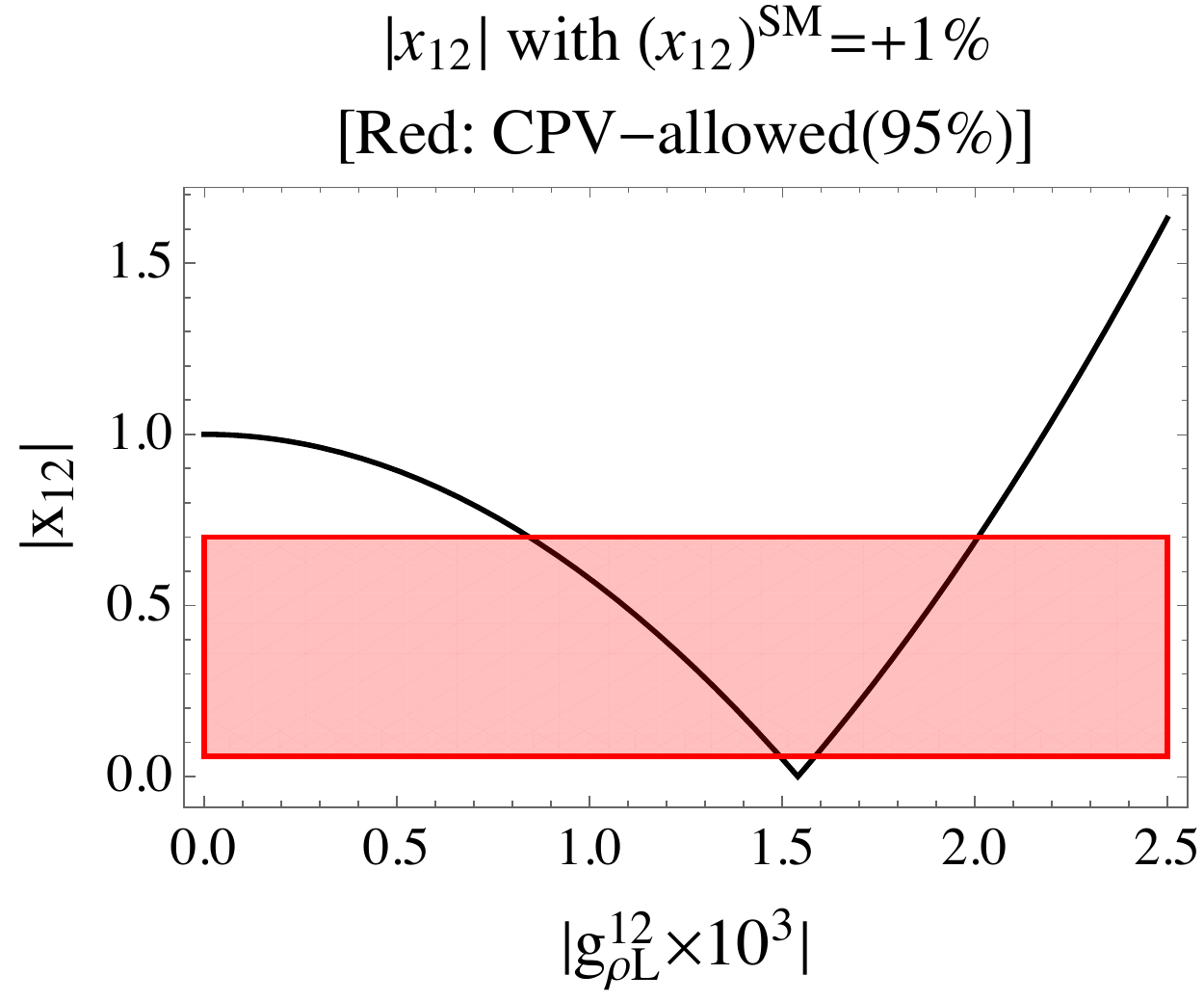} \quad
\includegraphics[width=0.475\columnwidth]{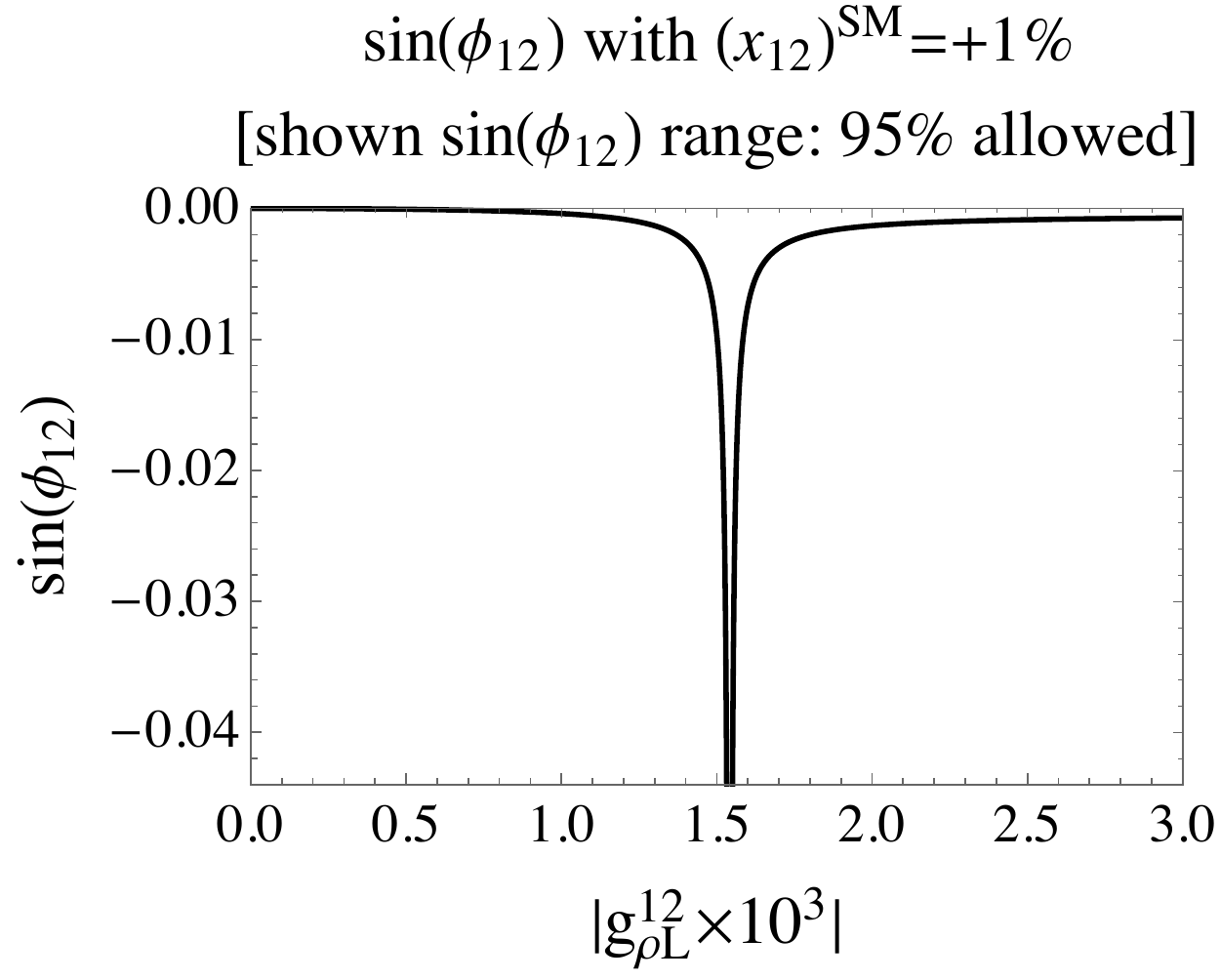} \\
\includegraphics[width=0.45\columnwidth]{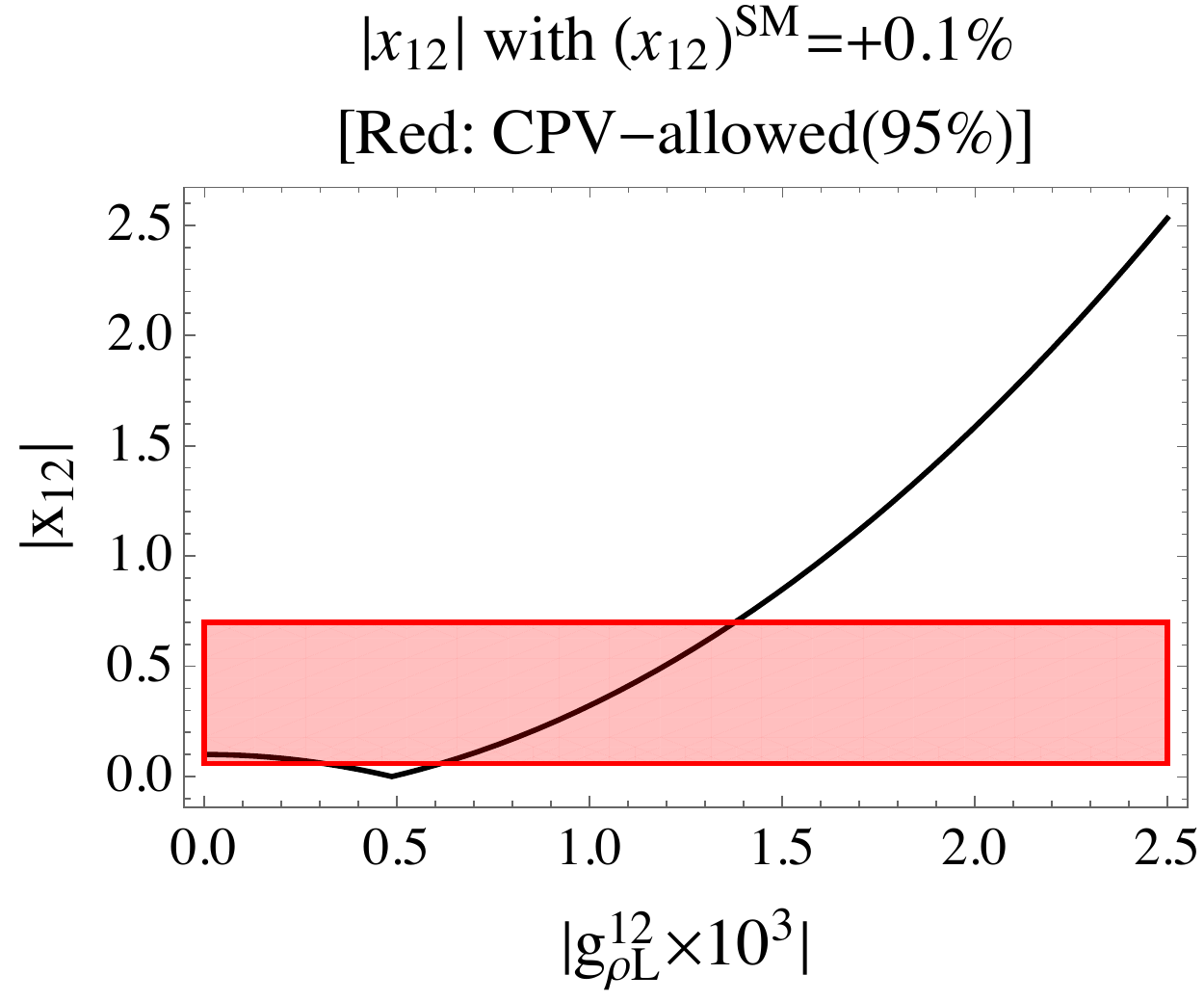} \quad
\includegraphics[width=0.475\columnwidth]{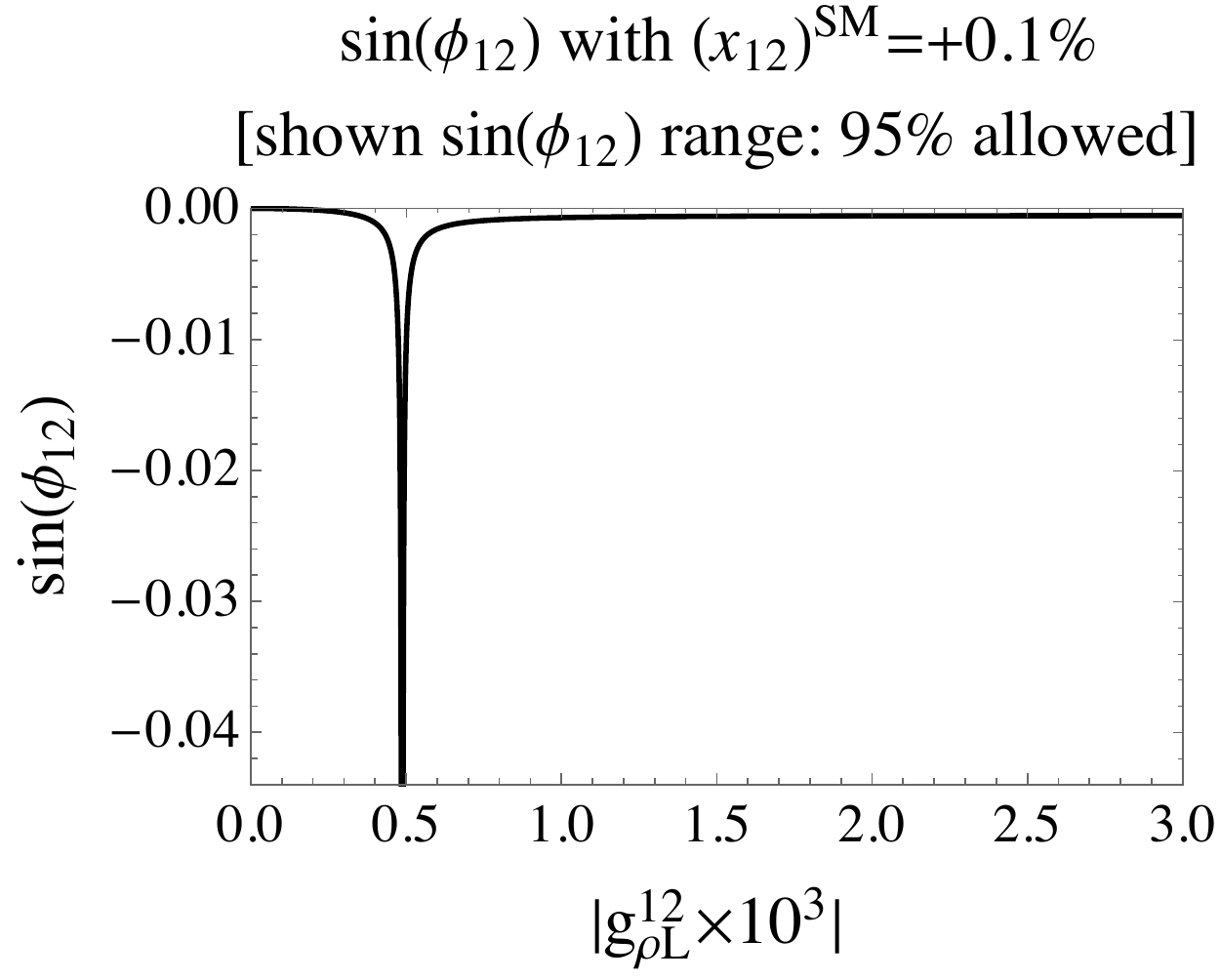}
\caption{Curves of $|x_{12}|$ and $\sin{\phi_{12}}$ as functions of $|g_{\rho L}^{12}|$ with $m_\rho = 1\,\text{TeV}$
for $x^\text{SM} = +1\%$ (upper panels) and $x^\text{SM} = +0.1\%$ (lower panels).
Note that the enhancements of $\sin\phi_{12}$ around $|g_{\rho L}^{12}| \sim 1.5 \times 10^{-3}$ (for $x^\text{SM} = +1\%$) 
and $\sim 0.5 \times 10^{-3}$ (for $x^\text{SM} = +0.1\%$) 
arises due to the vanishing ${\rm Re}[x_{12}]$ by the cancellation between the 
CFV and the SM contributions.
}
\label{fig:x12_and_sinphi12}
\end{figure}

In addition to those SM terms, the CFV scenario gives the NP contribution through
the operator which contributes to the $D^0$-$\bar{D}^0$ mixing,
\al{
\mathcal{H}^{D^0\text{-}\bar{D}^0} \simeq - \frac{7}{32} \frac{(g_{\rho L}^{12})^2}{m_\rho^2} (1 + 0.00054 \, i ) \times
	(\bar{c}'_L \gamma_\mu u'_L) (\bar{c}'_L \gamma^\mu u'_L),
}
where we have taken into account the tiny imaginary part via the CKM CP phase through the relation 
$V_{\text{CKM}} \simeq U^\dagger D$ with $\theta_D = 2 \, \pi \times 10^{-3}$.
The QCD running effect from 
$1 \,\text{TeV}$ (identical to the CFV mass scale $m_\rho$) down to $\mu_c = 1.3$ GeV for the operator
$\mathcal{O}_1^{\Delta C=2}$ can be extracted from~\cite{Golowich:2007ka} as
\al{
C_1(\mu = 1.3\,\text{GeV}) \simeq 0.72 \times C_1(\mu = 1\,\text{TeV}).
}
We have checked the constraints 
on $x_{12}$ and $\phi_{12}$ in a wide range for $|g_{\rho L}^{12}|$ with 
the reference points at $x^{\rm SM} = + 1\%$ and $x^{\rm SM}=+0.1\%$ chosen.
See Fig.~\ref{fig:x12_and_sinphi12}.
Thus we see that $\phi_{12}$ provides us with a milder bound than that from $|x_{12}|$ for $m_\rho = 1\,\text{TeV}$.



\subsubsection{Constraints from Kaon and $D$ systems}

\begin{figure}[t]
\begin{center}
\includegraphics[width=0.5\columnwidth]{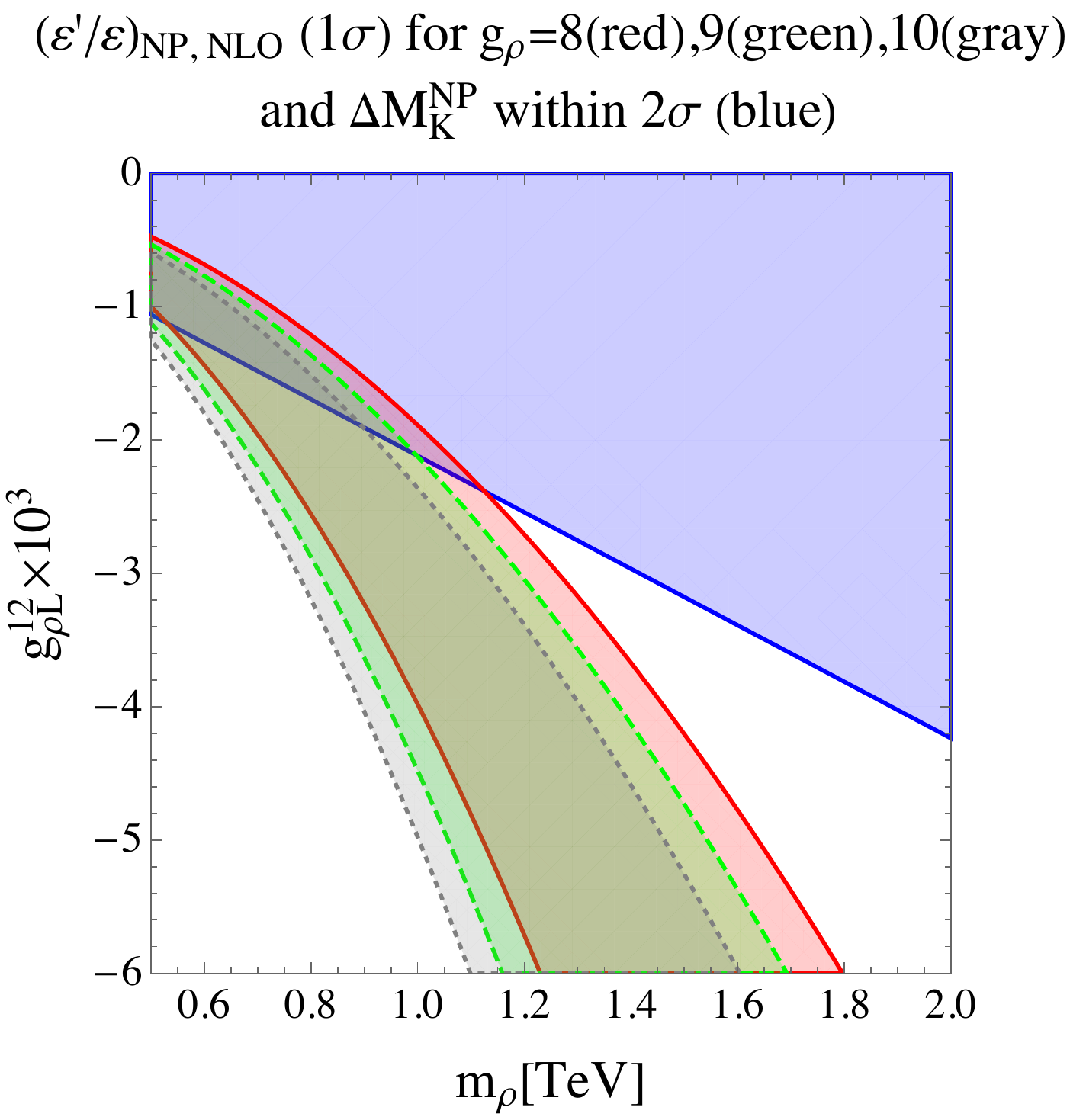}
\caption{
The $\Delta M_K^{\rm NP}$ and $(\epsilon'/\epsilon)_{\rm NP}$ constraints 
on the $(m_\rho, g_{\rho L}^{12})$ plane {for the three benchmark values of $g_\rho$}. 
}
\label{epsp-deltaMK-constraint}
\end{center}
\end{figure}

\begin{figure}[ht]
\begin{center}
\includegraphics[width=0.5\columnwidth]{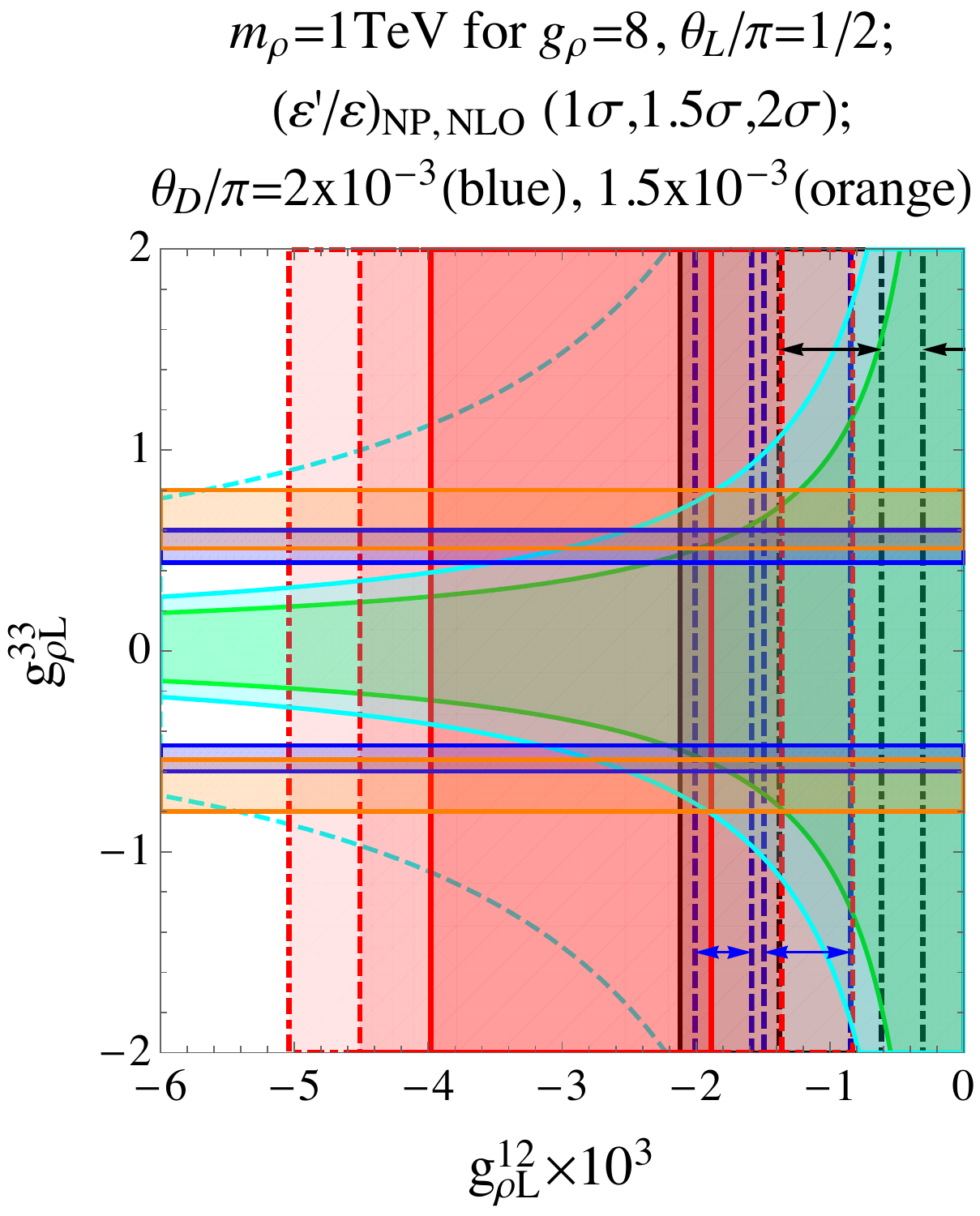}
\caption{
The combined constraint plot on 
{$(g_{\rho L}^{12}, g_{\rho L}^{33})$} for $m_\rho = 1$ TeV, $g_\rho = 8$, {$\theta_L/\pi = 1/2$ and $\theta_D/\pi = 2 \times 10^{-3}$ (horizontal band in blue) or $1.5 \times 10^{-3}$ (in orange)}, 
where the shaded regions are allowed. 
The red and pale-black vertical domains respectively 
correspond to the allowed regions set by {the $1\sigma$ (surrounded by solid line boundaries), $1.5\sigma$ (by dashed ones), $2\sigma$ 
(by dot-dashed ones) ranges} for $(\epsilon'/\epsilon)_{\rm NP}$,   
and the $2\sigma$ range for $\Delta M_K$. 
The $2\sigma$-allowed range for Br[$K^+ \to \pi^+ \nu \bar{\nu}$] 
and the 90\% C.L. upper bound for Br[$K_L \to \pi^0 \nu \bar{\nu}$] 
have been reflected in domains wrapped by green and cyan regions, respectively.  
The upper bound on Br[$K_L \to \pi^0 \nu \bar{\nu}$] was updated by the KOTO experiment at ICHEP in July 2018. 
In the figure we have also shown 
the previous boundary based on the previous bound by the dashed cyan curves 
(See also footnote~\ref{footnote:update}).
The regions surrounded by horizontal lines 
[in blue (for $\theta_D/\pi = 2 \times 10^{-3}$) or orange (for $\theta_D/\pi = 1.5 \times 10^{-3}$)] are allowed by the $B-\tau$ system constraint 
in Fig.~\ref{B-tau-sys-cons-SU(8)inv-grhoL-rev}, 
in which 
the lower bounds on the magnitude of $g_{\rho L}^{33}$ 
come from the requirement to account for the 
$R_{K^{(*)}}$ anomaly within the $2\sigma$ level, while the upper ones originate from circumventing the bound from 
$\Delta M_{B_s}$ at the $2\sigma$ level, respectively.
The vertical domains identified by the blue and black horizontal arrows correspond
to the $95\%$ C.L. intervals when $x^\text{SM} = +1\%$ and $+0.1\%$, respectively.
} 
\label{K-sys-cons-SU(8)inv-grhoL-rev}
\end{center}
\end{figure}

From {Eqs.(\ref{epp}), (\ref{semi})} 
and (\ref{MK}), 
we place the Kaon system limits on the CFV couplings 
$(g_{\rho L}^{33})$ and $(g_{\rho L}^{12})$ with $g_\rho$ chosen to be {$\sim 10$} 
and 
$m_\rho$ fixed to be on the 
order of ${\cal O}({\rm TeV})$. 
As to the $K^+ \to \pi^+ \nu \bar{\nu}$ 
we allow the model parameters in the $2\sigma$ range for the 
experimentally observed values~\cite{Artamonov:2008qb}:
\begin{align} 
{\rm Br}[K^+ \to \pi^+ \nu \bar{\nu}]  
&= 
(17.3^{+11.5}_{-10.5}) \times 10^{-11} 
\label{eq:Kp_to_pipnunubar__experimental-value}
\,, 
\end{align}
for the $K_L \to \pi^0 \nu \bar{\nu}$ we adopt 
the 90\% C.L. upper bound at present~\cite{Ahn:2018mvc},\footnote{
{This bound was updated by the KOTO experiment very recently at ICHEP 2018 (on July 7, 2018), 
which has been more stringent than 
the previous one, ${\rm Br}[K_L \to \pi^0 \nu \bar{\nu}] < 2.6 \times 10^{-8}$~\cite{Ahn:2009gb}. 
\label{footnote:update}}
}
\begin{align} 
{\rm Br}[K_L \to \pi^0 \nu \bar{\nu}] < 3.0 \times 10^{-9}
 \,. 
\end{align}
Regarding the NP contribution to 
$\epsilon'/\epsilon$,  
we take the {$1\sigma$, $1.5\sigma$, and $2\sigma$ ranges} for  
the difference between the experimental value and the SM prediction 
($(\epsilon'/\epsilon)_{\rm NP} \equiv  (\epsilon'/\epsilon)_{\rm exp} 
 -  (\epsilon'/\epsilon)_{\rm SM}$), 
as done in Ref.~\cite{Endo:2016tnu} {(for the $1\sigma$ range)},
\begin{align} 
 1.00 \times 10^{-3} < &\left. (\epsilon'/\epsilon)_{\rm NP}\right|_{1\sigma} < 2.11 \times 10^{-3} 
 \,, \notag \\
 0.72 \times 10^{-3} < &\left. (\epsilon'/\epsilon)_{\rm NP}\right|_{1.5\sigma} < 2.39 \times 10^{-3} 
 \,, \notag \\
 0.44 \times 10^{-3} < &\left. (\epsilon'/\epsilon)_{\rm NP}\right|_{2\sigma} < 2.67 \times 10^{-3} 
 \,.
 \label{eps:range}
 \end{align}

For the $\Delta M_K$ in Eq.(\ref{MK}), 
as was prescribed in Ref.~\cite{Endo:2016tnu},  
we may derive the limit simply by allowing 
the NP effect to come within the 2$\sigma$ uncertainty 
of the current measurement 
($ \Delta M_K^{\rm exp} = (3.484 \pm 0.006) \times 10^{-15}$ 
GeV~\cite{Patrignani:2016xqp}), such as  
\begin{align} 
 |\Delta M_K^{\rm NP} | < 3.496 \times 10^{-15}\,{\rm GeV} 
\,.   \label{MKbound:2sigma}
\end{align}

First, we constrain the model parameter space by  
the current bound on $\Delta M_K$ 
in Eq.(\ref{MKbound:2sigma}) and $\epsilon'/\epsilon$ in Eq.(\ref{eps:range}), 
which is shown in Fig.~\ref{epsp-deltaMK-constraint}
\footnote{
{Though the coefficient vector 
$\langle \vec{Q}_{\epsilon'} ( \mu)^T\rangle \hat{U} \left(\mu, m_{\rho} \right)$
in Eq.(\ref{epp}) is scale dependent, 
we have checked the dependence on the NP scale ($m_\rho$)~\cite{Kitahara:pri} is negligibly small enough among the focused range, compared to the required accuracy for the $\epsilon'/\epsilon$ of $\mathcal{O}(10^{-3})$.}
}.
The figure implies that as long as the $g_\rho$ takes the value in a range 
such as in Eq.(\ref{grho:size}), {$g_\rho \sim 10$}, 
the CFV mass $m_\rho$ is severely bounded to be around $\sim 1$ TeV, 
which is actually consistent with the $B-\tau$ system analysis described 
in the previous subsection. 
To address the discrepancy in ${\epsilon'}/{\epsilon}$ 
satisfying the constraint from $\Delta M_K$, 
the following conditions for $g^{12}_{\rho L}$ are required: 
the sign of $g^{12}_{\rho L}$ should be negative to enhance $\epsilon'/\epsilon$, 
and the magnitude of $g^{12}_{\rho L}$ is constrained to be at around of $\mathcal{O}(10^{-3})$.

Second, we take account of the constraints from the $D^0$-$\bar{D}^0$ mixing, where
large uncertainties are inevitable in contributions from the SM part as we have discussed in section~\ref{sec:D0D0bar}.
We exemplify the $95\%$ C.L. intervals in the assumed values for $x^\text{SM}$, $+ 1\%$ and $+0.1\%$.

Taking into account all the CFVs contributions to the $K$ and $D$ systems,   
in Fig.~\ref{K-sys-cons-SU(8)inv-grhoL-rev} 
we show the constraints on the coupling space $(g_{\rho L}^{12}, g_{\rho L}^{33})$ 
for $m_\rho = 1$ TeV, $g_\rho = 8$, {$\theta_L/\pi = 1/2$ and $\theta_D/\pi = 
(2 \text{ or } 1.5)
\times 10^{-3}$} as {benchmarks}}.  
Asymmetries for the $K^+ \to \pi^+ \nu \bar{\nu}$ (denoted in green) 
and the {$K_L \to \pi^0 \nu \bar{\nu}$} (in cyan) regarding the sign of the coupling 
$g_{\rho L}^{33}$ 
have been somewhat generated
due to the flavor-universal coupling $(1/g_\rho)$ term 
in Eq.(\ref{semi}). 
Interestingly enough, 
those Kaon decay rates have strong dependencies on 
the $g_{\rho L}^{33}$, where 
the pairs of neutrinos in the two processes are inclusively 
summed up, hence are significantly constrained by 
the $B - \tau$ system (horizontal lines in orange, in the figure), 
particularly, from the $\Delta M_{B_s}$ 
(placing the upper bound on the magnitude of $g_{\rho L}^{33}$ 
at $\theta_L \simeq \pi/2$) 
and the consistency with the {$R_{K^{(*)}}$} (setting the lower bound). 
This is the characteristic consequence derived from the present CFV scenario 
based on the one-family $SU(8)$ symmetry.\\[5pt]
\indent
A part of the region where $\epsilon'/\epsilon$ can be addressed
seems to be excluded by the constraint from the $D^0$-$\bar{D}^0$ mixing.
However, as shown in Fig.~\ref{K-sys-cons-SU(8)inv-grhoL-rev}, the $95\%$ C.L.
interval highly depends on the choice of the uncertain input $x^\text{SM}$ (refer to the discussion in section~\ref{sec:D0D0bar}).
Taking account of the uncertainty for $x^\text{SM}$, we can conclude that no definite bound is put on the red vertical domains
(for $\epsilon'/\epsilon$)
in Fig.~\ref{K-sys-cons-SU(8)inv-grhoL-rev}.
Overall Fig.~\ref{K-sys-cons-SU(8)inv-grhoL-rev} tells us that
there exist the parameter spaces for 
the present CFV scenario to simultaneously account for the two anomalies 
in $R_{K^{(\ast)}}$ and $\epsilon'/\epsilon$
within $2\sigma$ or $1\sigma$ C.L.

\begin{figure}[H]
\begin{center}
\includegraphics[width=6.5cm]{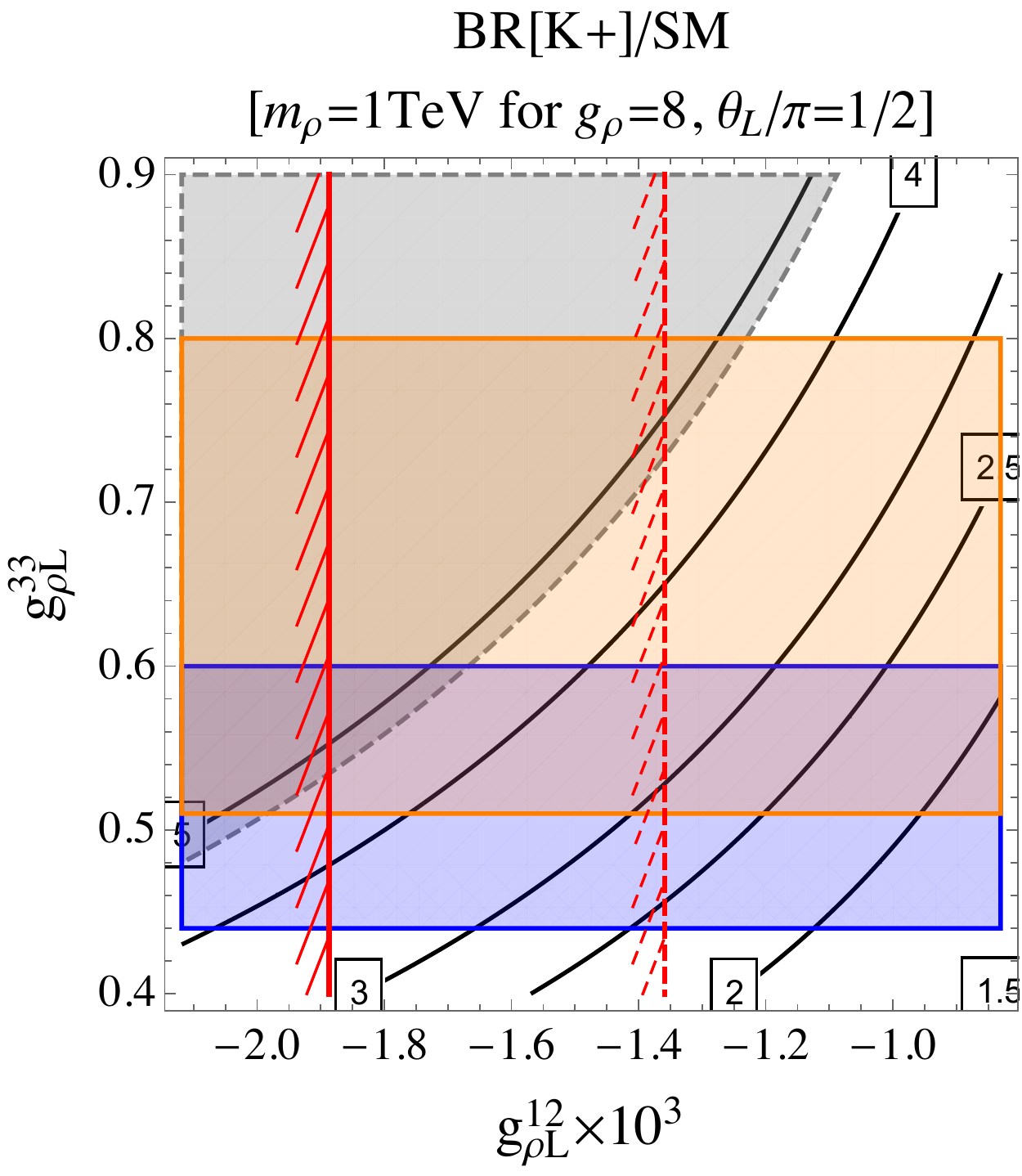} \quad
\includegraphics[width=6.5cm]{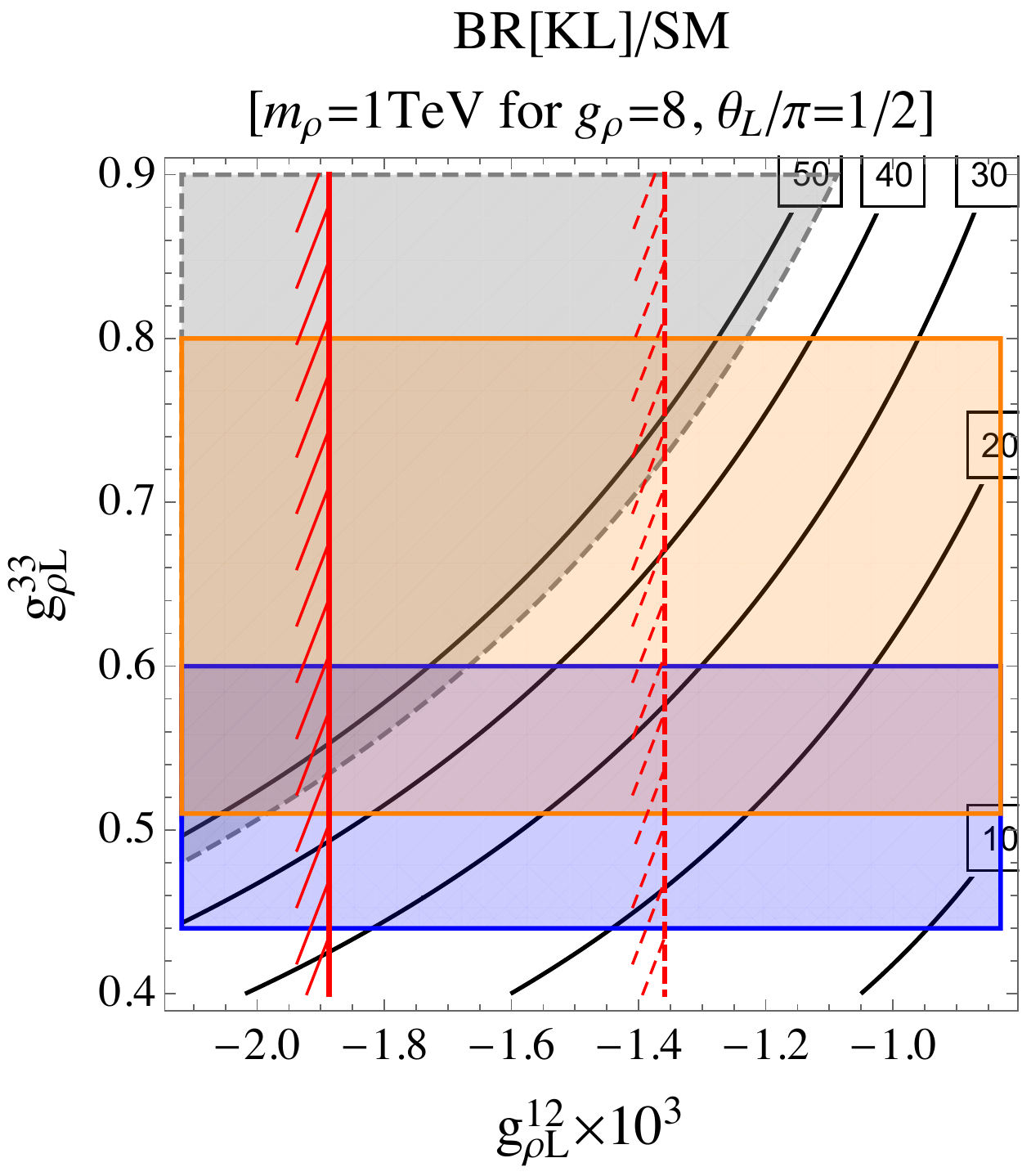} \\[10pt]
\includegraphics[width=6.5cm]{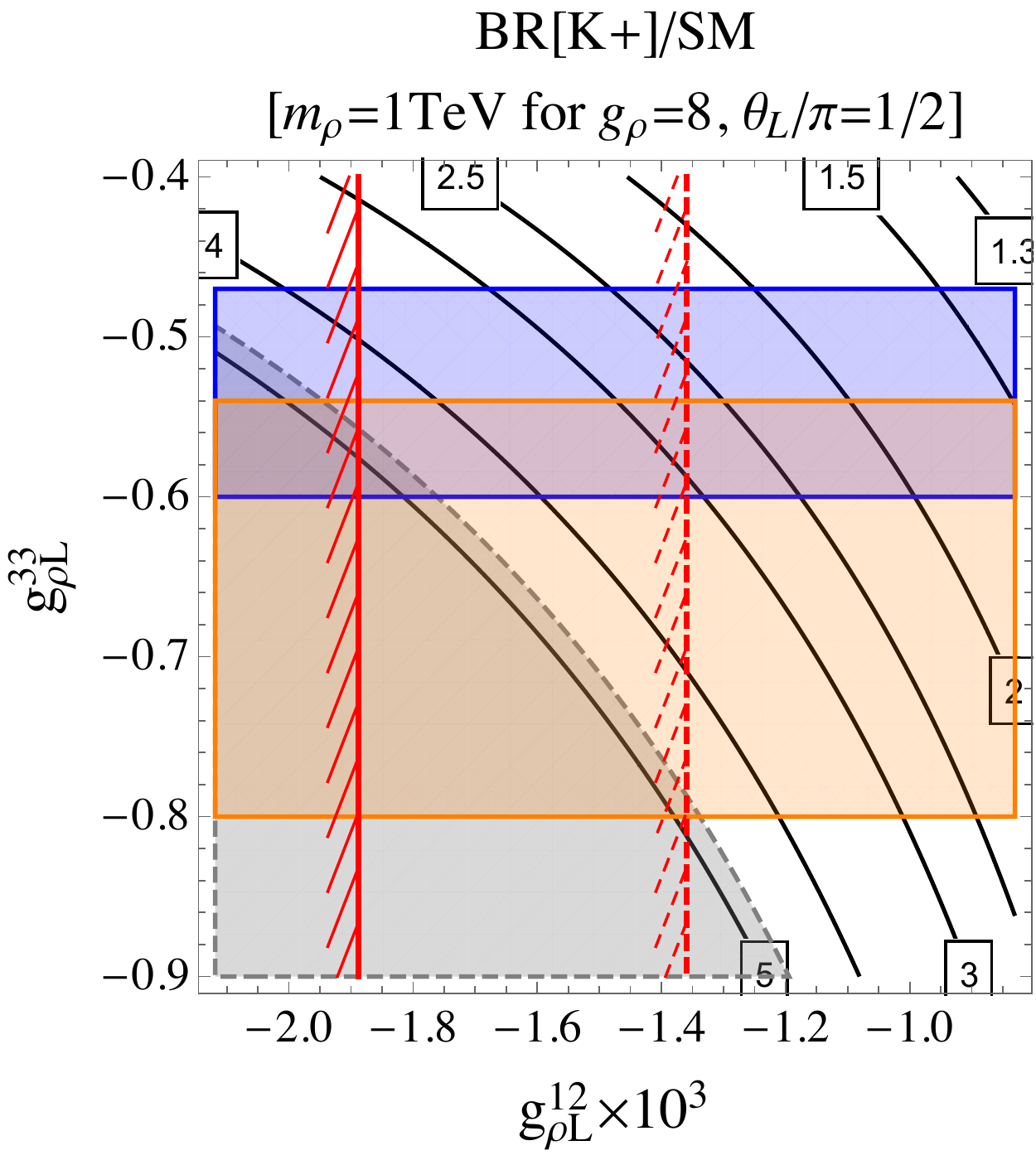} \quad
\includegraphics[width=6.5cm]{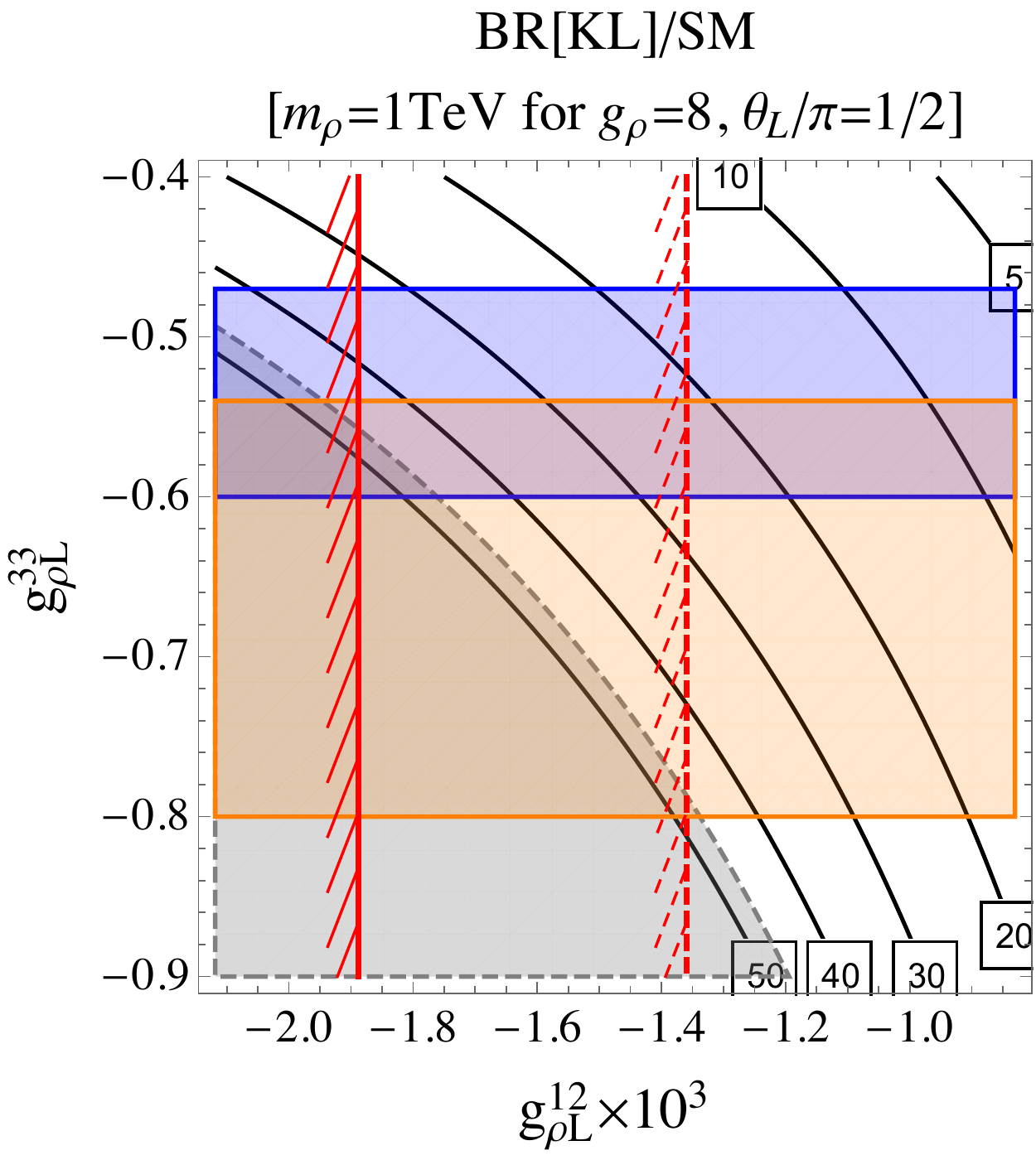}
\caption{
The plots of the predicted curves 
for the ${\rm Br}[K^+ \to \pi^+ \nu \bar{\nu}]$ 
(top- and bottom-left panels) 
and ${\rm Br}[K_L \to \pi^0 \nu \bar{\nu}]$ 
(top- and bottom-right panels), normalized to the SM values, 
in the $(g_{\rho L}^{12}, g_{\rho L}^{33})$ plane with $m_\rho =$ 
1 TeV, $g_\rho=8$ {and $\theta_L = \pi/2$ fixed}. 
The numbers attached on the curves denote the values of 
evaluated branching ratios over the SM predictions.  
The plotted ranges for $g_{\rho L}^{12}$ and $g_{\rho L}^{33}$ 
have been zoomed in on a viable parameter space
extracted from Fig.~\ref{K-sys-cons-SU(8)inv-grhoL-rev}, 
which fully satisfies the $\Delta M_K$ bound and 
are separated into two, depending on the sign of
$g_{\rho L}^{33}$ 
(shown in top and bottom panels for positive 
and negative cases, respectively).
The two red-vertical lines in each panel depict boundaries set by   
the $1\sigma$ (solid) and $1.5\sigma$ (dashed) ranges 
allowed by the $\epsilon'/\epsilon$ constraint.  
The parameter spaces inside the blue and orange regions 
enable us to address the $R_{K^{(\ast)}}$ anomaly 
with $\theta_D/\pi = 2.0 \times 10^{-3}$ and $1.5 \times 10^{-3}$, respectively.
The shaded regions colored in gray have already been excluded by 
the current bound on $\text{Br} \left[K^+ \to \pi^+ \nu \bar{\nu} \right]$ 
at the $2\sigma$ level~[see Eq.(\ref{eq:Kp_to_pipnunubar__experimental-value})].
We skipped to depict the $95\%$ C.L. intervals drawn by $x^\text{SM} = +1\%$ and $+0.1\%$
shown in Fig.~\ref{K-sys-cons-SU(8)inv-grhoL-rev}
(where a large uncertainty exists in the choice of $x^\text{SM}$).
}
\label{KOTO-NA62-LHC-B-tauphilic-prospect-v2}
\end{center}
\end{figure}

\section{Future prospects}
\label{sec:future}

\subsection{NA62 and KOTO experiments} 

The CFVs predict 
the deviation from the SM for 
the ${\rm Br}[K^+ \to \pi^+ \nu \bar{\nu}]$ 
and ${\rm Br}[K_L \to \pi^0 \nu \bar{\nu}]$, 
which can be tested in 
the upcoming data from the NA62 experiment 
at CERN~\cite{NA62:2017rwk} 
and KOTO experiment at J-PARC~\cite{Ahn:2016kja}. 
The SM predictions are read off, say, from Ref.~\cite{Endo:2016tnu}, 
as  
$ 
{\rm Br}[K^+ \to \pi^+ \nu \bar{\nu}] |_{\rm SM}
 = (8.5 \pm 0.5) \times 10^{-11}
 $ and 
 $ 
 {\rm Br}[K_L \to \pi^0 \nu \bar{\nu}]|_{\rm SM} 
 = (3.0 \pm 0.2) \times 10^{-11} 
$.  
Remarkable to note is that 
as long as 
the $R_{K^{(*)}}$ anomaly persists
beyond the SM, 
the CFVs necessarily give the larger values for those branching ratios 
than the SM predictions. 
In correlation with 
the $R_{K^{(*)}}$ anomaly,
the size of the deviations for the $K^+$ and $K_L$ decay rates 
significantly depends on the flavorful coupling $g_{\rho L}^{33}$ 
as seen from 
Fig~\ref{K-sys-cons-SU(8)inv-grhoL-rev}.
Suppose that 
the $R_{K^{(*)}}$ anomaly
will go away in the future.  
In that case, the lower bound on the $g_{\rho L}^{33}$ will not be placed, 
so we then expect 
from the formula in Eq.(\ref{semi})
that by adjusting the couplings $g^{12}_{\rho L}$ and $g_{\rho L}^{33}$,  
the CFV contributions to the 
$K^+ \to \pi^+ \nu \bar{\nu}$ decay can be vanishing or 
even make the branching ratio slightly smaller than the SM prediction. 
According to the literature~\cite{NA62:2017rwk}, 
by the end of 2018 the NA62 experiment will measure  
the $K^+ \to \pi^+ \nu \bar{\nu}$ with about 10\% accuracy 
of the SM prediction, 
while the Belle II experiment is expected to 
measure the deviation on the $R_{K^{(*)}}$ at about 3\% level 
with $\sim 10\, {\rm ab}^{-1}$ data up until 2021~\cite{Ishikawa}. 
(The current accuracy for the $\delta R_{K^{(*)}} 
\equiv R_{K^{(*)}}^{\rm exp} -R_{K^{(*)}}^{\rm SM}$ is at least about 
$30\% - 40\%$~\cite{Aaij:2017vbb,Aaij:2014ora}.) 
The KOTO experiment also plans to report new results on the 
data analysis on the $K_L \to \pi^0 \nu \bar{\nu}$ in the near future, 
to be expected to reach the level of $< 10^{-9}$ for the branching ratio, 
corresponding to 2015 - 2018 data taking~\cite{KOTO2018}.   
The CFV scenario will therefore be very soon tested first by 
the NA62 and KOTO, 
which will constrain the size of the  flavorful coupling $g_{\rho L}^{33}$ 
and the allowed deviation for the  $R_{{K^{(*)}}}$, 
and then will be confirmed or excluded by the upcoming Belle II data. 

Assuming that the 
$R_{K^{(*)}}$ anomaly persists
in the future, 
in Fig.~\ref{KOTO-NA62-LHC-B-tauphilic-prospect-v2} 
we display the plots 
showing the predicted curves  
for the ${\rm Br}[K^+ \to \pi^+ \nu \bar{\nu}]$ (top- and  bottom-left panels) 
and ${\rm Br}[K_L \to \pi^0 \nu \bar{\nu}]$ (top- and bottom-right panels)
in the $(g_{\rho L}^{12}, g_{\rho L}^{33})$ plane with 
$g_\rho=8$, {$m_\rho =1$ TeV and $\theta_L = \pi/2$} fixed.
{For $\theta_D$, we chose the two relevant values as reference points,  
which lead to simultaneous explanations for the anomalies in
$R_{K^{\ast}}$ (within the $2\sigma$ C.L.) and $\epsilon'/\epsilon$ (within the $1\sigma$ C.L.).}
The parameter spaces displayed in the figure have taken into account  
currently available all flavor limits 
together with 
the $R_{K^{(*)}}$ anomaly
 (see Fig.~\ref{K-sys-cons-SU(8)inv-grhoL-rev}).  
The figure implies that 
the significantly large branching ratios for 
the  ${\rm Br}[K^+ \to \pi^+ \nu \bar{\nu}]$ (by about a few times larger amount)
and ${\rm Br}[K_L \to \pi^0 \nu \bar{\nu}]$ (by about several ten times larger 
amount) are predicted in the presence of 
the $R_{K^{(*)}}$ anomaly.

\subsection{LHC searches} 

Since the CFVs having the mass of around 1 TeV 
couple to the SM fermions in the flavorful form as well as 
in the flavor-universal form,  they can potentially 
have a large enough sensitivity to be detected also at the LHC.  
As has been discussed so far, 
the flavor-physics analysis implies a muon-philic structure 
{[}$\theta_L \sim \pi/2$ in Eq.(\ref{3-cons}){]} and the CFVs are 
allowed to couple to the $u$ and $d$ quarks through 
the mixing with the SM gauge bosons {[}see Eq.(\ref{indirect:rho-coupling}){]}, 
so the most dominant discovery channel will be a resonant dimuon process 
$pp \to {\rm CFVs} \to 
\mu^+ \mu^-$, in which the neutral CFVs mixing with the EW gauge bosons 
(i.e. $\rho_{(1)}^3$ and $\rho_{(1)^\prime}^0$) are generated by $u \bar{u}$ 
and $d \bar{d}$ via Drell-Yan process~\footnote{
Actually, the bottom quark pair can also produce 
the neutral CFVs via the third-generation-philic coupling 
$g_{\rho L}^{33}$, though the contribution would be small due to the small bottom luminosity 
inside proton. We have checked that 
this $pp \to b \bar{b} \to $CFVs production is slightly subdominant to be by about 
a few factor smaller than the $pp \to u\bar{u}/d \bar{d} \to $ CFVs production, 
even for the suppressed {($g^2_{s,W,Y}/g_{\rho}$)} coupling. 
Even with inclusion of the $b \bar{b} \to $CFVs production, however, 
our discussion on the LHC searches would not substantially be altered, 
as will be manifested in Fig.~\ref{with-Gamma-1TeV-rev}.   
}.

\begin{figure}[t]
\begin{center}
\includegraphics[width=0.6\columnwidth]{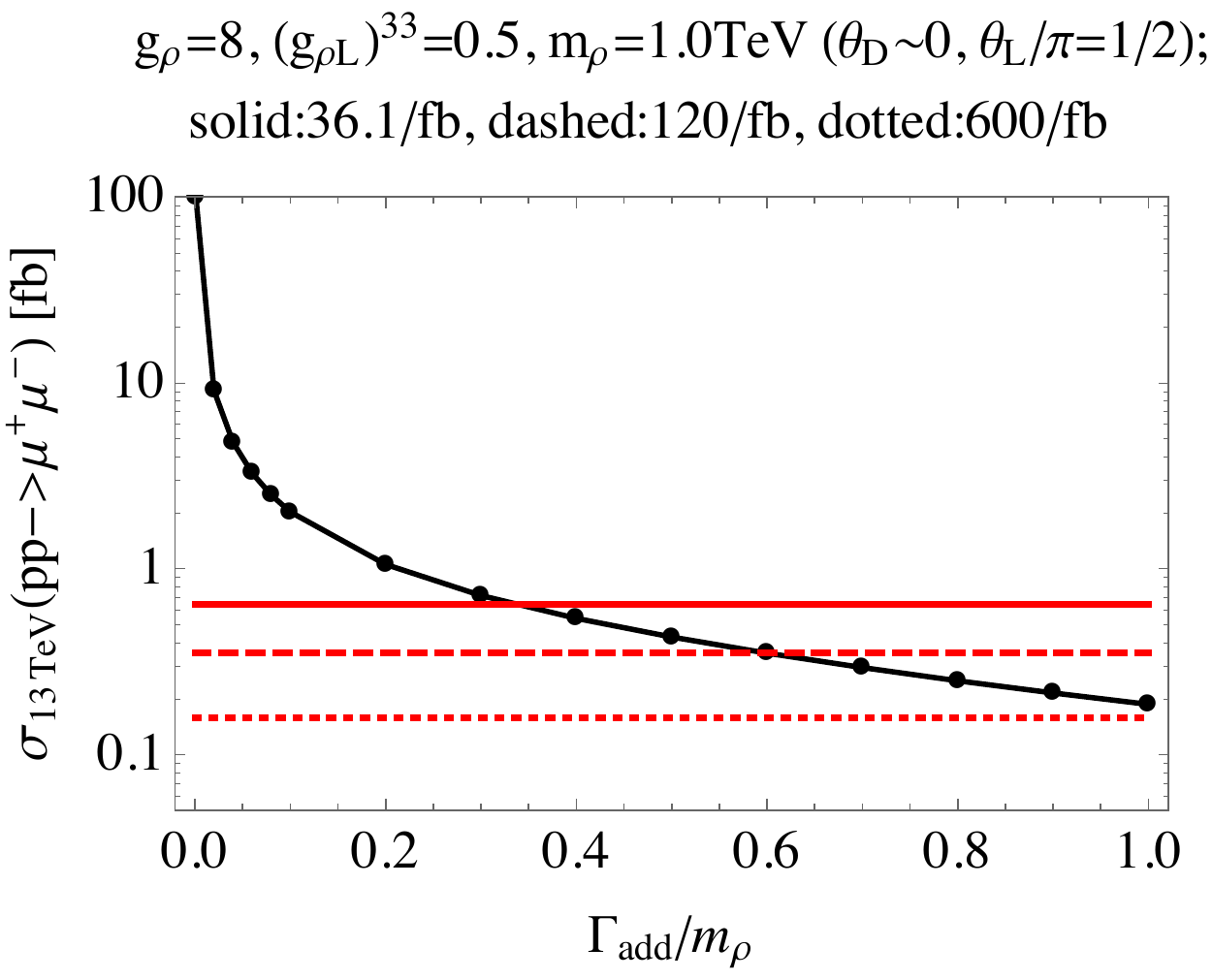}
\caption{
The dimuon resonant production cross section for the target 
CFVs ($\rho_{(1)}^3$ and $\rho_{(1)^\prime}^0$) at LHC with $\sqrt{s}=$13 TeV 
as a function of a possibly added width term (common for two CFVs) 
normalized to the mass $m_\rho$,  
for $m_\rho$= 1 TeV, $g_\rho=8$, {$g_{\rho L}^{33}=0.5$ (and $\theta_D \sim 0$, $\theta_L = \pi/2$)}. 
The horizontal {solid, dashed and dotted} lines (in red)
respectively correspond to the current 95\% C.L upper limit 
placed by the ATLAS group with the integrated luminosity 
${\cal L}=36.1 \,{\rm fb}^{-1}$~\cite{ATLAS:2017wce}, 
and the expected upper bounds at ${\cal L}=120 \,{\rm fb}^{-1}$ 
and {${\cal L}=600\,{\rm fb}^{-1}$}   
estimated just by {simply} scaling the luminosity.
The LHC cross section has been computed by implementing 
the {\tt CTEQ6L1} {parton distribution function~(PDF)}~\cite{Pumplin:2002vw} 
in {\tt Mathematica} with the help of a PDF parser package, 
{\tt ManeParse\_2.0}~\cite{Clark:2016jgm}, and setting 
{$\tau_0\equiv 4 m^2_{\rm threshold}/s= 10^{-6}$} 
as the minimal value of the Bjorken $x$ in the {\tt CTEQ6L1} PDF set, 
where the PDF scale is set to $m_\rho$.
The {\tt CUBA} package~\cite{Hahn:2004fe} {has been}
utilized for numerical integrations.
}
\label{with-Gamma-1TeV-rev}
\end{center}
\end{figure}

In Fig.~\ref{with-Gamma-1TeV-rev} we show the dimuon 
production cross section at 13 TeV 
for a viable parameter space with 
$m_\rho$= 1 TeV, $g_\rho=8$, {$g_{\rho L}^{33}=0.5$ (and $\theta_D \sim 0$, $\theta_L = \pi/2$) 
[see Fig.~\ref{K-sys-cons-SU(8)inv-grhoL-rev}]}. 
It is seen from the figure that 
the present scenario, in which the CFVs are 
allowed to couple only to the SM fermions,  seems to have a strong tension with 
the current dimuon data (solid red line) 
provided by 
the ATLAS group~\cite{ATLAS:2017wce}  
(see the black-dot point {$\simeq {114}\,{\rm fb}$} 
at $\Gamma_{\rm add}/m_\rho=0$).

Assuming a possible extension from 
the simplest setup and incorporating 
an additional width (other than the widths for 
SM fermions decays) into the CFV propagators, 
we may vary the extra width term to control the 
dimuon cross section {(see e.g.,~\cite{Boucenna:2016qad,Cline:2017lvv,Asadi:2018wea,Greljo:2018ogz})}. 
Such an additional width would be present 
when the CFV can dominantly couple to a 
hidden dark sector including a dark matter candidate, 
or a pionic sector realized 
as in a hidden QCD with a setup similar to the present CFV 
content~\cite{Matsuzaki:2017bpp}. 
In a hidden QCD embedding, for instance, 
the additional width almost fully saturated by the 
decay channel to hidden pion pairs
would be expected like 
$\Gamma/m_\rho \sim |g_\rho|^2/(48 \pi) \simeq 
0.2(0.4)$ for $g_\rho=6(8)$, based on 
a simple scaling from the QCD case  
 (up to possible group factors 
depending on the number of hidden-sector flavors).

The additional effect is monitored also in Fig.~\ref{with-Gamma-1TeV-rev}, 
where the extra width (denoted by $\Gamma_{\rm add}$) is assumed 
to be in common for the target $\rho_{(1)}^3$ and $\rho_{(1)^\prime}^0$
 and  
is normalized to their common mass $m_\rho$. 
The figure shows that the CFVs with {$\Gamma_{\rm add}/m_\rho \gtrsim 30\%$} 
can survive for the current dimuon bound. 
(The dimuon widths for $\rho_{(1)}^3$ and $\rho_{(1)^\prime}^0$ 
are {$\simeq 1.75$ GeV and $\simeq 1.39$ GeV for $(g_{\rho L}^{33}, g_\rho) = (0.5, 8)$}, 
so the total widths are indeed dominated by the extra sector.) 
The size of the additional width can be explored up to around {60\%} at 
the LHC with ${\cal L}=120\,{\rm fb}^{-1}$ 
and will be further done in the phase of a high-luminosity LHC 
with {${\cal L}=600\,{\rm fb}^{-1}$}, 
which would constrain modeling of a concrete CFV scenario 
with the CFVs coupled to a hidden sector.

\section{Summary and {discussion}} 
\label{sec:summary} 

In this paper, we have proposed 
 flavorful and 
chiral vector bosons as the new physics constitution at around TeV scale, 
to address the presently reported significant flavor anomalies 
in the Kaon and $B$ meson systems such as the CP violating Kaon
decay $\epsilon'/\epsilon$ and lepton-flavor violating 
$B$ meson decays. 
We have introduced 
the chiral-flavorful vectors (CFVs) 
as a 63-plet of 
the global $SU(8)$ symmetry, identified as 
the one-family symmetry for left-handed quarks and leptons 
in the standard model (SM) forming the 8-dimensional vector. 
Thus the CFVs include massive gluons (of $G'$ type), vector leptoquarks, and 
$W',Z'$-type bosons, which are allowed to have 
flavorful couplings with left-handed quarks and leptons, and flavor-universal 
couplings to right-handed ones, where the latter arises from 
mixing with the SM gauge bosons. 
The characteristic feature 
in the present CFV scenario based on the one-family $SU(8)$ 
symmetry is seen in the predictions 
derived necessarily with a significant correlation 
between the $2 \leftrightarrow 3$ and $1 \leftrightarrow 2$ 
transition processes{,}
while the next-to nearest-generation transitions 
like $1 \leftrightarrow 3$ processes turn out to be highly 
suppressed in a correlation with 
the $1 \leftrightarrow 2$ transition constraints in the $K$ system: 
based on the proposed flavor texture, 
the current $\epsilon'/\epsilon$ (in the Kaons) and $R_{K^{(\ast)}}$ (in the $B$ mesons) anomalies can simultaneously 
be interpreted by the presence of CFVs, where 
it is predicted that the $R_{D^{(*)}}$ anomaly does not survive
due to the (approximate) one-family $SU(8)$ symmetry (see section~\ref{sec:RD-anomaly} for details).  
Remarkably, we found that 
as long as both the $\epsilon'/\epsilon$ and the $R_{K^{(\ast)}}$ anomalies 
persist beyond the SM,   
the CFVs 
predict the enhanced $K^+ \to \pi^+ \nu \bar{\nu}$ and $K_L \to \pi^0 \nu\bar{\nu}$ 
decay rates compared to the SM values, which will readily be explored  
by the NA62 and KOTO experiments, and would also be explored 
in the dimuon resonant channel at the LHC with a higher luminosity.

In closing, we shall give several comments: 
 
\begin{itemize}

\item
Since (some of) the CFVs can mix with the SM gauge bosons at tree level, 
one might think that the present scenario gets severely constrained 
by the EW precision tests, as well as the flavor observables 
and direct searches at LHC, as has so far been discussed. 
However, it turns out not to be the case for the CFV model  
which is due to the vectorlike gauging and the formulation based the hidden local symmetry approach~\cite{Bando:1984ej,Bando:1985rf,Bando:1987ym,Bando:1987br,Harada:2003jx}:     
{as was discussed in~\cite{Matsuzaki:2017bpp} 
where the mixing with the SM gauge boson has been 
determined by the hidden gauge symmetry structure 
in the same way as in the present CFV model,  
the EW precision test constraints, which mainly 
come from the flavor-dependent $Z$-mass shell observables 
(such as forward-backward asymmetries for charged leptons 
and $Z \to \bar{b}b$ decay),  
would place the bound for the flavorful coupling 
$g_{\rho L}^{33}$ at $95\%$ C.L.~\cite{Matsuzaki:2017bpp}, 
$-0.72 \times \left( \frac{m_\rho}{1\,\text{TeV}} \right)^2 
< g_{\rho L}^{33}/g_\rho 
< 0.25 \times \left( \frac{m_\rho}{1\,\text{TeV}} \right)^2
$,   
with the third-generation philic and the second-generation philic 
textures assumed for quarks and leptons, respectively. 
This bound is easily fulfilled in a typical benchmark such as 
$m_\rho = 1\,\text{TeV}$, $g_{\rho L}^{33} = +0.5$, and $g_\rho = 8$.  

Though the tree-level contributions are somewhat insignificant,  
EW renormalization group effects on the semi-leptonic current operators 
could give non-negligible  
corrections to 
the $W$ and $Z$ boson currents at the NLO level (referred to as EW-NLO below, in short), 
hence could be severely constrained by 
the EW precision measurements 
and lepton-flavor violating observables~\cite{Feruglio:2016gvd,Feruglio:2017rjo} (see also~\cite{Falkowski:2017pss,Buttazzo:2017ixm,Bordone:2017bld,Bordone:2018nbg}).  
Of interest is first to note  
the almost complete cancellation in the $SU(2)$-triplet(-contracted) semi-leptonic operators  
as the consequence of the $SU(8)$-one family symmetry structure, 
leading to null correction to the $W$-boson current from such EW-NLO effects. 
On the other hand, EW-NLO corrections to the $Z$-boson currents can be 
induced to give rise to nontrivial shifts for vector/axial-vector $Z$-couplings 
to the lepton $\ell$ $(v_\ell/a_\ell)$ 
and for the number of (active) neutrinos 
from the invisible $Z$ decay width ($N_\nu$). 
Following the formulation in~\cite{Feruglio:2016gvd,Feruglio:2017rjo}  
and taking the benchmark-parameter set in the above 
(with the favored mixing structure $\theta_D \sim 0$ and $\theta_L \sim \pi/2$ 
taken into account), 
we estimate the shifts for those couplings to find 
$ 
\left(\frac{v_\mu}{v_e} - 1\right) \sim 5 \times 10^{-3}, 
\left(\frac{a_\mu}{a_e} - 1\right) \sim 4 \times 10^{-4}, 
\left(N_\nu - 3\right) \sim 8 \times 10^{-4}
$. 
which are (maximally by one order of magnitude) small enough and 
totally safe 
when compared with the current bounds~\cite{Patrignani:2016xqp}  
${v_\mu}/{v_e}|_{\text{exp}} = 0.961(61)$, ${a_\mu}/{a_e}|_{\text{exp}} = 1.0002(13)$, 
and $N_\nu|_{\text{exp}} = 2.9840 \pm 0.0082$. 
This is actually mainly due to the almost vanishing contributions 
from the $SU(2)$-triplet(-contracted) semi-leptonic operators: 
if they were sizable, the shifts would get as large as 
the current accuracies for the $v_\mu/v_e, a_\mu/a_e$ and $N_\nu$, 
to have a strong tension with the $B$ anomalies, as emphasized in~\cite{Feruglio:2016gvd,Feruglio:2017rjo}. 

Besides those potentially nontrivial NLO corrections as above, 
similar effects on flavor-dependent $Z$-pole observables 
could arise from renomalization evolutions for 
fully quarkonic and leptonic operators, which 
is to be pursued elsewhere.}

\item 
As discussed recently in Ref.~\cite{Aloni:2017eny}, 
flavorful couplings for vector leptoquarks and $W'/Z'$-type vectors such as those 
in the present CFV model 
are potentially 
sensitive also to the LFU of $\Upsilon(nS)$ ($n=1,2,3$) decays, 
when those couplings are large enough to account for 
the $B$ anomalies. 
Note that in the present CFV model 
both semi-leptonic and fully-leptonic decay processes for $b$ 
are controlled by the same Wilson coefficient of type $C_{qqll}^{[1]}$ 
(see Eq (\ref{eq:effective_operators}) and recall the consequence of 
the approximate $SU(8)$ invariance leading to $C_{qqll}^{[3]} \simeq 0$), where the vector-leptoquarks ($\rho_{(3)}^{0, \alpha}$) 
dominantly contribute (see Eq.(\ref{C1:qqll})). 
Hence the bound on the CFV contribution 
can roughly be translated in terms of a isotriplet-vector leptoquark scenario 
when we talk about the correlated limit from 
the LFU of $\Upsilon(nS)$ decays and the $B$ anomalies 
(by roughly saying that the sensitivities to the flavorful coupling 
are the same for the $R_{D^{(*)}}$, which was taken into account in 
Ref.~\cite{Aloni:2017eny}, and the $R_{K^{(*)}}$, which has been addressed 
in the present model). 

In fact, just following the formulae available in Ref.~\cite{Aloni:2017eny}, 
we have explicitly evaluated 
the CFVs contributions to the LFU of $\Upsilon(1S,2S,3S)$ decays, 
characterized by the ratio of the decay rates, 
$R_{\tau/\mu}^{\Upsilon(nS)} \equiv \Gamma[\Upsilon(nS) \to 
\tau^+ \tau^-]/\Gamma[\Upsilon(nS) \to \mu^+ \mu^-]$ 
for the viable parameter set explored by the present analysis 
($\theta_L\simeq \pi/2, \theta_D \simeq 0, g_{\rho L}^{33} \simeq 0.5, g_\rho =8, 
m_\rho=1$ TeV),  
to find that 
$R_{\tau/\mu}^{\Upsilon(1S, 2S, 3S)} \simeq (0.992, 0.994, {0.994})$, 
in accord with the numbers estimated in the reference.  
With the SM predictions subtracted, 
the numbers actually turn out to be {smaller than} 
the current 1$\sigma$ uncertainties in experiments~\cite{Besson:2006gj,delAmoSanchez:2010bt}. 
Thus, the CFVs are at present fairly insensitive to the LFU of $\Upsilon(nS)$ decays, 
which would possibly be explored in Belle II experiment.

\item 

For the $\Delta M_K$ in Eq.(\ref{MK}), 
we could quantify the NP effect for the deviation from the SM 
by adopting the recent lattice result on the SM prediction~\cite{Bai:2014cva}, 
$\Delta M_K^{\rm SM}|_{\rm lattice} = (3.19 \pm 1.04) \times 10^{-15}$ GeV, 
which includes significant 
long distance contributions, 
$ \Delta M_K^{\rm NP} \equiv 
 \Delta M_K^{\rm exp} - \Delta M_K^{\rm SM}|_{\rm lattice} 
 = (0.29 \pm 1.04) \times 10^{-15} \,{\rm GeV}
$, 
by which the $2\sigma$ allowed range is found to be 
$ 
- 1.79 \times 10^{-15}\,{\rm GeV} 
\le \Delta M_K^{\rm NP}|_{2\sigma} \le 
2.37 \times 10^{-15}\,{\rm GeV} 
$. 
With this bound used, which gets {to be more shrunk} than 
that we have adopted in the present analysis, 
the NP contribution of $(\epsilon'/\epsilon)_{\rm NP}$ 
would be necessary to be allowed up to {$\gtrsim 1.5\sigma$ range}, 
to be consistent with the $\Delta M_K$, 
while the predicted values for the rare $K$-semi-leptonic decays 
would not be {substantially} affected in magnitude. 
Improvement on the accuracy for lattice estimates on the form factors $B_6^{1/2}$ and $B_8^{3/2}$, 
crucial for the SM prediction to $\epsilon'/\epsilon$,  
would give a more definite constraint on the present CFV scenario, 
in correlation with the fate of $B$ anomalies.

\item 
In addition to the single-neutral CFV ($\rho_{(1)}^3$ and $\rho_{(1)^\prime}^0$) 
production at LHC as has been discussed so far,  
one might think that the vector-leptoquark 
pair ($\rho_{(3)}^{\alpha,0}$) production would also severely be constrained 
by the LHC searches. 
However, it would not be the case: 
when the CFV model is formulated based on 
the hidden local symmetry 
approach~\cite{Bando:1984ej,Bando:1985rf,Bando:1987ym,Bando:1987br,Harada:2003jx}, 
the gluon-gluon fusion coupling to the vector leptoquark pair, 
which has been severely constrained by the null-results at the LHC,  
can be set to be suppressed as discussed in~\cite{Matsuzaki:2017bpp} with a similar 
model-setup for the vector bosons.  
Instead of such a direct coupling, the color-octet CFV ($\rho_{(8)}^0$)  
exchange would then dominate to couple the gluon to 
the vector-leptoquarks (so-called ``vector meson dominance''), 
so that the effective vertex goes like a form factor 
$\sim m_{\rho}^2/(m_\rho^2 - \hat{s})$,  
where $\hat{s}$ denotes the square of the transfer momentum. 
Because of the almost degenerate mass structure for CFVs supported 
by the $SU(8)$ symmetry, 
the on-shell production for the leptoquark pair 
through the $\rho_{(8)}^0$ exchange is kinematically forbidden, 
hence only the off-shell production is allowed to be 
highly suppressed by the form factor structure for
a  high energy event at LHC, like $\sim m_\rho^2/\hat{s}$. 
A similar argument is also applicable to the Drell-Yan process 
($q \bar{q} \to \rho_{(3)} \bar{\rho}_{(3)}$ along with 
the similar form factor structure).       
Thus, the presently placed bound on the vector leptoquark pair 
production is not directly applicable to the CFVs detection 
(even if the vector-leptoquarks sub-dominantly decay to quarks and leptons 
due to a presence of some hidden sector). 
It would need other event topologies to
detect the CFV-leptoquarks with the characteristic final state reflecting the {second-generation- and third-generion-philic properties} for leptons and quarks, respectively,\footnote{
{For the case of a scalar leptoquark candidate $(S_1)$ for explaining the $R_{D^{(\ast)}}$ anomaly, e.g., see~\cite{Dumont:2016xpj} for details on flavor tagging.}
}
which would be accessible in a high-luminosity LHC through a single production process \cite{MoriondCMS} with tagging muons, instead of tau leptons.


\item
Other flavor-changing processes like
$\bar{B} \to K^{(\ast)} \mu^\pm \tau^\mp$, $\Upsilon \to \mu^\pm \tau^\mp$ and $J/\psi \to \mu^\pm \tau^\mp$,
have been discussed to place limits on
the scenarios for addressing the $R_{K^{(\ast)}}$ anomaly, e.g. in~\cite{Kumar:2018kmr}.
It turns out that no additional bound from them emerges within 
our focused parameter space in Fig.~\ref{B-tau-sys-cons-SU(8)inv-grhoL-rev},
where $g_{\rho L}^{33}$ and $\theta_L/\pi$ vary in a range  $[-1.5, 1.5]$ and $[0, 0.5]$, while the other relevant parameters are fixed
as $m_\rho = 1 \,\text{TeV}$, $g_\rho = 8$, $\theta_D/\pi = 2 \times 10^{-3}$.
In the following part, we briefly look at this parameter space as a reference point for 
discussing the above flavor-changing processes in terms of our scenario.
\begin{itemize}
\item
$\bar{B} \to K^{(\ast)} \mu^\pm \tau^\mp$:

By referring to~\cite{Calibbi:2015kma,Kumar:2018kmr}, the following branching ratios are formulated
in terms of our notation as
$\text{Br}[\bar{B} \to K^{(\ast)} \mu^- \tau^+] \sim \left( \left| C_9^{\mu\tau} (\text{NP}) \right|^2 + \left| C_{10}^{\mu\tau} (\text{NP}) \right|^2 \right)
\times 10^{-8}$,
and the conjugated processes $\bar{B} \to K^{(\ast)} \mu^+ \tau^-$ are obtained by the replacement of
$C_{9,10}^{\mu\tau} (\text{NP}) \to C_{9,10}^{\tau\mu} (\text{NP})$.
The current experimental results for $B^+ \to K^+ \mu \tau$ are available:
$\text{Br}[B^+ \to K^+ \mu^+ \tau^-] = 0.8^{+1.9}_{-1.4} \times 10^{-5}$ or $< 4.5 \times 10^{-5}\,(90\%\,\text{C.L.})$;
$\text{Br}[B^+ \to K^+ \mu^- \tau^+] = (-0.4)^{+1.4}_{-0.9} \times 10^{-5}$ or $< 2.8 \times 10^{-5}\,(90\%\,\text{C.L.})$~\cite{Lees:2012zz}.
Our scenario is thus apparently insignificant for those processes because 
$|C_{9,10}^{\mu\tau} (\text{NP})| \sim |C_{9,10}^{\mu\mu} (\text{NP})| \to \, \lesssim 1 \, (\sim 2\sigma\,\text{upper bound})$ 
with fixing $m_\rho = 1\,\text{TeV}$, $\theta_D/\pi = 2.0\times10^{-3}$ and varying $g_{\rho L}^{33}$ in the range $[-1.5,1.5]$.
\item
$\Upsilon(nS) \to \mu^\pm \tau^\mp$:

The formulae of the branching ratios are available in~\cite{Kumar:2018kmr},
and from the reference we can read off a consistent condition for $m_\rho = 1\,\text{TeV}$ as
${7}/{16} \, (g_{\rho L}^{33})^2 (\sin{\theta_L} \cos{\theta_L}) \lesssim 4$.\footnote{
We used experimental data for $\text{Br}[\Upsilon(2S,\,3S) \to \mu^\pm \tau^\mp]$
as $\text{Br}[\Upsilon(2S) \to \mu^\pm \tau^\mp] = 0.2^{+1.5+1.0}_{-1.3-1.2} \times 10^{-6}$ or
$< 3.3 \times 10^{-6} \,(90\%\,\text{C.L.})$, and
$\text{Br}[\Upsilon(3S) \to \mu^\pm \tau^\mp] = (-0.8)^{+1.5+1.4}_{-1.5-1.3} \times 10^{-6}$ or
$< 3.1 \times 10^{-6} \,(90\%\,\text{C.L.})$~\cite{Lees:2010jk}.
Here the associated parameters used as inputs are
$f_{\Upsilon(2S)} = (496 \pm 21)\,\text{MeV}$,
$f_{\Upsilon(3S)} = (430 \pm 21)\,\text{MeV}$,
$m_{\Upsilon(2S)} = 10.02326\,\text{GeV}$,
$m_{\Upsilon(3S)} = 10.35526\,\text{GeV}$,
$\Gamma_{\Upsilon(2S)} = (31.98 \pm 2.63)\,\text{KeV}$, and
$\Gamma_{\Upsilon(3S)} = (20.32 \pm 1.85)\,\text{KeV}$, respectively~\cite{Kumar:2018kmr,Tanabashi:2018oca}.
}
Within the focused range of $g_{\rho L}^{33}$ as $[-1.5,1.5]$, this condition is satisfied and
hence the focused spot remains viable in the range.
\item
$J/\psi \to \mu^\pm \tau^\mp$:

The formulae for these decay processes take the form similar to
 that for $\Upsilon(nS) \to \mu^\pm \tau^\mp$
and available in~\cite{Kumar:2018kmr}.
We find that the bound coming from this process (and its conjugate one) is much weaker than that from $\Upsilon(nS) \to \mu^\pm \tau^\mp$
(and the focused range is completely safe) since (i) much more sizable $g_{\rho L}^{33}$ is required for
putting a constraint on the focused range compared with $\Upsilon(nS) \to \mu^\pm \tau^\mp$;
(ii) $c\bar{c}$ coupling is suppressed when $\theta_D \ll 1$.
\end{itemize}

\item 
We make comments on other potentially-sensitive flavor-changing processes in the presence of 
the $W', Z', G'$ and vector leptoquark candidates under our flavor texture.
Here, we focus on typical digits on the benchmark point:
$m_\rho = 1\,\text{TeV}$, $g_\rho = 8$, $\theta_D/\pi = 2 \times 10^{-3}$, $\theta_L/\pi \sim 1/2$, $g_{\rho L}^{33} \sim +0.6 \text{ or } -0.6$,
and $|g_{\rho L}^{12}| \sim 2 \times 10^{-3}$, where 
the best-fit value for $b \to s \mu^+ \mu^-$ is achieved and
the $\epsilon'/\epsilon$ anomaly can be addressed within $1\sigma$ C.L. consistently (see Figs.~\ref{B-tau-sys-cons-SU(8)inv-grhoL-rev}
and \ref{K-sys-cons-SU(8)inv-grhoL-rev}).
\begin{itemize}
\item
$\mu \to e \gamma$:

This process is not produced at the tree level due to the gauge invariance of the QED, but the dipole operator
$\Delta_{\mu \to e \gamma} \, (\bar{e}'_R \sigma_{\mu \nu} \mu'_L) F^{\mu \nu}$ ($F_{\mu\nu}$ being the field strength tensor of the photon) is 
induced by the vector leptoquark exchange even when $\theta_L \sim \pi/2$ at the one loop level.
The coefficient $\Delta_{\mu \to e \gamma}$ is roughly estimated as 
$\sim {e \, m_\mu (g_{\rho L}^{12}) \, (g_{\rho L}^{33}) \, \sin\theta_D}/({(4\pi)^2 m_\rho^2})$.
At the focused point, we have $\text{Br}[\mu \to e \gamma] \sim 7 \times 10^{-16}$ with
$\tau_\mu = 2.197 \times 10^{-6}\,\text{s}$, $m_\mu = 0.1057\,\text{GeV}$~\cite{Tanabashi:2018oca}
used as inputs,
while the latest bound is $< 4.2 \times 10^{-13}\,(90\%\,\text{C.L.})$~\cite{TheMEG:2016wtm}.
No competitive bound may come in the near future
(where the targeting sensitivity of the MEG-I\!I experiment 
is expected to reach the level of 
$6 \times 10^{-14}$~\cite{Baldini:2018nnn}).

\item
$\mu \to 3e$:

This process is induced at the tree level, but the branching ratio is proportional to $\cos^2\theta_L$, 
which is vanishing at our benchmark point including $\theta_L \sim \pi/2$.
At the one loop level, even in the case of $\theta_L \sim \pi/2$, the operator $\Delta_{\mu \to 3e} \, 
(\bar{e}'_L \gamma_\mu \mu'_L) (\bar{e}'_L \gamma^\mu e'_L)$ 
is induced through box diagrams with vector leptoquark exchanges, the coefficient of which ($\Delta_{\mu \to 3e}$)
is roughly estimated as
$\sim ({m_c^2}/{m_\rho^2}) \times {(g_{\rho L}^{12})^3 \, (g_{\rho L}^{33}) \, {\sin\theta_D}}/({(4\pi)^2 m_\rho^2})$.
At the focused point,
we thus have {$\text{Br}[\mu \to 3e] \sim 1 \times 10^{-34}$}, where
the latest bound is fairly above from the above estimated magnitude as $< 1.0 \times 10^{-12}\,(90\%\,\text{C.L.})$~\cite{Bellgardt:1987du}.

\item
$K_L \to \mu^{\mp} e^{\pm}$:

Similar to $\mu \to 3e$, our scenario does not yield any contribution when $\theta_L \sim \pi/2$ at the tree level.
A crude estimation on a typical size of the coefficients for the relevant operators, 
$\Delta_{K_0 \to e^- \mu^+} \, (\bar{e}'_L \gamma_\mu \mu'_L)(\bar{s}'_L \gamma^\mu s'_L)$ 
and other three at the one loop level, 
gives
$\sim ({m_c^2}/{m_\rho^2}) \times {g_W^2 V_{cs} V_{cd}^\ast (g_{\rho L}^{12}) \, (g_{\rho L}^{33}) \, \sin\theta_D}/({(4\pi)^2 m_\rho^2})$
(see {e.g.} \cite{Becirevic:2017jtw}).
We then find $\text{Br}[K_L \to \mu^{\mp} e^{\pm}] \sim 1 \times 10^{-23}$ at the focused point
with the inputs,
$V_{cs} = 0.974$,
$V_{cd} = 0.224$,
$f_{K} = 155.0(1.9)\,\text{MeV}$~\cite{Carrasco:2014poa},
$m_{K^0} = 497.6\,\text{MeV}$,
$\tau_{{K_L}} = 5.18 \times 10^{-8}\,\text{s}$~\cite{Tanabashi:2018oca}.
Thus this is totally safe
[c.f.~the latest bound $< 4.7 \times 10^{-12}\,(90\%\,\text{C.L.})$~\cite{Ambrose:1998us}].

\item
$\text{Br}[J/\psi \to \mu^+ \mu^-]/\text{Br}[J/\psi \to e^+ e^-] \ (\equiv R_{J/\psi \to 2\ell})$:

The general formula of the decay width for
$J/\psi \to 2\ell$ is read off from~\cite{Aloni:2017eny} and 
our estimation at the focused point goes like $R_{J/\psi \to 2\ell} \simeq R_{J/\psi \to 2\ell}|_\text{SM} \simeq 1$.
From the experimental results for the branching ratios, $\text{Br}[J/\psi(1S) \to e^+ e^-] = 0.05971 \pm 0.00032$ and
$\text{Br}[J/\psi(1S) \to \mu^+ \mu^-] = 0.05961 \pm 0.00033$~\cite{Tanabashi:2018oca},
we may estimate $R_{J/\psi \to 2\ell}|_\text{exp} = 0.9983 \pm 0.0078$.
Thus we see that no significant tension occurs from this observable.

\item
$\tau \to \mu \nu \bar{\nu}$ and $\mu \to e \nu \bar{\nu}$:

The SM can contribute to the process $\tau \to \mu \nu \bar{\nu}$ via the $W$-boson exchange at the tree level.
Also in our scenario, non-vanishing contribution shows up at the tree level even for $\theta_L \sim \pi/2$.
The additional contribution to the branching ratio for $\tau \to \mu \nu \bar{\nu}$ at the focused point is evaluated as $\sim 2 \times 10^{-9}$, while
no tree-level correction to $\mu \to e \nu \bar{\nu}$ remains when $\theta_L \sim \pi/2$.

According to~\cite{Pich:2013lsa,Kumar:2018kmr}, a tighter bound comes from the measurement of the following ratio:
\al{
R^{(\tau \to \mu) / (\mu \to e)}_{\nu \bar{\nu}} &\equiv 
\frac{ \text{Br}[\tau \to \mu \nu \bar{\nu}]/\text{Br}[\tau \to \mu \nu \bar{\nu}]_\text{SM} }
       { \text{Br}[\mu \to e \nu \bar{\nu}]/\text{Br}[\mu \to e \nu \bar{\nu}]_\text{SM} }, \notag
}
where the $2\sigma$ deviation from the SM was observed as $1.0060 \pm 0.0030$
with apparently $R^{(\tau \to \mu) / (\mu \to e)}_{\nu \bar{\nu}}|_\text{SM}$ being unity. 
By use of $\text{Br}[\tau \to \mu \nu \bar{\nu}]_\text{SM} \simeq \text{Br}[\tau \to \mu \nu \bar{\nu}]_\text{obs} = 0.1739$~\cite{Tanabashi:2018oca}, 
we reach
\al{
\frac{ \text{Br}[\tau \to \mu \nu \bar{\nu}] }{ \text{Br}[\tau \to \mu \nu \bar{\nu}]_\text{SM} } &=
1 + \frac{ \delta \text{Br}[\tau \to \mu \nu \bar{\nu}] }{ \text{Br}[\tau \to \mu \nu \bar{\nu}]_\text{SM} }
\sim 1 + 1 \times 10^{-8}, \notag \\
&\Rightarrow \ R^{(\tau \to \mu) / (\mu \to e)}_{\nu \bar{\nu}} \sim 1 + 1 \times 10^{-8}, \notag
}
which implies an extremely tiny effect, not to reach a tractable range in the future.

\item
Other lepton-flavor violating $\tau$ decays:

For example, the rare decays $\tau \to eee$ and $\tau \to e \mu \mu$ can be produced at the tree level
in a way similar to $\tau \to 3 \mu$ 
which were discussed in section~\ref{subsec:tauto3mu}. 
The magnitude of these two decay modes is much less 
than that for $\tau \to 3 \mu$ due to the smallness of the coupling 
$g_{\rho L}^{12}$ of $\mathcal{O}(10^{-3})$.
Thus, the CFV scenario is insensitive also to these rare decay processes.

\end{itemize}

\end{itemize}

\acknowledgments 

We are grateful to Teppei Kitahara for providing us with a numeric code for the NLO calculations, 
and Satoshi Mishima for giving us several useful comments.
K.N. thanks Chao-Qiang Geng, Hiroyuki Ishida and  Ryoutaro Watanabe for fruitful discussions.
This work was supported in part by the JSPS Grant-in-Aid for Young Scientists (B) \#15K17645 (S.M.), 
the JSPS KAKENHI 16H06492 and 18J01459 (K.Y.).
K.N. has been supported by the European Union through the European Regional Development Fund -- the Competitiveness and 
Cohesion Operational Programme (KK.01.1.1.06),
the European Union's Horizon 2020 research and
innovation program under the Twinning grant agreement No.~692194, RBI-T-WINNING,
and the grant funded from the European Structural and Investment Funds,
RBI-TWINN-SIN.
The authors thank the Yukawa Institute for Theoretical Physics at Kyoto University, where this work was initiated 
during the YITP-W-17-07 on ``Progress in Particle Physics 2017''.

\appendix

\section{Explicit expressions for CFV couplings to quarks and leptons}
\label{rho-repr} 
In this appendix we present the way of embedding CFVs into the $SU(8)$ 
multiplet and the CFV couplings to SM fermions.

The CFVs are embedded in the adjoint representation 
of the $SU(8)$ flavor symmetry, which are defined as 
\begin{eqnarray} 
\sum_{A=1}^{63} \rho^A \cdot T^A 
&=& 
\sum_{{\alpha}=1}^3 \sum_{a=1}^8 \rho^{{\alpha}}_{(8)a} \cdot T_{(8)a}^{{\alpha}} 
+ 
\sum_{a=1}^8 \rho_{(8)a}^0 \cdot T_{(8) a}  
\nonumber \\ 
&& 
+ 
\sum_{c=r,g,b} \sum_{{\alpha}=1}^3
\left[ {\rho_{(3)c}^{[1]\alpha} \cdot T_{(3) c}^{[1]\alpha} + 
       \rho_{(3)c}^{[2]\alpha} \cdot T_{(3) c}^{[2]\alpha}} \right] 
+
\sum_{c=r,g,b}
\left[ {\rho_{(3)c}^{[1]0} \cdot T_{(3) c}^{[1]} + 
       \rho_{(3)c}^{[2]0} \cdot T_{(3) c}^{[2]}} \right] 
\nonumber \\ 
&&
+ 
\sum_{{\alpha}=1}^3 \rho^{{\alpha}}_{(1)} \cdot T_{(1)}^{{\alpha}}  
+ 
\sum_{{\alpha}=1}^3 \rho^{{\alpha}}_{(1)'} \cdot T^{{\alpha}}_{(1)'} + \rho^0_{(1)'} \cdot T_{(1)'} 
\,, \label{A1}
\end{eqnarray} 
with  the $SU(8)$ generators, 
\begin{eqnarray} 
 T_{(8) a}^{{\alpha}}
 &=& 
\frac{1}{\sqrt{2}}  \left( 
\begin{array}{c|c} 
  \tau^{\alpha} \otimes \lambda^{{a}}  &  \\ 
  \hline 
   & {\bf 0_{2 \times 2}} 
\end{array}
\right)  \,, 
\qquad 
T_{(8)a} 
= 
\frac{1}{2 \sqrt{2}}
\left( 
\begin{array}{c|c} 
  {\bf 1}_{2\times 2} \otimes \lambda^{{a}}  &  \\ 
  \hline 
   & {\bf 0_{2 \times 2}}  
\end{array}
\right)  \,, \nonumber \\
{T^{[1]\alpha}_{(3) c}} 
&=&  
{
\frac{1}{\sqrt{2}}
\left( 
\begin{array}{c|c} 
 & \tau^{\alpha} \otimes {\bf e}_c    \\ 
  \hline 
\tau^{\alpha} \otimes {\bf e}_c^\dag  &  
\end{array}
\right) 
\,, \qquad
T^{[2]\alpha}_{(3) c} 
 =  
\frac{1}{\sqrt{2}}
\left( 
\begin{array}{c|c} 
 & -i  \tau^{\alpha} \otimes {\bf e}_c    \\ 
  \hline 
 i  \tau^{\alpha} \otimes {\bf e}_c^\dag  &  
\end{array}
\right)
\,,} \nonumber \\
{T^{[1]}_{(3) c}} 
&=&  
{
\frac{1}{2\sqrt{2}}
\left( 
\begin{array}{c|c} 
 & {\bf 1}_{2 \times 2} \otimes {\bf e}_c    \\ 
  \hline 
{\bf 1}_{2 \times 2} \otimes {\bf e}_c^\dag  &  
\end{array}
\right) 
\,, \qquad
T^{[2]}_{(3) c} 
 =  
\frac{1}{2\sqrt{2}}
\left( 
\begin{array}{c|c} 
 & -i {\bf 1}_{2 \times 2} \otimes {\bf e}_c    \\ 
  \hline 
 i {\bf 1}_{2 \times 2} \otimes {\bf e}_c^\dag  &  
\end{array}
\right)
\,,} \nonumber \\
 T_{(1)}^{{\alpha}} 
 &=& 
\frac{1}{2} \left( 
\begin{array}{c|c} 
  \tau^{{\alpha}} \otimes {\bf 1}_{3\times 3}  &  \\ 
  \hline 
   & \tau^{{\alpha}} 
\end{array}
\right)  \,, 
\qquad 
T^{{\alpha}}_{(1)'} 
= 
\frac{1}{2 \sqrt{3}}
\left( 
\begin{array}{c|c} 
  \tau^{{\alpha}} \otimes {\bf 1}_{3\times 3}  &  \\ 
  \hline 
   & -3 \cdot \tau^{{\alpha}} 
\end{array}
\right)  
\,, \nonumber \\
T_{(1)'}
&=& 
\frac{1}{4 \sqrt{3}}
\left( 
\begin{array}{c|c} 
  {\bf 1}_{6 \times 6}  &  \\ 
  \hline 
   & -3\cdot {\bf 1}_{2 \times 2} 
\end{array}
\right) 
\,, 
\end{eqnarray}
where $\tau^{{\alpha}} = \sigma^{{\alpha}}/2$ ($\sigma^{{\alpha}}$: Pauli {matrices}), {$\lambda^a$ and ${\bf e}_c$ represent the Gell-Mann matrices and three-dimensional unit vectors in color space, respectively,}
and the generator $T^A$ is normalized as ${\rm tr}[T^A T^B]=\delta^{AB}/2$.
For color-triplet components (leptoquarks), we define the following eigenforms which discriminate ${\bf 3}$ and ${\bf \bar{3}}$ states of the $SU(3)_c$ gauge group,
\al{
T^{\alpha}_{(3) c} &\equiv \frac{1}{\sqrt{2}} \left( T^{[1]\alpha}_{(3) c} +i T^{[2]\alpha}_{(3) c} \right) 
=
\left( 
\begin{array}{c|c} 
 & \tau^{\alpha} \otimes {\bf e}_c    \\ 
  \hline 
 {\bf 0_{2 \times 6}} &  
\end{array}
\right)\,, & 
T^{\alpha}_{(\bar{3}) c} &\equiv \left( T^{\alpha}_{(3) c} \right)^\dagger, \notag \\
T_{(3) c} &\equiv \frac{1}{\sqrt{2}} \left( T^{[1]}_{(3) c} +i T^{[2]}_{(3) c} \right) 
=
\frac{1}{2}
\left( 
\begin{array}{c|c} 
 & {\bf 1_{2 \times 2}} \otimes {\bf e}_c    \\ 
  \hline 
 {\bf 0_{2 \times 6}} &  
\end{array}
\right)\,, & 
T_{(\bar{3}) c} &\equiv \left( T_{(3) c} \right)^\dagger,  \label{eq:LQ_relation_1}
}
\vspace{-6mm}
\al{
\rho^\alpha_{(3) c} &\equiv \frac{1}{\sqrt{2}} \left( \rho^{[1]\alpha}_{(3) c} -i \rho^{[2]\alpha}_{(3) c} \right), \quad
\rho^0_{(3) c} \equiv \frac{1}{\sqrt{2}} \left( \rho^{[1]0}_{(3) c} -i \rho^{[2]0}_{(3) c} \right), \quad
\bar{\rho}^\alpha_{(3) c} \equiv \left( \rho^\alpha_{(3) c} \right)^\dagger, \quad
\bar{\rho}^0_{(3) c} \equiv \left( \rho^0_{(3) c} \right)^\dagger.
\nonumber
}

As in the text, the CFV fields ($\rho$) can be expressed by a couple of 
 sub-block matrices as 
\begin{equation} 
{\rho = 
\left( 
\begin{array}{cc} 
(\rho_{QQ})_{6\times 6} & {(\rho_{QL})_{6 \times 2}} \\ 
{(\rho_{LQ})_{2 \times 6}} & (\rho_{LL})_{2 \times 2} 
\end{array}
\right)}
\,, \label{rho:para}  
\end{equation}  
where the entries are read off from the above decomposition form 
as 
\begin{align}
 \rho_{QQ} = 
 &
 \left[ \sqrt 2 \rho_{(8)a}^{{\alpha}} \left( {\tau^\alpha} \otimes {\lambda^{{a}} \over 2} \right) 
 + {1 \over \sqrt 2} \rho_{(8)a}^0 \left( {\bf 1}_{2\times2} \otimes {\lambda^{{a}} \over 2} \right) \right] \notag \\[0.5em]
 & 
 + \left[ {1 \over 2} \rho_{(1)}^{{\alpha}} \left( {\tau^\alpha} \otimes {\bf 1}_{3\times3} \right) 
 + {1 \over 2\sqrt 3} \rho_{(1)'}^{{\alpha}} \left( {\tau^\alpha} \otimes {\bf 1}_{3\times3} \right) 
 + {1\over 4 \sqrt 3} \rho_{(1)'}^0 \Big( {\bf 1}_{2\times2} \otimes {\bf 1}_{3\times3} \Big) \right] \,, \notag \\[1em] 
 \rho_{LL} = 
 &
 {1 \over 2} \rho_{(1)}^{{\alpha}} \left( {\tau^\alpha} \right) - {\sqrt 3 \over 2} \rho_{(1)'}^{{\alpha}} \left( {\tau^\alpha} \right) - {\sqrt 3 \over 4} \rho_{(1)'}^0 \Big( {\bf 1}_{2\times2} \Big) \,, \notag \\[1em]
 \rho_{QL} = 
 &
 \rho_{(3){c}}^{{\alpha}} \left( {\tau^\alpha} \otimes {\bf e}_c \right) + {1 \over 2} \rho_{(3) {c}}^0 \Big( {\bf 1}_{2\times2} \otimes {\bf e}_c \Big) \,, \notag \\[1em]
 {\rho_{LQ}} = & \Big( {\rho_{QL}} \Big)^\dag \,.
 \label{rho:assign}  
\end{align}
{We reintroduce the above form for convenience, which appeared as Eq.(\ref{rho:assignment}) in the main body of this manuscript.}

Thus, by expanding coupling terms in Eq.(\ref{rhoLff}) and 
Eq.(\ref{V-rho}) with the SM gauge fields in Eq.(\ref{Vcal}) 
 in terms of the $\rho_{(8)}$s, $\rho_{(3)}$s and $\rho_{(1),(1)'}$s,   
the CFV couplings are found 
as listed below:  
\begin{itemize} 
\item 
flavorful (direct) couplings: \\
\\ 
for color-singlet neutral ($Z'$ type) CFVs ($\rho_{(1)}^{3,0}, \rho_{(1)'}^{3,0}$): 
\begin{align}
 &
 \label{eq:decrhouu1}
 \mathcal L_{\rho_{1}uu} =
{ -g_{\rho L}^{ij} } 
\bar u_L^i \gamma_\mu u_L^j \, \left[ +{1 \over 4}~\rho_{(1)3}^\mu +{1 \over 4\sqrt 3}~\rho_{(1)'3}^\mu +{1 \over 4\sqrt 3}~\rho_{(1)'0}^\mu \right] \,, \\[0.5em]
 &
 \mathcal L_{\rho_{1}dd} = { -g_{\rho L}^{ij} }  
 \bar d_L^i \gamma_\mu d_L^j \, \left[ -{1 \over 4}~\rho_{(1)3}^\mu -{1 \over 4\sqrt 3}~\rho_{(1)'3}^\mu +{1 \over 4\sqrt 3}~\rho_{(1)'0}^\mu \right] \,, 
\label{rho1-s-d}
 \\[0.5em]
 &
 \mathcal L_{\rho_{1}\ell\ell} ={ -g_{\rho L}^{ij} }  \bar e_L^i \gamma_\mu e_L^j \, \left[ -{1 \over 4}~\rho_{(1)3}^\mu +{\sqrt 3 \over 4}~\rho_{(1)'3}^\mu -{\sqrt 3 \over 4}~\rho_{(1)'0}^\mu \right] \,, \\[0.5em]
 &
 \mathcal L_{\rho_{1}\nu\nu} ={ -g_{\rho L}^{ij} }  \bar \nu_L^i \gamma_\mu \nu_L^j \, \left[ +{1 \over 4}~\rho_{(1)3}^\mu -{\sqrt 3 \over 4}~\rho_{(1)'3}^\mu -{\sqrt 3 \over 4}~\rho_{(1)'0}^\mu \right] \, ;   
\end{align}
for color-singlet charged ($W'$ type) CFVs ($\rho_{(1)}^{\pm}, \rho_{(1)'}^{\pm}$):  
\begin{align}
 &
 \mathcal L_{\rho_{1}ud} = { -g_{\rho L}^{ij} }  \bar u_L^i \gamma_\mu d_L^j \, \left[ {1\over 2 \sqrt 2} \rho_{(1)+}^\mu + {1\over 2 \sqrt 6} \rho_{(1)'+}^\mu  \right] + \text{h.c.} \,, \\[0.5em] 
 &
 \mathcal L_{\rho_{1}\nu\ell} = { -g_{\rho L}^{ij} }  \bar \nu_L^i \gamma_\mu e_L^j \, \left[ {1\over 2 \sqrt 2} \rho_{(1)+}^\mu - {\sqrt 3 \over 2 \sqrt 2} \rho_{(1)'+}^\mu  \right] + \text{h.c.} \,;  
\end{align}
for color-triplet (vector-leptoquark type) CFVs ($\rho_{(3)}^{\pm,3,0}$){:}
\begin{align}
 &
 \mathcal L_{\rho_{3}d\ell} = { -g_{\rho L}^{ij} }  \bar d_L^i \gamma_\mu e_L^j \, \left[ -{1 \over 2}~\rho_{(3)3}^\mu +{1 \over 2}~\rho_{(3)0}^\mu \right] + \text{h.c.} \,,  \\[0.5em]
 &
 \mathcal L_{\rho_{3}u\ell} = { -g_{\rho L}^{ij} } \bar u_L^i \gamma_\mu e_L^j \, \left[ +{1 \over \sqrt 2}~\rho_{(3)+}^\mu \right] + \text{h.c.} \,, \\[0.5em]
 &
 \mathcal L_{\rho_{3}d\nu} = { -g_{\rho L}^{ij} }  \bar d_L^i \gamma_\mu \nu_L^j \, \left[ + {1 \over \sqrt 2}~\rho_{(3)-}^\mu \right] + \text{h.c.} \,, \\[0.5em] 
 &
 \mathcal L_{\rho_{3}u\nu} = { -g_{\rho L}^{ij} } \bar u_L^i \gamma_\mu \nu_L^j \, \left[ +{1 \over 2}~\rho_{(3)3}^\mu +{1 \over 2}~\rho_{(3)0}^\mu \right] + \text{h.c.} \,,  
\end{align}
where $\rho_{(3)0}$ and $\rho_{(3)3}$ have $+{2 \over 3}$ electric charges whereas $\rho_{(3)\pm} = (\rho_{(3)1} \mp i\rho_{(3)2})/\sqrt 2$ have $+{5 \over 3}$ and $-{1 \over 3}$, respectively;  \\
\\
for color-octet ($G'$-type) CFVs $(\rho_{(8)}^{\pm,3,0})${:}
\begin{align}
 &
 \mathcal L_{\rho_{8}uu} = { -g_{\rho L}^{ij} }  \bar u_L^i \gamma_\mu \left(\lambda^a \over 2\right) u_L^j \left[ + {1\over\sqrt 2} \rho_{(8)3}^{a\mu} + {1\over \sqrt 2} \rho_{(8)0}^{a\mu} \right] \,, \\[0.5em]
 &
 \mathcal L_{\rho_{8}dd} = { -g_{\rho L}^{ij} } \bar d_L^i \gamma_\mu \left(\lambda^a \over 2\right) d_L^j \left[ - {1\over\sqrt 2} \rho_{(8)3}^{a\mu} + {1\over \sqrt 2} \rho_{(8)0}^{a\mu} \right] \,, \\[0.5em]
 &
 \label{eq:decrhouu8}
 \mathcal L_{\rho_{8}ud} = { -g_{\rho L}^{ij} } \bar u_L^i \gamma_\mu \left(\lambda^a \over 2\right) d_L^j \left[ + \rho_{(8)+}^{a\mu} \right] \,. 
\end{align}

\item
Indirect couplings induced from mixing with the SM gauge bosons{:} \\    
\\
Coupling Eq.(\ref{mass-mixing}) to the SM fermions via the SM gauge boson exchanges 
with the square of the transfer momentum $q^2 = m_\rho^2$ taken,  
we may evaluate   
the CFV-on-shell couplings to the SM fermions:
\begin{eqnarray} 
{\cal L}^{\rm indirect} 
&\approx& 
- \frac{1}{g_\rho}\Bigg[
\sqrt{2} g_s^2 \left(\bar{q} \gamma_\mu \rho_{(8)}^{\mu 0} q \right) 
+ 2 g_W^2 \left( \bar{q}_L \gamma_\mu \rho_{(1)}^\mu q_L \right) 
\nonumber \\ 
&& 
+ \frac{g_Y^2}{3\sqrt{3}} \left( \bar{q}_L \gamma_\mu \rho_{(1)'}^{\mu 0} q_L \right) 
+ \frac{2 g_Y^2}{\sqrt{3}} \left( \bar{q}_R \gamma_\mu Q_{\rm em}^q  \rho_{(1)'}^{\mu 0} q_R \right) 
\nonumber \\ 
&& 
+ 2 g_W^2 \left( \bar{l}_L \gamma_\mu \rho_{(1)}^\mu l_L \right) 
- \frac{g_Y^2}{\sqrt{3}} \left( \bar{l}_L \gamma_\mu \rho_{(1)'}^{\mu 0} l_L \right) 
+ \frac{2 g_Y^2}{\sqrt{3}} \left( \bar{l}_R \gamma_\mu Q_{\rm em}^l  \rho_{(1)'}^{\mu 0} l_R \right) 
\Bigg] 
\,, \label{indirect:rho-coupling}
\end{eqnarray}
where we have neglected terms suppressed by a factor of 
${\cal O}(m_{W/Z}/m_\rho)^2$. 
{Here we suppress the generation indices (in the gauge eigenbases) for simplicity since they are manifestly generation independent.}
{Also, the doublet-like notations are introduced for clarity; $q_R \equiv (u_R, d_R)^T$ and $l_R \equiv (\nu_R, e_R)^T$.}
\end{itemize}

\section{Effective four-fermion operators induced from CFV exchanges} 
\label{4-fermi}

In this appendix we derive effective four-fermion operators induced from 
the CFVs exchanges, relevant to discussing the flavor physics contributions.

Integrating out the CFVs coupled to the SM fermions with the $g_{\rho L}^{{ij}}$ in Eq.{(\ref{rhoLff})} together with Eq.(\ref{rho:assignment}) (or Eq.(\ref{rho:assign})) generate  
the following four-fermion operators at the mass scales of CFVs
\al{
-\mathcal{L}^{\text{(8)}}_\text{eff} &=
	\left(\sqrt{2}\right)^2 \frac{g^{{ij}}_{\rho L} g^{{kl}}_{\rho L}}{({M_{\rho^{{\alpha}}_{(8)}}})^2}
	{\Delta^{ik;jl}}
	\left[ (\overline{q}^{{i}}_L \gamma_\mu \tau^{\alpha} T^a q^{{j}}_L)
	       (\overline{q}^{{k}}_L \gamma^\mu \tau^{\alpha} T^a q^{{l}}_L) \right] \notag \\
&\, +
	\left(\frac{1}{\sqrt{2}}\right)^2 \frac{g^{{ij}}_{\rho L} g^{{kl}}_{\rho L}}{({M_{\rho^0_{(8)}}})^2}
	{\Delta^{ik;jl}}
	\left[ (\overline{q}^{{i}}_L \gamma_\mu T^a q^{{j}}_L)
	       (\overline{q}^{{k}}_L \gamma^\mu T^a q^{{l}}_L) \right], 
	\label{Lagrho8} \\[0.5em]
-\mathcal{L}^{\text{(1)}}_\text{eff} &=
	\left(\frac{1}{2}\right)^2 \frac{g^{{ij}}_{\rho L} g^{{kl}}_{\rho L}}{({M_{\rho^{{\alpha}}_{(1)}}})^2}
	{\Delta^{ik;jl}}
	\left[ (\overline{q}^{{i}}_L \gamma_\mu \tau^{\alpha} q^{{j}}_L)
	       (\overline{q}^{{k}}_L \gamma^\mu \tau^{\alpha} q^{{l}}_L) \right] \notag \\ 
&\, +
	\left(\frac{1}{2}\right)^2 \frac{g^{{ij}}_{\rho L} g^{{kl}}_{\rho L}}{({M_{\rho^{{\alpha}}_{(1)}}})^2}
	{\Delta^{ik;jl}}
	\left[ (\overline{l}^{{i}}_L \gamma_\mu \tau^{\alpha} l^{{j}}_L)
	       (\overline{l}^{{k}}_L \gamma^\mu \tau^{\alpha} l^{{l}}_L) \right] \notag \\
&\, +
	\left(\frac{1}{2}\right)^2 \frac{g^{{ij}}_{\rho L} g^{{kl}}_{\rho L}}{({M_{\rho^{{\alpha}}_{(1)}}})^2}
	\left[ (\overline{q}^{{i}}_L \gamma_\mu \tau^{\alpha} q^{{j}}_L)
	       (\overline{l}^{{k}}_L \gamma^\mu \tau^{\alpha} l^{{l}}_L) \right] \notag \\
&\, +
	\left(\frac{1}{2\sqrt{3}}\right)^2 \frac{g^{{ij}}_{\rho L} g^{{kl}}_{\rho L}}{({M_{\rho^{{\alpha}}_{{(1)'}}}})^2}
	{\Delta^{ik;jl}}
	\left[ (\overline{q}^{{i}}_L \gamma_\mu \tau^{\alpha} q^{{j}}_L)
	       (\overline{q}^{{k}}_L \gamma^\mu \tau^{\alpha} q^{{l}}_L) \right] \notag \\ 
&\, +
	\left(\frac{-\sqrt{3}}{2}\right)^2 \frac{g^{{ij}}_{\rho L} g^{{kl}}_{\rho L}}{({M_{\rho^{{\alpha}}_{{(1)'}}}})^2}
	{\Delta^{ik;jl}}
	\left[ (\overline{l}^{{i}}_L \gamma_\mu \tau^{\alpha} l^{{j}}_L)
	       (\overline{l}^{{k}}_L \gamma^\mu \tau^{\alpha} l^{{l}}_L) \right] \notag \\
&\, +
	\left(\frac{1}{2\sqrt{3}}\right)
	\left(\frac{-\sqrt{3}}{2}\right) \frac{g^{{ij}}_{\rho L} g^{{kl}}_{\rho L}}{({M_{\rho^{{\alpha}}_{{(1)'}}}})^2}
	\left[ (\overline{q}^{{i}}_L \gamma_\mu \tau^{\alpha} q^{{j}}_L)
	       (\overline{l}^{{k}}_L \gamma^\mu \tau^{\alpha} l^{{l}}_L) \right] \notag \\
&\, +
	\left(\frac{1}{4\sqrt{3}}\right)^2 \frac{g^{{ij}}_{\rho L} g^{{kl}}_{\rho L}}{({M_{\rho^0_{{(1)'}}}})^2}
	{\Delta^{ik;jl}}
	\left[ (\overline{q}^{{i}}_L \gamma_\mu q^{{j}}_L)
	       (\overline{q}^{{k}}_L \gamma^\mu q^{{l}}_L) \right] \notag \\
&\, +
	\left(\frac{-\sqrt{3}}{4}\right)^2 \frac{g^{{ij}}_{\rho L} g^{{kl}}_{\rho L}}{({M_{\rho^0_{{(1)'}}}})^2}
	{\Delta^{ik;jl}}
	\left[ (\overline{l}^{{i}}_L \gamma_\mu l^{{j}}_L)
	       (\overline{l}^{{k}}_L \gamma^\mu l^{{l}}_L) \right] \notag \\
&\, +
	\left(\frac{1}{4\sqrt{3}}\right)
	\left(\frac{-\sqrt{3}}{4}\right) \frac{g^{{ij}}_{\rho L} g^{{kl}}_{\rho L}}{({M_{\rho^0_{{(1)'}}}})^2}
	\left[ (\overline{q}^{{i}}_L \gamma_\mu q^{{j}}_L)
	       (\overline{l}^{{k}}_L \gamma^\mu l^{{l}}_L) \right], 
	\label{Lagrho1} \\[0.5em]
-\mathcal{L}^{\text{(3)}}_\text{eff} &=
	\frac{g^{{ij}}_{\rho L} g^{{kl}}_{\rho L}}{({M_{\rho^{{\alpha}}_{(3)}}})^2}
	\left[ (\overline{q}^{{i}}_L \gamma_\mu \tau^{\alpha} l^{{j}}_L)
	       (\overline{l}^{{k}}_L \gamma^\mu \tau^{\alpha} q^{{l}}_L) \right] +
	\left(\frac{1}{2}\right)^2
	\frac{g^{{ij}}_{\rho L} g^{{kl}}_{\rho L}}{({M_{\rho^0_{(3)}}})^2}
	\left[ (\overline{q}^{i}_L \gamma_\mu l^{j}_L)(\overline{l}^{k}_L \gamma^\mu q^{l}_L) \right]{,}
	\label{Lagrho3}
}
with $T^a \equiv \lambda^a/2$.
Here, {$\Delta^{ik;jl}$ is} a combinatorics factor, which satisfies 
\al{
{
\Delta^{ik;jl} \left(= \Delta^{ki;jl} = \Delta^{ik;lj} = \Delta^{ki;lj}\right) = 
	\begin{cases}
	1/2 & \text{for } i=k \text{ and } j=l, \\
	1   & \text{for } \text{others}.
	\end{cases}}
	\label{eq:Delta_factor}
} 
After Fiertz transformations, 
we have 
\al{
-\mathcal{L}_\text{eff}^{(8)+(3)+(1)} \Bigg|_{\rm Fiertz} 
&= 
	C^{[3]}_{q_{{i}} q_{{j}} q_{{k}} q_{{l}}}
	(\overline{q}^{{i}}_L \gamma_\mu \sigma^{{\alpha}} q^{{j}}_L)
	(\overline{q}^{{k}}_L \gamma^\mu \sigma^{{\alpha}} q^{{l}}_L) +
	C^{[3]}_{l_{{i}} l_{{j}} l_{{k}} l_{{l}}}
	(\overline{l}^{{i}}_L \gamma_\mu \sigma^{{\alpha}} l^{{j}}_L)
	(\overline{l}^{{k}}_L \gamma^\mu \sigma^{{\alpha}} l^{{l}}_L) \notag \\
&\, +
	C^{[3]}_{q_{{i}} q_{{j}} l_{{k}} l_{{l}}}
	(\overline{q}^{{i}}_L \gamma_\mu \sigma^{{\alpha}} q^{{j}}_L)
	(\overline{l}^{{k}}_L \gamma^\mu \sigma^{{\alpha}} l^{{l}}_L) +
	C^{[1]}_{q_{{i}} q_{{j}} q_{{k}} q_{{l}}}
	(\overline{q}^{{i}}_L \gamma_\mu q^{{j}}_L)
	(\overline{q}^{{k}}_L \gamma^\mu q^{{l}}_L) \notag \\
&\, +
	C^{[1]}_{l_{{i}} l_{{j}} l_{{k}} l_{{l}}}
	(\overline{l}^{{i}}_L \gamma_\mu l^{{j}}_L)
	(\overline{l}^{{k}}_L \gamma^\mu l^{{l}}_L) +
	C^{[1]}_{q_{{i}} q_{{j}} l_{{k}} l_{{l}}}
	(\overline{q}^{{i}}_L \gamma_\mu q^{{j}}_L)
	(\overline{l}^{{k}}_L \gamma^\mu l^{{l}}_L). 
	\label{B5}
}
Including the indirect coupling contributions arising from Eq.(\ref{indirect:rho-coupling}) 
(excluding the relatively small right-handed fermion couplings), the Wilson coefficients are evaluated as 
\al{
C^{[3]}_{q_{{i}} q_{{j}} q_{{k}} q_{{l}}} &= 
	{\Delta^{ik;jl}} \Bigg\{
	{\frac{1}{2}}
	\left[ \frac{1}{2} \alpha^{{il;kj}} - \frac{1}{6} \alpha^{{ij;kl}} \right] \frac{1}{({M_{\rho^\alpha_{(8)}}})^2} +
	{\frac{1}{16}} \frac{ [\alpha^{{ij;kl}}]_{\rho_{(1)}^\alpha} }{({M_{\rho^\alpha_{(1)}}})^2} +
	{\frac{1}{48}} \frac{\alpha^{{ij;kl}}}{({M_{\rho^\alpha_{{(1)'}}}})^2}
	\Bigg\}, \\[0.5em]
C^{[3]}_{l_{{i}} l_{{j}} l_{{k}} l_{{l}}} &=
	{\Delta^{ik;jl}} \Bigg\{
	{\frac{1}{16}} \frac{ [\alpha^{{ij;kl}} ]_{\rho_{(1)}^\alpha } }{({M_{\rho^\alpha_{(1)}}})^2} +
	{\frac{3}{16}} \frac{\alpha^{{ij;kl}}}{({M_{\rho^\alpha_{{(1)'}}}})^2}
	\Bigg\}, \\[0.5em]
C^{[3]}_{q_{{i}} q_{{j}} l_{{k}} l_{{l}}} &=
	{\frac{1}{16}} \frac{ [\alpha^{{ij;kl}}]_{\rho_{(1)}^\alpha } }{({M_{\rho^\alpha_{(1)}}})^2} -
	{\frac{1}{16}} \frac{\alpha^{{ij;kl}}}{({M_{\rho^\alpha_{{(1)'}}}})^2} -
	{\frac{1}{8}} \frac{\beta^{{il;kj}}}{({M_{\rho^\alpha_{(3)}}})^2} +
	{\frac{1}{8}} \frac{\beta^{{il;kj}}}{({M_{\rho^0_{(3)}}})^2}, \\[0.5em]
C^{[1]}_{q_{{i}} q_{{j}} q_{{k}} q_{{l}}} &= 
	{\Delta^{ik;jl}} \Bigg\{
	{\frac{1}{2}}
	\left[ \frac{1}{2} [\alpha^{{il;kj}}]_{\rho_{(8)}^0 } - \frac{1}{6} [\alpha^{{ij;kl}}]_{\rho_{(8)}^0 } \right] \frac{1}{({M_{\rho^0_{(8)}}})^2} +
	{\frac{1}{48}} \frac{ [\alpha^{{ij;kl}}]_{\rho_{(1)'}^0 } }{({M_{\rho^0_{{(1)'}}}})^2}
	\Bigg\}, \\[0.5em]
C^{[1]}_{l_{{i}} l_{{j}} l_{{k}} l_{{l}}} &=
	{\Delta^{ik;jl}} \Bigg\{
	{\frac{3}{16}} \frac{ [\alpha^{{ij;kl}}]_{\rho_{(1)'}^0 } }{({M_{\rho^0_{{(1)'}}}})^2}
	\Bigg\}, \\[0.5em]
C^{[1]}_{q_{{i}} q_{{j}} l_{{k}} l_{{l}}} &=
	{- \frac{1}{16}}
	\frac{ [\alpha^{{ij;kl}}]_{\rho_{(1)'}^0 } }{({M_{\rho^0_{{(1)'}}}})^2} +
	{\frac{3}{8}} \frac{\beta^{{il;kj}}}{({M_{\rho^{{\alpha}}_{(3)}}})^2} +
	{\frac{1}{8}} \frac{\beta^{{il;kj}}}{({M_{\rho^0_{(3)}}})^2}, \label{C1:qqll}
}
where 
\al{
\alpha^{{ij;kl}} &\equiv g_{\rho L}^{{ij}} g_{\rho L}^{{kl}}, \\
\beta^{{ij;kl}}  &\equiv g_{\rho L}^{{ij}} (g_{\rho L}^\dagger)^{{kl}}, \\ 
[\alpha^{{ij;kl}} ]_{\rho_{(1)}^\alpha } &\equiv 
\left( g_{\rho L}^{{ij}} + \frac{4 g_W^2}{g_\rho} \delta^{ij} \right) \left( g_{\rho L}^{{kl}} + \frac{4 g_W^2}{g_\rho} \delta^{kl}  \right) \,, \\  
[\alpha^{{ij;kl}} ]_{\rho_{(1)'}^0 } & \equiv 
\left( g_{\rho L}^{{ij}} + \frac{4 g_Y^2}{3 g_\rho} \delta^{ij} \right) \left( g_{\rho L}^{{kl}} + \frac{4 g_Y^2}{3 g_\rho} \delta^{kl}  \right)  \,, \\ 
[\alpha^{{ij;kl}} ]_{\rho_{(8)}^\alpha } &\equiv  
\left( g_{\rho L}^{{ij}} + \frac{2 g_s^2}{g_\rho} \delta^{ij} \right) \left( g_{\rho L}^{{kl}} + \frac{2 g_s^2}{g_\rho} \delta^{kl}  \right)  \,. 
}
In terms of the $(u,d)_L^i$ and $(\nu, e)_L^i$ fields, 
we thus find 
\al{
-\mathcal{L}_\text{eff}^{(8)+(3)+(1)} \Bigg|_{\rm Fiertz}  
& \supset
	\left( C^{[1]}_{q_{{i}} q_{{j}} l_{{k}} l_{{l}}} +
	       C^{[3]}_{q_{{i}} q_{{j}} l_{{k}} l_{{l}}} \right)
	(\overline{d}^{{i}}_L \gamma_\mu d^{{j}}_L)
	(\overline{e}^{{k}}_L \gamma^\mu e^{{l}}_L) 
	\notag\\ 
&	+
	\left( C^{[1]}_{q_{{i}} q_{{j}} l_{{k}} l_{{l}}} - 
	       C^{[3]}_{q_{{i}} q_{{j}} l_{{k}} l_{{l}}} \right)
	(\overline{d}^{{i}}_L \gamma_\mu d^{{j}}_L)
	(\overline{\nu}^{{k}}_L \gamma^\mu \nu^{{l}}_L) \notag \\
&+
	2 C^{[3]}_{q_{{i}} q_{{j}} l_{{k}} l_{{l}}} \left(
	(\overline{u}^{{i}}_L \gamma_\mu d^{{j}}_L)
	(\overline{e}^{{k}}_L \gamma^\mu \nu^{{l}}_L) +
	(\overline{d}^{{i}}_L \gamma_\mu u^{{j}}_L)
	(\overline{\nu}^{{k}}_L \gamma^\mu e^{{l}}_L) \right) \notag \\
&+
	\left( C^{[1]}_{q_{{i}} q_{{j}} q_{{k}} q_{{l}}} + 
	       C^{[3]}_{q_{{i}} q_{{j}} q_{{k}} q_{{l}}} \right)
	(\overline{d}^{{i}}_L \gamma_\mu d^{{j}}_L)
	(\overline{d}^{{k}}_L \gamma^\mu d^{{l}}_L) 
	\notag\\ 
	&+
	\left( C^{[1]}_{l_{{i}} l_{{j}} l_{{k}} l_{{l}}} + 
	       C^{[3]}_{l_{{i}} l_{{j}} l_{{k}} l_{{l}}} \right)
	(\overline{e}^{{i}}_L \gamma_\mu e^{{j}}_L)
	(\overline{e}^{{k}}_L \gamma^\mu e^{{l}}_L).
	\label{eq:effective_operators}
}

\section{Global significance of a benchmark point \label{sec:global-fit}}
  
We estimate the global significance of our benchmark point where
$m_\rho = 1\,\text{TeV}$, $g_\rho = 8$, $\theta_D/\pi = 2 \times 10^{-3}$, $\theta_L/\pi \sim 1/2$, $g_{\rho L}^{33} \sim +0.6 \text{ or } -0.6$,
and $|g_{\rho L}^{12}| \sim 2 \times 10^{-3}$, at which 
the best-fit value for $b \to s \mu^+ \mu^-$ is achieved and
the $\epsilon'/\epsilon$ anomaly can be addressed within $1\sigma$ C.L. consistently (see Figs.~\ref{B-tau-sys-cons-SU(8)inv-grhoL-rev}
and \ref{K-sys-cons-SU(8)inv-grhoL-rev}).
Note also that at this benchmark point, 
the CFV effect on violation of LFU is minimized, while the CFV couplings to dimuon for addressing the 
$b \to s \mu^+ \mu^-$ anomaly are maximized.

To facilitate the analysis, it is convenient to introduce the following quantities:
\al{
R_X^{l_1/l_2} &\equiv \frac{ \text{Br}[\bar{B} \to X l_1 \bar{\nu}_{l_1}] }{ \text{Br}[\bar{B} \to X l_2 \bar{\nu}_{l_2}] }
\quad (X = D,\,D^\ast,\,\text{or}\, J/\psi), \\
R^{(\tau \to \mu) / (\mu \to e)}_{\nu \bar{\nu}} &\equiv 
\frac{ \text{Br}[\tau \to \mu \nu \bar{\nu}]/\text{Br}[\tau \to \mu \nu \bar{\nu}]_\text{SM} }
       { \text{Br}[\mu \to e \nu \bar{\nu}]/\text{Br}[\mu \to e \nu \bar{\nu}]_\text{SM} }.
       \label{eq:def_lepton_doubleratio}
}
We calculate the $\chi^2$ values for the 20 observables listed in Table~\ref{tab:chi_square},
where a part of them is related to $e$-flavor violation and their details are provided in section~\ref{sec:summary}.
Several comments on our $\chi^2$ test are in order:
\begin{itemize}
\item
We choose the leptonic angle $\theta_L$, 
the down-quark angle $\theta_D$ and $g_{\rho L}^{33}$
as relevant fit parameters and the others are treated to be fixed.
Thereby, the number of the degrees of freedom (d.o.f.) is $17$.
\item
As adopted in~\cite{Kumar:2018kmr}, for the variables to which only the $90\%\,\text{C.L.}$ bounds are available,
we assume those central values to be zero and define the $1
\sigma$ error from the $90\%\,\text{C.L.}$  upper bounds by simply assuming the Gaussian form.
\item
The quadrature form is used in combining errors of variables.
When the error is asymmetric, we simply adopt the average of the (combined) errors to estimate $\chi^2$.
\item
Actually, the present CFV scenario yields  somewhat large partial $\chi^2$ values, 9.0, 5.8, 0.64 and 2.9, 
with respect to the $R_{D^{(*)}}$-associate experimental results,
$(R^{\tau/\ell}_{D^\ast})/(R^{\tau/\ell}_{D^\ast})_\text{SM}$ ($1.18 \pm 0.06$~\cite{Lees:2013uzd,Huschle:2015rga,Aaij:2015yra,Abdesselam:2016xqt}),
$(R^{\tau/\ell}_{D})/(R^{\tau/\ell}_{D})_\text{SM}$ ($1.36 \pm 0.15$~\cite{Lees:2013uzd,Huschle:2015rga,Aaij:2015yra,Abdesselam:2016xqt}),
$(R^{e/\mu}_{D^\ast})/(R^{e/\mu}_{D^\ast})_\text{SM}$ ($1.04 \pm 0.05$~\cite{Abdesselam:2017kjf})
and
$(R^{\tau/\mu}_{J/\psi})/(R^{\tau/\mu}_{J/\psi})_\text{SM}$ ($2.51 \pm 0.97$~\cite{Aaij:2017tyk}), respectively.
Such large numbers have been obtained due to the approximate $SU(8)$ flavor symmetry which prohibits significantly exceeding from the SM predictions (see section~\ref{sec:RD-anomaly} for details). Therefore, we have removed those observables from our global fit. 
\item
Unavoidable theoretical {uncertainties remain} in many of observables
in the $K$ and $D$ meson physics.
We simply ignore them in the estimate on the global significance.
\end{itemize}
Thus, we can easily get
\al{
\left. \frac{\chi^2_\text{total}}{(\text{d.o.f.})} \right|_\text{at the benchmark} \simeq \frac{14.4}{17} \simeq 0.85,
}
which is less than unity and a completely acceptable goodness-of-fit at the benchmark point ($p$-value $\simeq 0.64$) is realized, 
even though sizable 
worse fitness comes from
$N_\nu$ and $R^{(\tau \to \mu) / (\mu \to e)}_{\nu \bar{\nu}}$, 
where this kind of worse fitness can be seen also in the SM case.

\begin{table}
\hspace{-50pt}
\begin{tabular}{|c|c|c|} \hline \hline
Observable & Experimental result & $\chi^2$ at the BP \\ \hline \hline
$B$ anomaly && \\ \hline
$b \to s \mu^+ \mu^-$ (combined) & $C^{\mu\mu}_9(\text{NP}) = - C^{\mu\mu}_{10}(\text{NP}) = -0.615 \pm 0.13$~\cite{Capdevila:2017bsm}
& $\sim 0$ \\ \hline \hline
$2 \leftrightarrow 3$ flavor transitions && \\
(see Sections~\ref{sec:2and3-transition} and \ref{sec:summary}) && \\ \hline
$\text{Br}[\bar{B} \to K^{(\ast)} \nu \bar{\nu}]/\text{Br}[\bar{B} \to K^{(\ast)} \nu \bar{\nu}]_\text{SM}$ &
$-13 \sum_{I=1}^{3} \text{Re}[C^{II}_L(\text{NP})] + \sum_{I,J=1}^{3} |C^{IJ}_L(\text{NP})|^2$ &  \\
& $\leq 473$ (90\%\,C.L.)~\cite{Bhattacharya:2016mcc} & $\lesssim 10^{-4}$ \\
$\text{Br}[\tau \to \phi \mu]$ & $< 8.4 \times 10^{-8}$ (90\%\,C.L.)~\cite{Miyazaki:2011xe} & $\sim 0$ \\
$\text{Br}[\tau \to 3 \mu]$ & $< 2.1\times10^{-8}$ (90\%\,C.L.)~\cite{Hayasaka:2010np} & $\sim 0$ \\
$\Delta M_{B_s}$ & $\Delta M_{B_s}^\text{exp}$ = $(17.757 \pm 0.021) \, \text{ps}^{-1}$~\cite{Amhis:2014hma} & $\sim 4$ \\
$\text{Br}[B^+ \to K^+ \tau^- \mu^+]$ & $0.8^{+1.9}_{-1.4} \times 10^{-5}; \quad < 4.5 \times 10^{-5}$ (90\%\,C.L.)~\cite{Lees:2012zz} & $0.22$ \\
$\text{Br}[B^+ \to K^+ \tau^+ \mu^-]$ & $(-0.4)^{+1.4}_{-0.9} \times 10^{-5}; \quad < 2.8 \times 10^{-5}$ (90\%\,C.L.)~\cite{Lees:2012zz} & $0.11$ \\
$\text{Br}[\Upsilon(2S) \to \mu^\pm \tau^\mp]$ & $0.2^{+1.5+1.0}_{-1.3-1.2} \times 10^{-6}; \quad < 3.3 \times 10^{-6}$ (90\%\,C.L.)~\cite{Lees:2010jk} & $0.012$ \\
$\text{Br}[\Upsilon(3S) \to \mu^\pm \tau^\mp]$ & $(-0.8)^{+1.5+1.4}_{-1.5-1.3} \times 10^{-6}; \quad < 3.1 \times 10^{-6}$ (90\%\,C.L.)~\cite{Lees:2010jk} & $0.16$ \\
$\text{Br}[J/\psi \to \mu^\pm \tau^\mp]$ & $< 2.0 \times 10^{-6}$ (90\%\,C.L.)~\cite{Ablikim:2004nn} & $\sim 0$ \\ \hline \hline
LFV associated with $e$ flavor && \\ 
(see Section~\ref{sec:summary}) && \\ \hline
$v_\mu/v_e$ & $0.961(61)$~\cite{Tanabashi:2018oca} & $\sim 0.5$ \\
$a_\mu/a_e$ & $1.0002(13)$~\cite{Tanabashi:2018oca} & $\sim 0.02$ \\
$N_\nu$ & $2.9840 \pm 0.0082$~\cite{Tanabashi:2018oca} & $\sim 4$ \\
$\Gamma(\Upsilon(1S) \to \tau^+ \tau^-)/\Gamma(\Upsilon(1S) \to \ell^+ \ell^-)$ & $1.005 \pm 0.013 \pm 0.022$~\cite{Besson:2006gj,delAmoSanchez:2010bt} & $\sim 0.3$ \\
$\Gamma(\Upsilon(2S) \to \tau^+ \tau^-)/\Gamma(\Upsilon(2S) \to \ell^+ \ell^-)$ & $1.04 \pm 0.04 \pm 0.05$~\cite{Besson:2006gj,delAmoSanchez:2010bt} & $\sim 0.6$ \\
$\Gamma(\Upsilon(3S) \to \tau^+ \tau^-)/\Gamma(\Upsilon(3S) \to \ell^+ \ell^-)$ & $1.05 \pm 0.08 \pm 0.05$~\cite{Besson:2006gj,delAmoSanchez:2010bt} & $\sim 0.4$ \\
$R^{(\tau \to \mu) / (\mu \to e)}_{\nu \bar{\nu}}$ & $1.0060 \pm 0.0030$~\cite{Pich:2013lsa} & $\sim 4$ \\
$\text{Br}[\mu \to 3 e]$ (at 1 loop) & $< 1.0 \times 10^{-12}\,(90\%\,\text{C.L.})$~\cite{Bellgardt:1987du} & $\sim 0$ \\
$\text{Br}[\mu \to e \gamma]$ (at 1 loop) & $< 4.2 \times 10^{-13}\,(90\%\,\text{C.L.})$~\cite{TheMEG:2016wtm} & $\sim 0$ \\ 
$\text{Br}[J/\psi \to \mu^+ \mu^-]/\text{Br}[J/\psi \to e^+ e^-]$ & $0.9983 \pm 0.0078$ & $\sim 0.05$ \\ 
& (evaluated for $J/\Psi(1S)$ from the data in~\cite{Tanabashi:2018oca}) & \\ \hline
\end{tabular}
\caption{List of observables used for our $\chi^2$ calculation and individual $\chi^2$ values at the benchmark point~(BP):
$m_\rho = 1\,\text{TeV}$, $g_\rho = 8$, $\theta_D/\pi = 2 \times 10^{-3}$, $\theta_L/\pi \sim 1/2$, $g_{\rho L}^{33} \sim +0.6 \text{ or } -0.6$,
and $|g_{\rho L}^{12}| \sim 2 \times 10^{-3}$, where 
the best-fit value {for} $b \to s \mu^+ \mu^-$ is achieved and
the $\epsilon'/\epsilon$ anomaly can be addressed within $1\sigma$ C.L. consistently (see Figs.~\ref{B-tau-sys-cons-SU(8)inv-grhoL-rev}
and \ref{K-sys-cons-SU(8)inv-grhoL-rev}).}
\label{tab:chi_square}
\end{table}

\bibliographystyle{JHEP}
\bibliography{eps-refs}

\end{document}